\documentclass[preprint,11pt]{elsarticle}

\usepackage{xcolor}
\usepackage{booktabs, multirow, longtable}
\usepackage{tikz-cd}

\usepackage{graphicx, epstopdf}
\usepackage{caption}
\usepackage{subcaption}
\usepackage{morefloats}

\usepackage{amsmath, amssymb, amsthm}
\usepackage{extarrows,mathtools}
\usepackage{bbm}
\usepackage{cases}
\usepackage{array}

\usepackage{natbib}  

\usepackage{hyperref}
\usepackage{cleveref}

\usepackage{chemmacros}

\usepackage{framed} 
\usepackage{multicol} 

\usepackage{nomencl} 
\makenomenclature
\renewcommand*\nompreamble{\begin{multicols}{2}}
\renewcommand*\nompostamble{\end{multicols}}

\newcommand{\R}{\mathbb{R}}
\newcommand{\ud}{\mathrm{d}}

\newcommand{\Par}{\mathrm{\partial}}
\newcommand{\E}{\mathbb{E}}
\newcommand{\T}{^\mathrm{T}}

\newcommand{\mbs}[1]{\ensuremath{\boldsymbol{#1}}}

\newcommand{\PDer}[2]{\frac{\Par #1}{\Par #2}}
\newcommand{\Der}[2]{\frac{\ud #1}{\ud #2}}

\newcommand{\Rstr}[1]{\left. #1 \right|}

\DeclareMathOperator*{\argmin}{arg\,min}
\DeclareMathOperator*{\argmax}{arg\,max}

\advance \topmargin by -\headheight
\advance \topmargin by -\headsep
\evensidemargin \oddsidemargin
\journal{Journal of Power Sources}

\begin{document}

\begin{frontmatter}



\title{Identifiability Study of Lithium-Ion Battery Capacity Fade
Using Degradation Mode Sensitivity
for a Minimally and Intuitively Parametrized Electrode-Specific Cell Open-Circuit Voltage Model}


\author[I2R]{Jing Lin}
\ead{lin_jing@ihpc.a-star.edu.sg}

\author[I2R]{Edwin Khoo\corref{EK}}
\ead{edwin_khoo@ihpc.a-star.edu.sg}
\cortext[EK]{Corresponding author.}

\affiliation[I2R]{
  organization={Institute for Infocomm Research (I2R),
  Agency for Science, Technology and Research (A*STAR)},
  addressline={1 Fusionopolis Way, \#21-01 Connexis South}, 
  postcode={Singapore 138632}, 
  country={Republic of Singapore},
}


\begin{abstract}
  When two electrode open-circuit potentials form a full-cell OCV (open-circuit voltage) model,
  cell-level SOH (state of health) parameters related to LLI (loss of lithium inventory)
  and LAM (loss of active materials) naturally appear.
  Such models have been used to
  interpret experimental OCV measurements and infer these SOH parameters
  associated with capacity fade.
  In this work, we first re-parametrize a popular OCV model formulation
  by the N/P (negative-to-positive) ratio and Li/P (lithium-to-positive) ratio,
  which have more symmetric and intuitive physical meaning,
  and are also pristine-condition-agnostic and cutoff-voltage-independent.
  We then study the modal identifiability of capacity fade
  by mathematically deriving the gradients of electrode slippage and cell OCV
  with respect to these SOH parameters,
  where the electrode differential voltage fractions,
  which characterize each electrode's relative contribution to the OCV slope,
  play a key role in passing the influence of a fixed cutoff voltage to the parameter sensitivity.
  The sensitivity gradients of the total capacity also reveal four characteristic regimes
  regarding how much lithium inventory and active materials are limiting the apparent capacity.
  We show the usefulness of these sensitivity gradients with an application
  regarding degradation mode identifiability from OCV measurements
  at different SOC (state of charge) windows.
  \end{abstract}

\begin{graphicalabstract}
\includegraphics[width=0.6\textwidth]{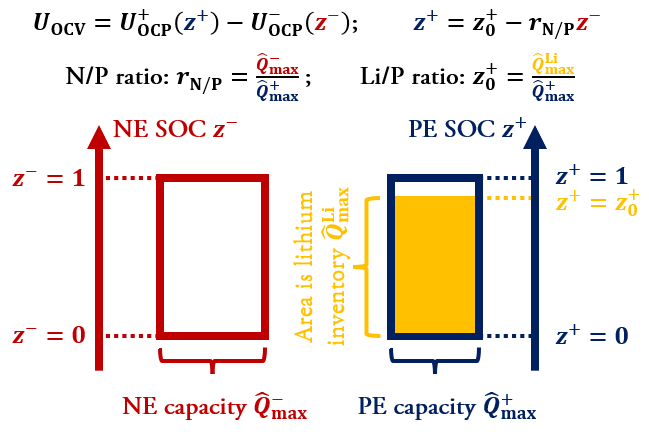}
\end{graphicalabstract}

\begin{highlights}
\item An electrode-specific OCV model parametrized by N/P and Li/P ratio.
\item Degradation modes and electrode SOC based on material-specific usable stoichiometry range.
\item Electrode differential voltage fractions indicating the limiting electrode.
\item Four regimes of degradation identifiability characterized by Li/N and Li/P ratio.
\item Informative SOC windows for degradation mode estimation by Fisher information.
\end{highlights}


\end{frontmatter}


\section{Introduction: Characterizing Degradation by SOH parameters}
\label{sec:intro}

Lithium-ion batteries degrade over time upon operation and storage
due to many complex and coupled physicochemical mechanisms and processes
\citep{edge_lithium_2021,attia_reviewknees_2022}.
Degradation causes a battery cell to lose capacity and rate capability,
where the former represents charge throughput
between the cell's fully charged and discharged state,
while the latter indicates the maximum charge/discharge C-rate
the cell can withhold within a short period.
These two aspects of aging are usually called capacity and power fade,
respectively \citep{birkl_degradation_2017}.

One technical subtlety here is that the fully charged and discharged state is defined by
the cell OCV (open-circuit voltage) reaching
a pre-specified upper and lower cutoff voltage $U_\mathrm{max}$ and $U_\mathrm{min}$, respectively,
and the charge throughput between these two states is named
the total capacity $\hat{Q}_\mathrm{max}$ and is what capacity fade specifically concerns
\citep{plett_battery_2015}.
However, under finite-current operation,
the observed terminal voltage shall reach a cutoff voltage before the underlying OCV does
due to polarization in non-equilibrium,
so the commonly reported charge/discharge capacity during cycling
can be significantly smaller than the total capacity,
unless both charging and discharging have a trailing CV (constant-voltage) phase
with a small enough cutoff current.
The key is that the cell must have reached quasi-equilibrium
at $U_\mathrm{max}$ and $U_\mathrm{min}$
before and after the Coulomb counting, respectively.
The total capacity depends only on the thermodynamic properties of the cell,
while the charge/discharge capacity also depends on the kinetic characteristics and operation protocols.
Therefore, the latter's reduction upon aging is due to both capacity and power fade.
It is not uncommon to see a charge/discharge capacity being confused with the total capacity
in literature,
but the distinction is important for the various modes of capacity fade.

A battery cell's voltage response to an input current profile
can be captured and predicted by a cell model,
be it physics-based and derived from first principles,
or semi-empirical and approximated by an equivalent circuit \citep{plett_battery_2015}.
Therefore, within such a model,
a cell's capacity and rate capability are completely encoded by certain model parameters,
which we will refer to as SOH (state of health) parameters,
and the granularity of this parametrization depends on the model's complexity.
Such SOH parameters interface between the underlying physicochemical degradation mechanisms
and the apparent cell performance metrics,
so the performance decay caused by the change of a particular parameter
is also called a degradation mode \citep{birkl_degradation_2017}.
If such SOH parameters can be identified from observed cell behaviors,
we can track how they evolve across a cell's lifetime under different operating conditions,
which might provide valuable clues about dominant aging mechanisms,
inspire the formulation of more accurate degradation models,
enable more effective aging-aware battery management and control,
and even guide battery design to attenuate degradation.

Although we treat SOH here as a loose term that denotes a cell's capacity and rate capability,
whose precise meaning depends on the context of a particular cell model,
SOH historically often refers to the total capacity $\hat{Q}_\mathrm{max}$
and equivalent internal resistance $R_0$ of a cell \citep{plett_battery_2016},
which form possibly the coarsest-grained description of capacity and power fade,
and correspond to a simple ECM (equivalent circuit model)
with only a monolithic OCV source $U_\mathrm{OCV}(z)$
assumed to be unchanged upon aging \citep{andre_advanced_2013}
and a series resistor of $R_0$.
Here $z$ denotes the cell SOC (state of charge), whose precise meaning will be clarified
in \cref{subsec:ocv},
and we will use the $\hat{Q}$ notation with a hat for any capacity
whose definition relies on specifying an upper and lower cutoff voltage or potential
for a full cell or an electrode, respectively.
There has been an enormous amount of research dedicated to estimating
$\hat{Q}_\mathrm{max}$ and $R_0$ from voltage and current measurements
\citep{plett_battery_2016,ng_predicting_2020},
but such parametrization is too coarse-grained
to shed light on possible underlying degradation mechanisms.
In this work, we will restrict ourselves to capacity fade only
and focus on a natural refinement of the monolithic OCV relation
by constructing the cell OCV as the difference between
the PE (positive electrode/cathode) and NE (negative electrode/anode) OCP (open-circuit potential):
\begin{equation}\label{eq:ocv-z+-}
  U_\mathrm{OCV}(z^-, z^+) = U^+_\mathrm{OCP}(z^+) - U^-_\mathrm{OCP}(z^-),
\end{equation}
where $z^\pm$ are the PE and NE SOC to be defined in \cref{subsec:ocp},
and three SOH parameters
lithium inventory $\hat{Q}^\mathrm{Li}_\mathrm{max}$,
PE active materials $\hat{Q}^+_\mathrm{max}$, and NE active materials $\hat{Q}^-_\mathrm{max}$,
all measured in charges,
will naturally surface in place of a single total capacity $\hat{Q}_\mathrm{max}$
as the SOH parameters,
which correspond to the three well-known capacity fade modes:
LLI (loss of lithium inventory), LAMp, and LAMn
(loss of active materials in PE/NE), respectively \citep{dubarry_synthesize_2012}.

The development history of the above electrode-specific OCV model,
though with different parametrization, has recently been reviewed by
\citet{olson_differential_2023} under the context of DV (differential voltage)
and IC (incremental capacity) analysis.
Indeed, this model was used to estimate the three modes of capacity fade
first by DV \citep{dahn_user-friendly_2012}
and IC \citep{dubarry_synthesize_2012,olson_differential_2023} analysis,
based on the parametrization
\begin{equation}\label{eq:ocv-LR-OFS}
  U_\mathrm{OCV}(z^-)
  = U^+_\mathrm{OCP}(1 - (b_\mathrm{OFS} + a_\mathrm{LR}z^-))
  - U^-_\mathrm{OCP}(z^-),
\end{equation}
where the loading ratio $a_\mathrm{LR}$ is the familiar NE-to-PE capacity ratio,
while $b_\mathrm{OFS}$ is interpreted as the loss of lithium inventory scaled by PE capacity.
However, DV and IC analysis is not yet a fully automatic technique,
and they require users to judge how the peaks and troughs of the full-cell DV and IC curves
are associated with electrode OCP features \citep{olson_differential_2023}.
Later, researchers attempted to estimate the three modes of LLI, LAMn, and LAMp
directly by fitting the model OCV curve to experimental measurements
using classical NLS (nonlinear least squares) \citep{birkl_degradation_2017,mohtat_towards_2019}.
In particular, \citet{birkl_degradation_2017} experimentally validate this approach
by assembling coin cells with a controlled amount of lithium inventory and active materials,
which serve as the ground truth against optimization results from NLS.
However, they parametrize their electrode-specific OCV
\begin{equation}\label{eq:ocv-z-zminmax}
  U_\mathrm{OCV}(z)
  = U^+_\mathrm{OCP}(z^+_\mathrm{max} - z(z^+_\mathrm{max} - z^+_\mathrm{min}))
  - U^-_\mathrm{OCP}(z^-_\mathrm{min} + z(z^-_\mathrm{max} - z^-_\mathrm{min}))
\end{equation}
by the extreme electrode SOC $z_\mathrm{max}^\pm$ and $z_\mathrm{min}^\pm$,
which are constrained by the upper and lower cutoff voltage
and have rather complicated dependence on the three modes.
\citet{mohtat_towards_2019} instead use the OCV in terms of the discharge throughput $Q_\mathrm{d}$
counted from the fully charged state:
\begin{equation}\label{eq:ocv-Qd}
  U_\mathrm{OCV}(Q_\mathrm{d})
  = U^+_\mathrm{OCP}(z^+_\mathrm{min} + Q_\mathrm{d}/\hat{Q}_\mathrm{max}^+)
  - U^-_\mathrm{OCP}(z^-_\mathrm{max} - Q_\mathrm{d}/\hat{Q}_\mathrm{max}^-),
\end{equation}
which is parametrized by $z^+_\mathrm{min}$, $z^-_\mathrm{max}$, and $\hat{Q}_\mathrm{max}^\pm$,
subject to the constraint
$U^+_\mathrm{OCP}(z^+_\mathrm{min}) - U^-_\mathrm{OCP}(z^-_\mathrm{max})= U_\mathrm{max}$.
They also derive the gradient of $U_\mathrm{OCV}(Q_\mathrm{d})$
with respect to the four parameters
and use Fisher information to quantify the parametric identifiability.
This OCV model has been incorporated in PyBaMM \citep{sulzer_python_2021},
an open battery modeling package implemented in Python,
and much subsequent work on model-based degradation mode estimation
has used different variants of this four-parameter formulation
\citep{pastor-fernandez_critical_2019,xu_data-driven_2023}.

In this work, we argue that the parametrization \eqref{eq:ocv-z-zminmax} or \eqref{eq:ocv-Qd}
based on electrode SOC at a certain cutoff voltage involves non-independent parameters,
of which the redundancy complicates their estimation.
Moreover, the cutoff voltage is an operational parameter that is not intrinsic to
the aging state of a cell,
so a parametrization that depends on it blurs the intrinsic degradation trend
and makes the SOH parameters less intuitive to interpret.
We also argue not to parametrize the OCV by the LLI, LAMn and LAMp percentage,
as these quantities rely on specifying the pristine-cell SOH parameters,
which are irrelevant to the current SOH and could be arbitrary.
Instead, these percentage losses should be defined in terms of the SOH parameters
and their values for a pristine cell \citep{olson_differential_2023},
and only calculated as derived quantities when needed and available.

Here, we slightly re-parametrize \eqref{eq:ocv-LR-OFS}
by the N/P (negative-to-positive) ratio
$r_\mathrm{N/P} = \hat{Q}^-_\mathrm{max}/\hat{Q}^+_\mathrm{max}$
and the Li/P (lithium-to-positive) ratio
$z^+_0=\hat{Q}^\mathrm{Li}_\mathrm{max}/\hat{Q}^+_\mathrm{max}$,
which have more symmetric and readily interpretable physical meaning,
leading to further insights not yet revealed in previous work using \eqref{eq:ocv-LR-OFS}.
This parametrization also implies that the discharge-count-based OCV
$U_\mathrm{OCV}(Q_\mathrm{d})$ depends only on
three SOH parameters:
lithium inventory $\hat{Q}^\mathrm{Li}_\mathrm{max}$,
NE active materials $\hat{Q}^-_\mathrm{max}$, and PE active materials $\hat{Q}^+_\mathrm{max}$,
all measured in equivalent charges.
These parameters are independent,
do not depend on any full-cell cutoff voltage or pristine-cell SOH,
and their changes directly and intuitively capture the three modes of LLI, LAMn, and LAMp.
Moreover, after the upper and lower cutoff voltage of a cell are specified
to define the cell SOC $z$,
the SOC-based OCV $U_\mathrm{OCV}(z)$ only depends on two independent dimensionless parameters
of Li/P and N/P ratio.
Along the way, we will also clarify several common misconceptions regarding the use of
an electrode-specific OCV model.

We will further analytically derive the gradients of
$U_\mathrm{OCV}(z)$, $U_\mathrm{OCV}(Q_\mathrm{d})$, and total capacity $\hat{Q}_\mathrm{max}$
with respect to the SOH parameters,
and demonstrate the further insights provided by this sensitivity analysis
beyond what has been shown in \cite{mohtat_towards_2019,lee_estimation_2020}.
In particular, we will introduce the electrode DV (differential voltage) fraction,
which is instrumental in channeling the effects of a fixed upper and lower cutoff voltage
into the sensitivity gradients.
The analytic results will also be visualized, verified, and corroborated by simulations.
Note that as shown in \cref{subsec:dv-fraction},
this DV fraction is closely related to
$\ud U^\pm_\mathrm{OCP}/\ud z^\pm$ and $\ud U_\mathrm{OCV}/\ud z$,
which are also the central objects in DV analysis
\citep{bloom_differential_2005,olson_differential_2023},
but to be clear,
this work is not related to DV analysis.

Finally, we will also present a direct application of our analytic results,
where the sensitivity gradients are fed to the Fisher information matrix
to quantify the identifiability of the Li/P and N/P ratio
from measurements of the SOC-based $U_\mathrm{OCV}(z)$
and to quantify the identifiability of the lithium inventory and active material amounts
from partial charge-based OCV measurements.

Note that there have been much work on adapting an OCV model
to degradation mode estimation under more realistic scenarios
such as with finite current or composite electrodes
\citep{dubarry_perspective_2022,schmitt_capacity_2023},
as well as extracting certain OCV features as better health indicators \citep{dubarry_state_2017},
but these are all built upon a baseline electrode-specific OCV model.
This work is not about a more advanced estimation algorithm,
but presents a more compact and intuitive parametrization
that can be substituted into any such algorithm.
Moreover, this parametrization also leads to further insights
into how the cell OCV and capacity depend on these SOH parameters and degradation modes,
which can guide us even in more complicated situations.

\section{Methodology and Analysis}
\label{sec:methodology}

Before diving into the degradation mode parametrization and identifiability,
we want to clarify a few concepts that could be ambiguous in literature.

\subsection{Electrode OCP: Stoichiometry, Electrode SOC and Active Material Amount}
\label{subsec:ocp}

For a lithium intercalation electrode,
the stoichiometry $x$ is literally the stoichiometric coefficient of lithium
in the active material chemical formula that indicates the lithiation state,
usually with $0\le x\le 1$ such as in graphite $\mathrm{Li}_x\mathrm{C}_6$
and LFP (lithium iron phosphate) $\mathrm{Li}_x\mathrm{FePO}_4$,
where $x=0$ and $x=1$ corresponds to the theoretical fully delithiated and lithiated state,
respectively.
The chemical formula also yields the theoretical electrode specific capacity,
such as $q_\mathrm{max}^-=372~\mathrm{mAh/g}$ for $\mathrm{Li}_x\mathrm{C}_6$
and $q_\mathrm{max}^+=170~\mathrm{mAh/g}$ for $\mathrm{Li}_x\mathrm{FePO}_4$,
where $q_\mathrm{max}$ denotes gravimetric specific capacity
of dimension charge per unit mass,
while $Q_\mathrm{max}$ for absolute capacity of charge dimension.
In the following, we will use the superscript $^+$ and $^-$ to indicate quantities
associated with PE and NE active materials, respectively.

An ideal open-circuit potential relation $U_\mathrm{OCP}^\pm(x^\pm)$ is with respect to
the stoichiometry $x^\pm$,
where the potential is measured against the standard lithium metal electrode.
However, due to material instability and electrochemical side reactions in practice,
we cannot fully lithiate or delithiate an electrode.
Besides, we also lack means to directly measure the electrode stoichiometry $x^\pm$,
unless we start from a pristine electrode with a precisely known amount of active materials,
which implies $x^-=0$ for NE materials like graphite or $x^+=1$ for PE materials like LFP initially,
and we can use Coulomb counting and the active material mass to deduce
subsequent lithiation states.

In most cases, we can only lithiate/delithiate an electrode (half cell)
across a usable potential window
and obtain $U_\mathrm{OCP}^\pm(Q^\pm)$ from voltage measurements and Coulomb counting.
We only know that the charge throughput $Q^\pm$ is linearly related to $x^\pm$,
but we have no access to the slope and intercept.
Hence, the best we can do is to define an electrode SOC $z^\pm\in[0, 1]$
such that $z^\pm=0$ corresponds to some unknown $x_\mathrm{min}^\pm$
at the top of the usable potential window,
while $z^\pm=1$ corresponds to some unknown $x_\mathrm{max}^\pm$ at the bottom.
This way, we can map $Q^\pm$ linearly to $z^\pm$ and obtain the most commonly seen
OCP relations $U_\mathrm{OCP}^\pm(z^\pm)$.
Here we define $Q^\pm$ and the electrode SOC $z^\pm$ to count positively
in the lithiation direction corresponding to a cathodic current.
We deviate from the convention in general electrochemistry
of treating anodic current as positive
to correlate $Q^\pm$ and $z^\pm$ with the amount of lithium intercalated,
which we find more intuitive in the case of an intercalation electrode.
There is also work that defines electrode SOC to be negatively correlated with stoichiometry
and positively correlated with electrode potential \citep{dubarry_synthesize_2012}.

The Coulomb counting also tells us the charge throughput associated with
the whole interval of $0\le z^\pm \le 1$,
which is the usable electrode capacity $\hat{Q}_\mathrm{max}^\pm$.
Note that
\begin{equation}
  \hat{Q}_\mathrm{max}^\pm = (x_\mathrm{max}^\pm - x_\mathrm{min}^\pm) Q_\mathrm{max}^\pm
\end{equation}
can be much smaller than the theoretical electrode capacity $Q_\mathrm{max}^\pm$
associated with $0\le x^\pm \le 1$,
and without knowing the active material mass, we cannot obtain $Q_\mathrm{max}^\pm$.
Moreover, since $z^\pm$ also relates linearly to $x^\pm$, we must have
\begin{equation}\label{eq:electrode-soc}
  z^\pm = \frac{x^\pm - x^\pm_\mathrm{min}}{x^\pm_\mathrm{max} - x^\pm_\mathrm{min}},
\end{equation}
although we do not know these stoichiometry limits in general.
See figure 1 in the Supporting Information for an illustration of the connection and distinction
between stoichiometry and electrode SOC.

Note also that mathematically, the choice of this upper and lower cutoff potential
for defining $z^\pm=0$ and $z^\pm=1$, respectively, is arbitrary,
as long as $\hat{Q}_\mathrm{max}^\pm$ is scaled accordingly and remains consistent with it,
and it will not affect the validity of the following full-cell analysis.
That is also why not knowing the stoichiometry $x^\pm$ is not an issue except
when we want to correlate electrode behaviors with certain lattice arrangement at a particular $x^\pm$.
However, we want $U_\mathrm{OCP}^\pm(z^\pm)$ to be defined for all of $0\le z^\pm \le 1$,
$z^\pm$ to intuitively indicate whether the electrode is in a safe operation range,
and $\hat{Q}_\mathrm{max}^\pm$ to faithfully reflect the usable electrode capacity,
so we do need to specify this potential window explicitly and reasonably.
We advocate that researchers do not use the term stoichiometry and electrode SOC interchangeably
and should report the upper and lower cutoff potential when presenting
an OCP curve $U_\mathrm{OCP}^\pm(z^\pm)$.

Since in general, the active material mass is not available,
we propose to use the usable electrode capacity $\hat{Q}_\mathrm{max}^\pm$
associated with a certain $U_\mathrm{OCP}^\pm(z^\pm)$
to characterize the amount of active materials in each electrode
and the corresponding degradation modes of LAMp and LAMn.

\subsection{Cell OCV: Cutoff Voltage, Cell SOC, and Degradation Mode Parametrization}
\label{subsec:ocv}

\begin{figure}[!htbp]
  \centering
  \begin{subfigure}{\textwidth}
    \centering
    \includegraphics[width=0.5\textwidth]{img/schematic/ocv-schematic.png}
    \caption{Schematic of an electrode-specific OCV model}
    \label{subfig:schematic}
  \end{subfigure}
  \begin{subfigure}{0.48\textwidth}
    \centering
    \includegraphics[width=\textwidth]{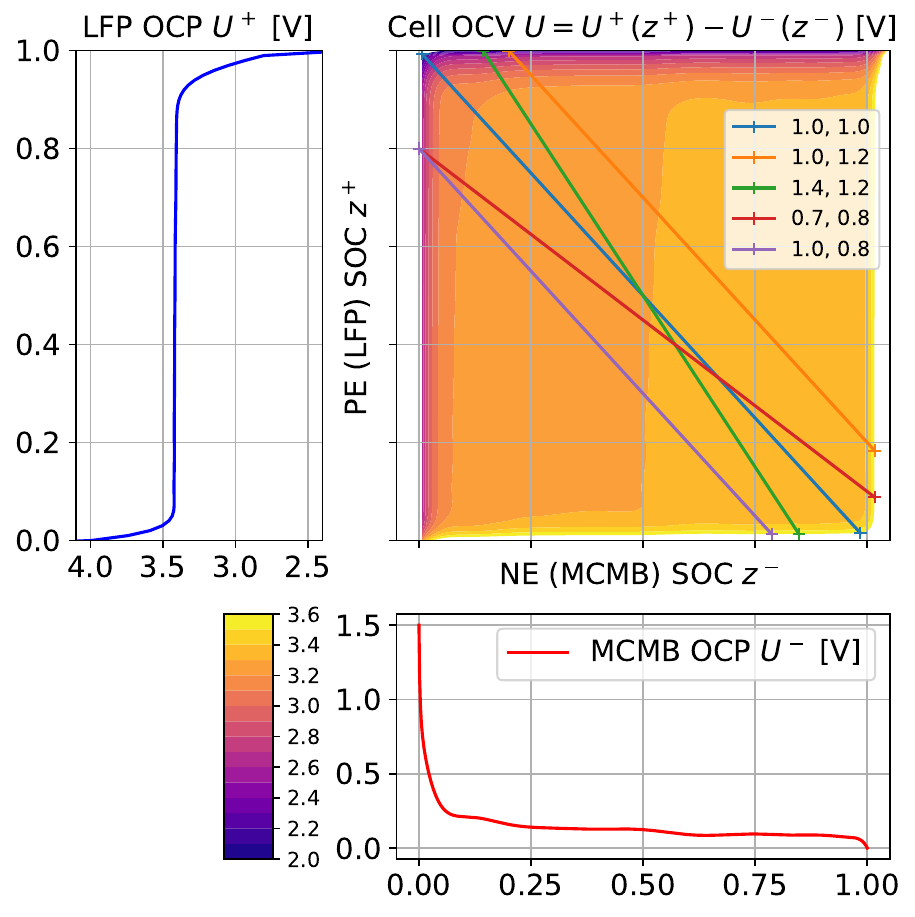}
    \caption{LFP/MCMB: $U_\mathrm{OCV}(z^-, z^+)$}
    \label{subfig:ocv-sq-lfp}
  \end{subfigure}
  \begin{subfigure}{0.48\textwidth}
    \centering
    \includegraphics[width=\textwidth]{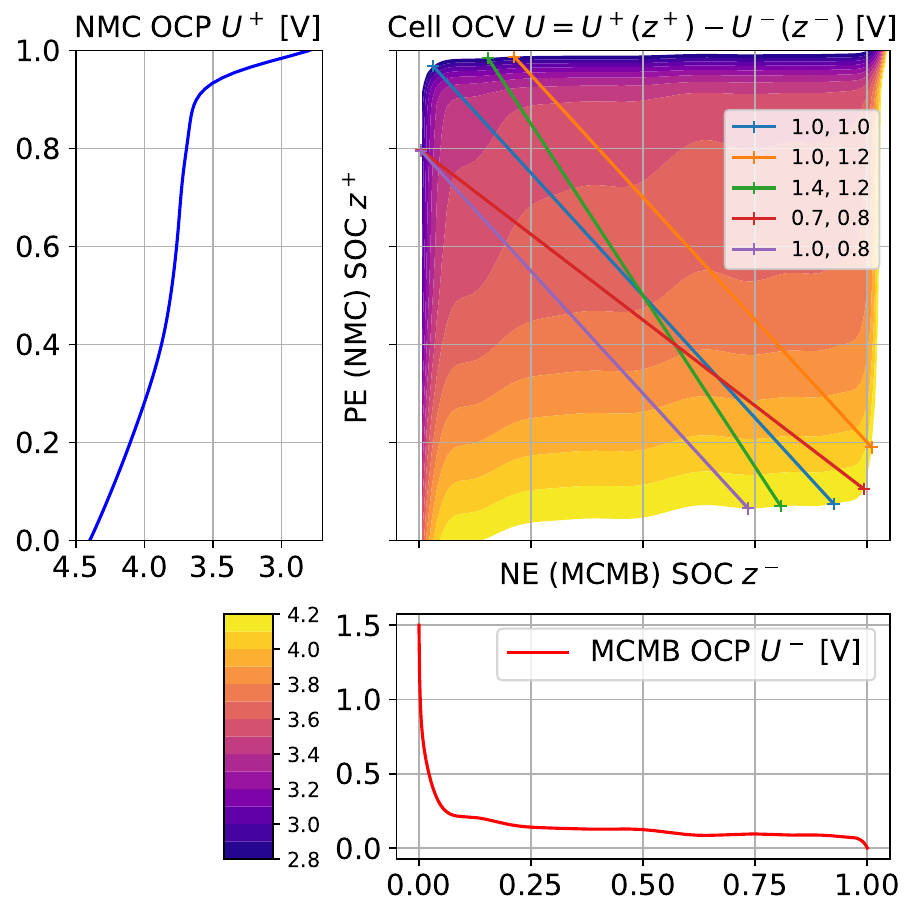}
    \caption{NMC/MCMB: $U_\mathrm{OCV}(z^-, z^+)$}
    \label{subfig:ocv-sq-nmc}
  \end{subfigure}
  \begin{subfigure}{0.48\textwidth}
    \centering
    \includegraphics[width=0.7\textwidth]{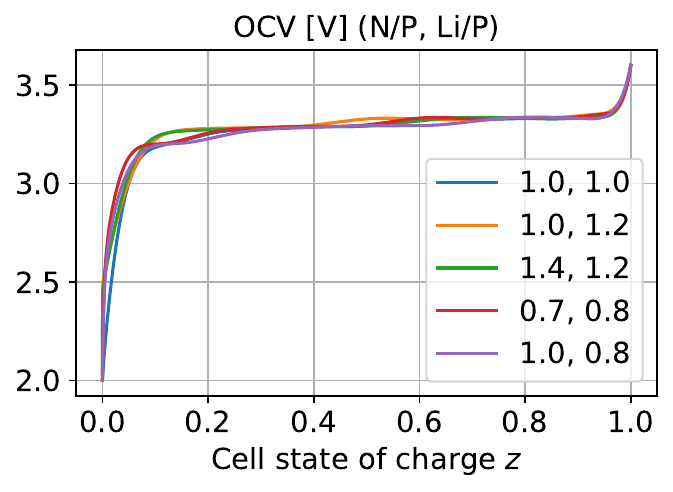}
    \caption{LFP/MCMB: $U_\mathrm{OCV}(z; r_\mathrm{N/P}, z^+_0)$}
    \label{subfig:ocv-line-lfp}
  \end{subfigure}
  \begin{subfigure}{0.48\textwidth}
    \centering
    \includegraphics[width=0.7\textwidth]{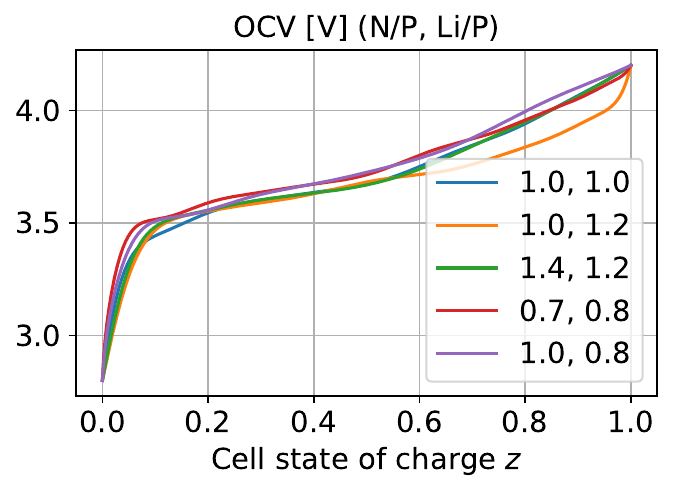}
    \caption{NMC/MCMB: $U_\mathrm{OCV}(z; r_\mathrm{N/P}, z^+_0)$}
    \label{subfig:ocv-line-nmc}
  \end{subfigure}
  \caption{
    {\bf (a)} A schematic of the OCV model.
    {\bf (b-e)} The assembly of electrode OCPs into a full-cell OCV,
    as well as how this assembly is affected by the N/P ratio $r_\mathrm{N/P}$ and Li/P ratio $z^+_0$.
    {\bf (b)} OCV square plots (advocated by \citet{olson_differential_2023}) of
    the LFP-graphite cell chemistry overlaid with five lines of $z^+(z^-)$
    corresponding to cells with different $(r_\mathrm{N/P}, z^+_0)$ pairs
    as indicated by the legends.
    Note that the color bar is associated with the contours.
    {\bf (d)} Various $z$-based OCV curves with different $(r_\mathrm{N/P}, z^+_0)$ pairs,
    which can be considered read off from the contour values on the respective lines in (b).
    {\bf (c,~e)} Counterparts of (b) and (d) for the NMC-graphite chemistry.
  }
  \label{fig:ocv-sq}
\end{figure}

When the PE and NE are assembled into a full cell,
their respective electrode SOC are no longer independent,
but are constrained by charge conservation.
Assuming no LLI or LAM, we have
\begin{equation}\label{eq:z+z-}
  z^+(z^-) = z_0^+ - r_\mathrm{N/P} z^-, \qquad
  r_\mathrm{N/P} = \frac{\hat{Q}_\mathrm{max}^-}{\hat{Q}_\mathrm{max}^+}, \qquad
  z_0^+ = \frac{\hat{Q}_\mathrm{max}^\mathrm{Li}}{\hat{Q}_\mathrm{max}^+},
\end{equation}
where the negative slope $r_\mathrm{N/P}$ is the commonly seen N/P (negative-to-positive) ratio,
while the intercept $z_0^+=z^+(z^-=0)$ is analogously called the Li/P ratio (lithium-to-positive) ratio,
since
\begin{equation}\label{eq:zp0}
  z_0^+ = \frac{z^+ \hat{Q}_\mathrm{max}^+ + z^- \hat{Q}_\mathrm{max}^-}{\hat{Q}_\mathrm{max}^+}
  = \frac{\hat{Q}_\mathrm{max}^\mathrm{Li}}{\hat{Q}_\mathrm{max}^+}.
\end{equation}
The lithium inventory $\hat{Q}_\mathrm{max}^\mathrm{Li}$ essentially accounts for
the lithium in both electrodes beyond $z^\pm=0$,
which is consistent with how $\hat{Q}_\mathrm{max}^\pm$ are defined.
Note also that the slope $r_\mathrm{N/P}$ and intercept $z_0^+$ fully characterize
the relative slippage between $z^\pm$.
\Cref{subfig:schematic} illustrates the physical meaning of these parameters.
Figure 2 in the Supporting Information provides further schematics illustrating
the fully discharged and charged state.

Substituting \cref{eq:z+z-} into \cref{eq:ocv-z+-},
we obtain the $z^-$-based full-cell OCV as
\begin{equation}\label{eq:full-ocv-z-}
  U_\mathrm{OCV}(z^-) = U_\mathrm{OCP}^+(z_0^+ - r_\mathrm{N/P} z^-)
  - U_\mathrm{OCP}^-(z^-).
\end{equation}
We can also equivalently relate $U_\mathrm{OCV}$ to $z^+$,
but $z^+$ is negatively correlated with the conventionally defined cell SOC $z$,
and it flips the well adopted N/P ratio to the denominator of the slope.

Assuming that the PE and NE OCP remain unchanged upon degradation,
the above electrode-specific OCV model is parametrized by
only two independent and dimensionless parameters
N/P ratio $r_\mathrm{N/P}$ and Li/P ratio $z_0^+$,
which have a simple and intuitive physical meaning that links to degradation modes.
Also, we have not yet introduced any cutoff voltage for the full cell,
so $U_\mathrm{OCV}(z^-)$ together with its two parameters are independent of that choice,
and we next show that this parametrization persists even at the presence of a cutoff voltage.

One caveat of assuming an invariant electrode OCP is that
crystal structural transformation undergone by certain active materials upon aging
that might affect the electrode potential is not captured.
Besides, the form of electrode OCP used in this work has also not
accounted for competing reactions that might become thermodynamically favorable
beyond certain electrode potential threshold,
of which the most prominent example is lithium plating
when $U^-_\mathrm{OCP}(z^-)$ reaches 0 V.
However, these layers of complexities can be laid upon the framework of analysis
proposed in this work when needed.
For example, see \cite{dubarry_best_2022} and references therein
for a review of techniques to incorporate some of such complexities into OCV modeling. 

Note that our N/P ratio $r_\mathrm{N/P}$ and Li/P ratio $z_0^+$
essentially correspond to the ``loading ratio'' and ``offset'' in \eqref{eq:ocv-LR-OFS}
used in \cite{dubarry_synthesize_2012,olson_differential_2023}
by $r_\mathrm{N/P}=a_\mathrm{LR}$ and $z_0^+ = 1 - b_\mathrm{OFS}$, respectively.
However, previous researchers using \eqref{eq:ocv-LR-OFS} typically
express $a_\mathrm{LR}$ and offset $b_\mathrm{OFS}$
in terms of at least three degradation modes
(e.g.~equation (5) and (8') in \cite{dubarry_synthesize_2012})
and sweep each degradation mode using the latter less compact parametrization.
Here we will perform parametric study directly on
the minimal and intuitive parametrization based on the N/P and Li/P ratio.

In principle, we can charge and discharge a cell until
one of the electrodes reaches $z^\pm =0$ or $z^\pm =1$,
but in practice, we do not typically get to monitor the two electrode potentials in a full cell
as with a three-electrode setup,
so instead,
we specify a terminal voltage range $[U_\mathrm{min}, U_\mathrm{max}]$
for a full cell to operate within,
such that either electrode is still prevented from over- or under-lithiation.
Hence, define $z_\mathrm{min}^-$ and $z_\mathrm{max}^-$,
which correspond to $z_\mathrm{max}^+$ and $z_\mathrm{min}^+$ by \cref{eq:z+z-}, respectively,
such that
\begin{equation}\label{eq:z-maxmin}
  U_{\mathrm{OCV}} (z^-_\mathrm{min}) = U_\mathrm{min}, \qquad
  U_{\mathrm{OCV}} (z^-_\mathrm{max}) = U_\mathrm{max},
\end{equation}
and call $U_\mathrm{OCV}=U_\mathrm{min}$ and $U_\mathrm{OCV}=U_\mathrm{max}$
the fully discharged and charged state, respectively.
For convenience, we can additionally define the full-cell SOC $z$ as
\begin{equation}\label{eq:z-zminus}
  z = \frac{z^- - z_\mathrm{min}^-}{z_\mathrm{max}^- - z_\mathrm{min}^-}
  = \frac{z^+ - z_\mathrm{max}^+}{z_\mathrm{min}^+ - z_\mathrm{max}^+},
\end{equation}
which now exactly traverses $[0\%,100\%]$ when $U_\mathrm{OCV}$ ranges
between $U_\mathrm{min}$ and $U_\mathrm{max}$.
Substituting $z^-(z)$ from \eqref{eq:z-zminus} into the $z^-$-based OCV \eqref{eq:full-ocv-z-},
we finally obtain the familiar $z$-based OCV:
\begin{equation}\label{eq:ocv-z}
  U_\mathrm{OCV}(z) = U_\mathrm{OCP}^+(z^+(z^-(z))) - U_\mathrm{OCP}^-(z^-(z)).
\end{equation}

Note that \cref{eq:z-maxmin} essentially comes from the inverse function $z^-(U_\mathrm{OCV})$
defined by \eqref{eq:full-ocv-z-} evaluated at $U_\mathrm{min}$ and $U_\mathrm{max}$
and is hence also parametrized by $r_\mathrm{N/P}$ and $z^+_0$.
Therefore, the $z^-(z)$ from \eqref{eq:z-zminus} and hence $U_\mathrm{OCV}(z)$
are parametrized by $r_\mathrm{N/P}$, $z^+_0$, $U_\mathrm{min}$, and $U_\mathrm{max}$,
where given $U_\mathrm{min}$ and $U_\mathrm{max}$ fixed,
$U_\mathrm{OCV}(z)$ is thus again parametrized
by the independent and dimensionless $r_\mathrm{N/P}$ and $z^+_0$ only.
This is the main difference from the parametrization of \eqref{eq:ocv-z-zminmax}
based on four electrode SOC limits and two cutoff-voltage constraints \citep{birkl_degradation_2017}.

\begin{table}[!htbp]
  \centering
  \begin{tabular}{cc|cccc}
  \toprule
  & & & MCMB & LFP & NMC111 \\
  \midrule
  $x^\pm=x^\pm_\mathrm{min}$ & $z^\pm=0$ & $U_\mathrm{max}^\pm$ [V] & 1.5 & 4.0 & 4.4 \\
  $x^\pm=x^\pm_\mathrm{max}$ & $z^\pm=1$ & $U_\mathrm{min}^\pm$ [V] & 0.0 & 2.5 & 2.8 \\
  \bottomrule
  \end{tabular}
  \caption{
    Upper and lower cutoff potential specified for each electrode chemistry used in this work.
  }
  \label{tab:ocp-cutoff}
\end{table}

\begin{table}[!htbp]
  \centering
  \begin{tabular}{ccc|ccc}
  \toprule
  & & & & LFP/MCMB & NMC/MCMB \\
  \midrule
  $z^-=z^-_\mathrm{max}$ & $z^+=z^+_\mathrm{min}$ & $z=1$ & $U_\mathrm{max}$ [V] & 3.6 & 4.2 \\
  $z^-=z^-_\mathrm{min}$ & $z^+=z^+_\mathrm{max}$ & $z=0$ & $U_\mathrm{min}$ [V] & 2.0 & 2.8 \\
  \bottomrule
  \end{tabular}
  \caption{
    Upper and lower cutoff potential specified for each electrode chemistry used in this work.
  }
  \label{tab:ocv-cutoff}
\end{table}

\Cref{subfig:ocv-sq-lfp,subfig:ocv-line-lfp,subfig:ocv-sq-nmc,subfig:ocv-line-nmc} illustrate
how $U_\mathrm{OCV}(z)$ is assembled from $U_\mathrm{OCP}^\pm(z^\pm)$
and depends on $r_\mathrm{N/P}$ and $z^+_0$
(see figure 3 in the Supporting Information for an illustration of such OCV assembly
using conventional line plots).
In this work, we will use the LFP/
and NMC (lithium nickel manganese cobalt oxide)/graphite cell chemistry
as two typical running examples
and perform simulations coded in Python to concretely illustrate our results,
where the electrode OCP relations are
from the LFP, NMC-111 and MCMB (mesocarbon microbeads, a form of synthetic graphite) OCP fits
provided in \citet[sec.~3.11.1]{plett_battery_2015},
with temperature set to $\mathrm{25^{\circ}C}$.
However, our results apply equally to all other chemistries.
The specifications of electrode cutoff potential and full-cell cutoff voltage are tabulated
in \cref{tab:ocp-cutoff,tab:ocv-cutoff}, respectively.

In \cref{subfig:ocv-sq-lfp} inspired by \citet{olson_differential_2023},
each line segment corresponds to a cell's $z^+(z^-)$ relation
with a particular $(r_\mathrm{N/P}, z^+_0)$ pair,
where the N/P ratio $r_\mathrm{N/P}$ is the negative slope
and the Li/P ratio $z^+_0$ (recall \cref{eq:zp0}) is the $z^+$-axis intercept.
The four pairs of N/P and Li/P ratio other than $r_\mathrm{N/P}=z^+_0=1$
cover the four characteristic regimes to be discussed in \cref{subsec:dq}.
These line segments are superposed on the 2D landscape of $U_\mathrm{OCV}(z^-, z^+)$.
The $U_\mathrm{min}$-contour marks
where these segments can start on the top left,
while the $U_\mathrm{max}$-contour marks their possible ends on the bottom right.
Then the slope $-r_\mathrm{N/P}$ and intercept $z^+_0$ determine
whether the segment starts close to the top or left side of the square,
and whether it ends near the bottom or right side.
Therefore, this indicates the electrode SOC limits at the fully discharged and charged state,
which implies which electrode is more limiting towards the end of discharging and charging,
as will be further discussed in \cref{subsec:dv-fraction}.
When we plot the OCV profile intersected by each line segment
and scale the segment to a unit interval,
we obtain the $z$-based OCV $U_\mathrm{OCV}(z)$ in \cref{subfig:ocv-line-lfp}.
The flatness of the LFP and MCMB OCP makes the variation in $U_\mathrm{OCV}(z)$ rather mild visually
with different $(r_\mathrm{N/P}, z^+_0)$ pairs,
which will be precisely quantified in \cref{subsec:dzpm-du,subsec:fisher-z}.
\Cref{subfig:ocv-sq-nmc,subfig:ocv-line-nmc} are simply the NMC/MCMB counterparts of
\cref{subfig:ocv-sq-lfp,subfig:ocv-line-lfp}.
The steeper slope of the NMC OCP curve is reflected
by the narrower contour spacing in \cref{subfig:ocv-sq-nmc}
and the larger variation between different OCV curves in \cref{subfig:ocv-line-nmc}.

\begin{figure}[!htbp]
  \centering
  \begin{subfigure}[b]{0.48\textwidth}
    \centering
    \includegraphics[width=\textwidth]{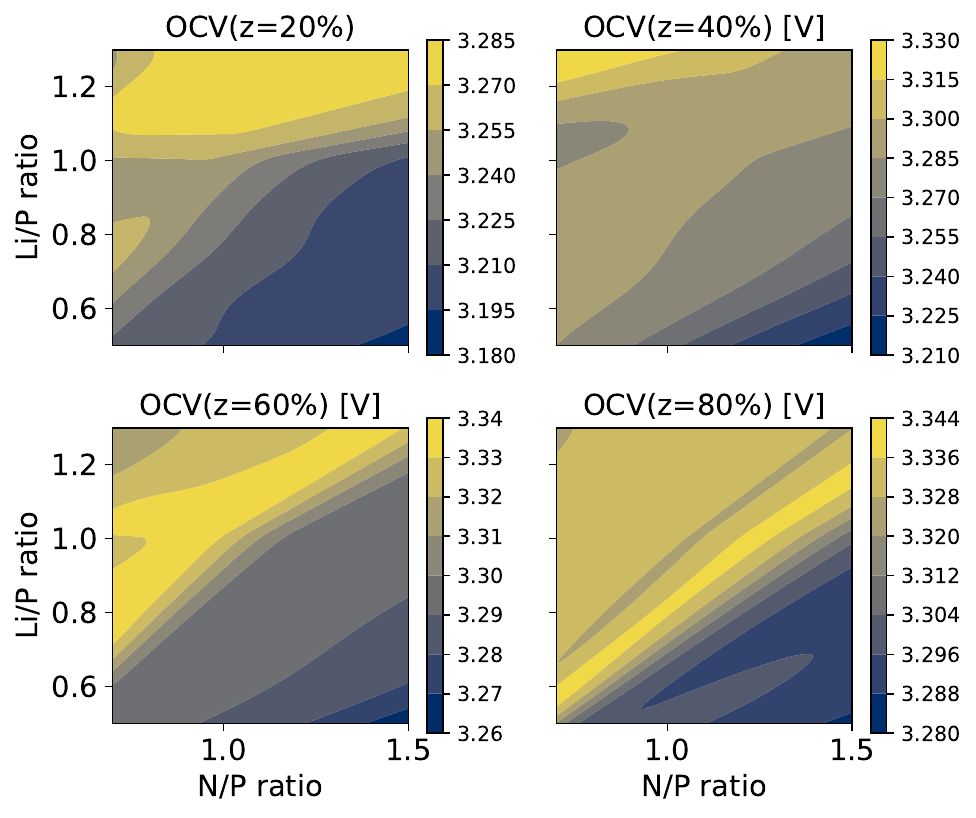}
    \caption{LFP: $U_\mathrm{OCV}(z; r_\mathrm{N/P}, z^+_0)$}
    \label{subfig:ocv-rz-u-lfp}
  \end{subfigure}
  \begin{subfigure}[b]{0.47\textwidth}
    \centering
    \includegraphics[width=\textwidth]{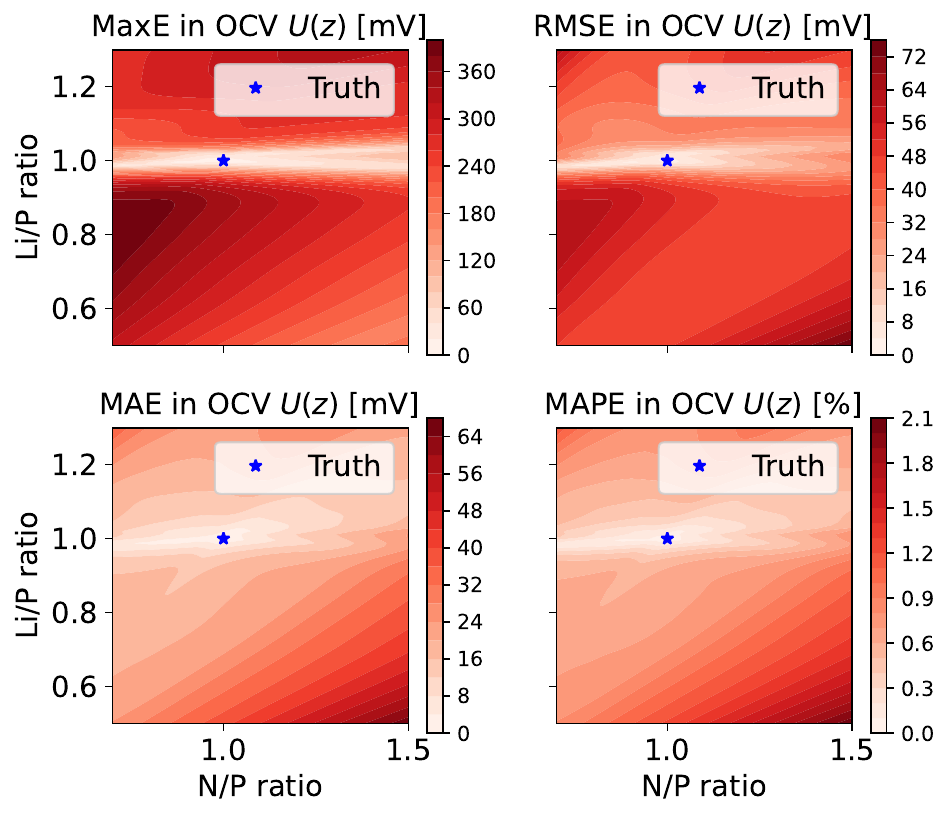}
    \caption{
      LFP: $\|U_\mathrm{OCV}(z; r_\mathrm{N/P}, z^+_0)
      - U_\mathrm{OCV}(z; 1, 1)\|$
    }
    \label{subfig:ocv-rz-ue-lfp}
  \end{subfigure}
  \begin{subfigure}[b]{0.48\textwidth}
    \centering
    \includegraphics[width=\textwidth]{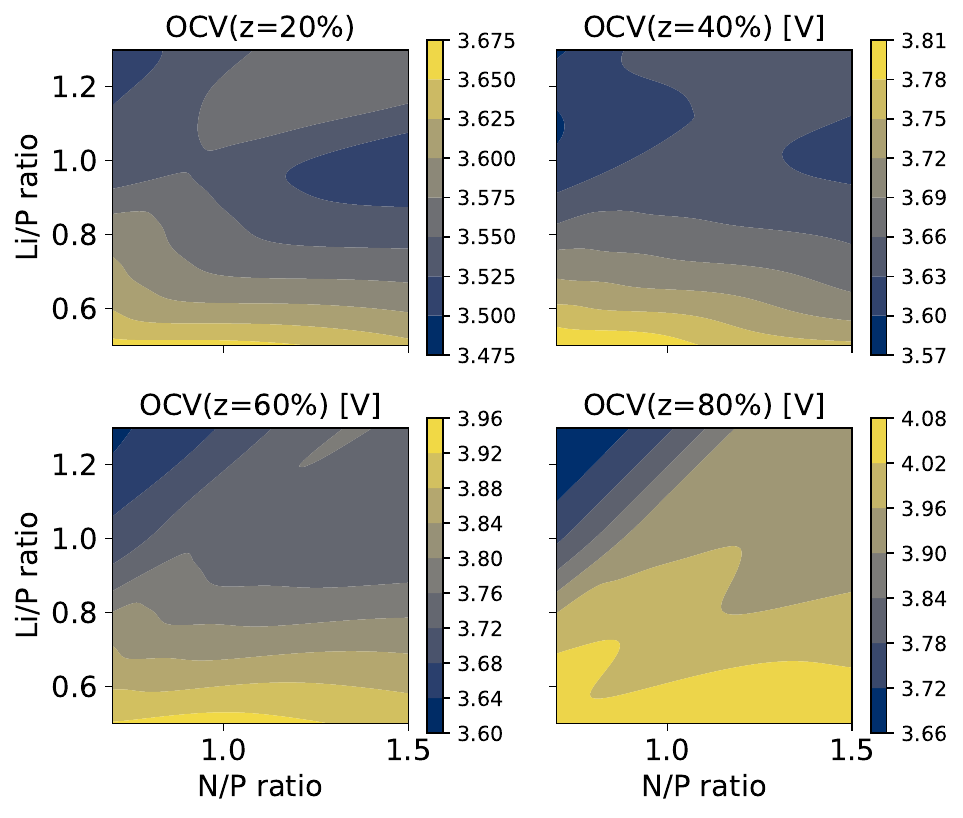}
    \caption{NMC: $U_\mathrm{OCV}(z; r_\mathrm{N/P}, z^+_0)$}
    \label{subfig:ocv-rz-u-nmc}
  \end{subfigure}
  \begin{subfigure}[b]{0.47\textwidth}
    \centering
    \includegraphics[width=\textwidth]{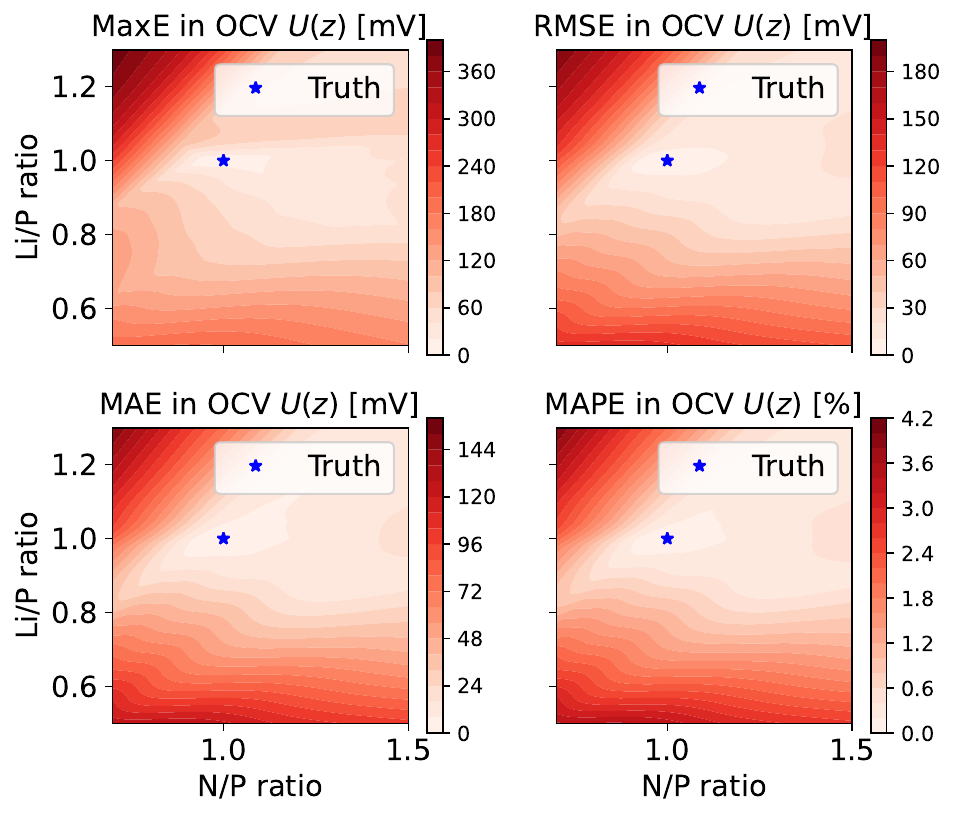}
    \caption{
      NMC: $\|U_\mathrm{OCV}(z; r_\mathrm{N/P}, z^+_0)
      - U_\mathrm{OCV}(z; 1, 1)\|$
    }
    \label{subfig:ocv-rz-ue-nmc}
  \end{subfigure}
  \caption{
    {\bf (a,~b)} LFP/MCMB.
    {\bf (c,~d)} NMC/MCMB.
    {\bf (a,~c)}
    Cell OCV at SOC $z=20\%$, 40\%, 60\%, and 80\% with different $(r_\mathrm{N/P}, z^+_0)$ pairs.
    Note the varied OCV sensitivity at different $z$'s.
    {\bf (b,~d)}
    Maximum errors (MaxE), Root-mean-square errors (RMSE), mean absolute errors (MAE),
    and mean absolute percentage errors (MAPE) between cell OCV
    with an arbitrary $(r_\mathrm{N/P}, z^+_0)$ pair
    and OCV with $r_\mathrm{N/P}=1$ and $z^+_0=1$ (marked as truth).
  }
  \label{fig:ocv-rz-u}
\end{figure}

One advantage of the $z$-based OCV $U_\mathrm{OCV}(z)$ is that it can be determined directly
by full-cell measurements without knowing anything about electrode OCP and SOC,
just as how $U_\mathrm{OCP}^\pm(z^\pm)$ can be determined agnostic of stoichiometry $x^\pm$.
Moreover, this function can be roughly assumed unchanged upon aging \citep{andre_advanced_2013}
because by definition, $U_\mathrm{OCV}(z)$ always monotonically increases from
$(0, U_\mathrm{min})$ to $(1, U_\mathrm{max})$,
as \cref{subfig:ocv-line-lfp,subfig:ocv-line-nmc} illustrate.
Hence, such electrode-agnostic monolithic cell OCV has been used in most ECMs in literature
and in prevailing battery management systems \citep{plett_battery_2015,plett_battery_2016}.
However, previous literature and the above have shown that $U_\mathrm{OCV}(z)$ does change appreciably
upon aging, as further corroborated by \cref{fig:ocv-rz-u} based on simulations.
\Cref{subfig:ocv-rz-u-lfp,subfig:ocv-rz-u-nmc} show
how $U_\mathrm{OCV}(z)$ at a particular $z$ varies with the Li/P and N/P ratio
ranging in $[0.5,~1.3]$ and $[0.7,~1.5]$, respectively.
The OCV variation is between $60~\mathrm{mV}$ and $100~\mathrm{mV}$ at these SOCs for LFP,
while it is as large as $200~\mathrm{mV}$ to $400~\mathrm{mV}$ for NMC.

Note that in a typical commercial lithium-ion cell,
the NE is manufactured as fully delithiated, while the PE as fully lithiated,
so upon formation cycles,
after some lithium is consumed to form the SEI (solid-electrolyte interface) layers,
we should expect the Li/P ratio to be less than 1.
Moreover, when graphite is used as the NE,
to avoid lithium plating and other side reactions during charging,
we typically design the N/P ratio to be slightly larger than 1.
However, when a cell degrades upon cycling or storage,
lithium inventory and active materials may not diminish at the same rate
so the N/P and Li/P ratio can potentially evolve to other non-typical values.
This is why in this work,
we study the cell OCV characteristics across a large range of these two parameters
to provide a big picture of a cell's potential behaviors throughout its life.

If we summarize the difference between two complete OCV curve $U_\mathrm{OCV}(z)$ by a norm,
and compare the $U_\mathrm{OCV}(z)$ of different $(r_\mathrm{N/P}, z^+_0)$ pairs
to the one with a fixed pair considered as the ground truth,
we obtain \cref{subfig:ocv-rz-ue-lfp,subfig:ocv-rz-ue-nmc}.
For the LFP case in \cref{subfig:ocv-rz-ue-lfp},
note in certain regions, RMSE and especially MaxE are much higher than MAE,
implying that the errors between two OCV curves are distributed very unevenly along $z$,
so only OCV values within certain SOC windows are informative about the parameters.
It is also evident in this case that the OCV is much more sensitive to Li/P than to N/P.
In summary, these results indicate that the OCV variation does encode
changes in $r_\mathrm{N/P}$ and $z^+_0$ and hence changes in degradation modes,
although the sensitivity can be highly chemistry-dependent
and vary across different $(r_\mathrm{N/P}, z^+_0)$ pairs.

The upper and lower cutoff voltage also define the cell total capacity $\hat{Q}_\mathrm{max}$,
which is the charge throughput associated with the SOC interval $0\le z \le 1$.
The electrode SOC limits in \eqref{eq:z-zminus} also relate the electrode capacity
to their full-cell counterpart by charge conservation:
\begin{equation}\label{eq:full-soc-capacity}
  \hat{Q}_\mathrm{max} = (z_\mathrm{max}^- - z_\mathrm{min}^-) \hat{Q}_\mathrm{max}^-
  = (z_\mathrm{max}^+ - z_\mathrm{min}^+) \hat{Q}_\mathrm{max}^+.
\end{equation}
We have already mentioned that the four electrode SOC limits are parametrized
by $r_\mathrm{N/P}$ and $z^+_0$ only,
so $\hat{Q}_\mathrm{max}$ also depends on these two dimensionless quantities,
but is additionally parametrized by the dimensional $\hat{Q}_\mathrm{max}^+$.
Of course $\hat{Q}_\mathrm{max}^+$ is not the only choice,
we choose it here for symmetry since it is the denominator of both $r_\mathrm{N/P}$ and $z^+_0$.
Equivalently, $\hat{Q}_\mathrm{max}$ can also be parametrized by
the three-dimensional SOH parameter $\hat{Q}_\mathrm{max}^\mathrm{Li}$, $\hat{Q}_\mathrm{max}^-$,
and $\hat{Q}_\mathrm{max}^+$.
We will discuss the trade-off between these two options.

Since what Coulomb counting records is the charge or discharge throughput,
sometimes we also want to express cell OCV in terms of, say charge throughput $\hat{Q}$,
instead of the dimensionless SOC $z$.
Assume $\hat{Q}$ counts from the fully discharged state at $U_\mathrm{min}$,
we readily have $z = \hat{Q}/\hat{Q}_\mathrm{max}$ and it ultimately yields
\begin{equation}\label{eq:ocv-Qc}
  U_\mathrm{OCV}(\hat{Q})
  = U^+_\mathrm{OCP}(z^+_\mathrm{max} - \hat{Q}/\hat{Q}_\mathrm{max}^+)
  - U^-_\mathrm{OCP}(z^-_\mathrm{min} + \hat{Q}/\hat{Q}_\mathrm{max}^-),
\end{equation}
which is essentially the same as \eqref{eq:ocv-Qd} used by \citep{mohtat_towards_2019},
except that $U_\mathrm{OCV}(\hat{Q})$
under our formulation here are parametrized by three independent SOH parameters
$r_\mathrm{N/P}$, $z^+_0$, and $\hat{Q}_\mathrm{max}^+$,
or equivalently by $\hat{Q}_\mathrm{max}^\mathrm{Li}$ and $\hat{Q}_\mathrm{max}^\pm$,
neither of which depends on the choice of cell cutoff voltage $U_\mathrm{max}$ or $U_\mathrm{min}$.

The cutoff-voltage independence also makes \eqref{eq:ocv-Qc} easily adaptable to
modelling partial OCV curve.
If the starting voltage $U_\mathrm{start}$ is known (e.g.,~reliably measured after a long rest),
the initial electrode SOC $z^\pm_\mathrm{start}$ are uniquely determined by
$U_\mathrm{start}$ and any three-parameter set above,
so after substituting $z^\pm_\mathrm{start}$ into $z^+\mathrm{max}$ and $z^-_\mathrm{min}$
in \eqref{eq:ocv-Qc},
this partial OCV model is still parametrized by the same three parameters.
However, if $U_\mathrm{start}$ is subject to the same uncertainty
as any subsequent OCV measurements are,
we will have to add an extra unknown parameter,
such as $z^-_\mathrm{start}$, to mark the electrode SOC to be matched with
the beginning of the partial OCV.
One popular parametrization for this case in literature is
with $z^\pm_\mathrm{start}$ and $\hat{Q}_\mathrm{max}^\pm$ \citep{schmitt_capacity_2023}.
Although it is mathematically equivalent to
using $z^-_\mathrm{start}$, $\hat{Q}_\mathrm{max}^\mathrm{Li}$ and $\hat{Q}_\mathrm{max}^\pm$,
the physical meaning of each parameter in the latter might be clearer and more intuitive.

In the above analysis, we have assumed no LLI or LAM during charging or discharging.
In practice, due to the slowness of degradation compared to charging and discharging,
this assumption is valid within a few cycles so the above OCV model will be a good approximation,
and we only need to gradually adjust the SOH parameters upon cell aging.

\subsection{Relating SOH Parameters to LLI and LAM Percentage}
\label{subsec:lli-lam-percent}

We have parametrized $U_\mathrm{OCV}(z)$ by the dimensionless N/P ratio $r_\mathrm{N/P}$
and Li/P ratio $z^+_0$,
and parametrized $U_\mathrm{OCV}(\hat{Q})$
by the dimensional lithium inventory $\hat{Q}_\mathrm{max}^\mathrm{Li}$
and PE and NE active material capacity $\hat{Q}_\mathrm{max}^\pm$.
All these SOH parameters are independent of cell cutoff voltage
and agnostic of the pristine-cell condition.

In cases where we can specify the initial lithium inventory and active material amount
in a pristine cell,
denoted by $\hat{Q}_\mathrm{max,ini}^\mathrm{Li}$ and $\hat{Q}_\mathrm{max,ini}^\pm$, respectively,
we can easily calculate the commonly used LLI, LAMn, and LAMp percentage as derived quantities:
\begin{equation}\label{eq:loss-percent}
  \mathrm{LLI}
  = 1 - \frac{\hat{Q}_\mathrm{max}^\mathrm{Li}}{\hat{Q}_\mathrm{max,ini}^\mathrm{Li}}, \qquad
  \mathrm{LAM}^-
  = 1 - \frac{\hat{Q}_\mathrm{max}^-}{\hat{Q}_\mathrm{max,ini}^-}, \qquad
  \mathrm{LAM}^+
  = 1 - \frac{\hat{Q}_\mathrm{max}^+}{\hat{Q}_\mathrm{max,ini}^+}.
\end{equation}
We can similarly correlate the N/P ratio $r_\mathrm{N/P}$ and Li/P ratio $z^+_0$
to their initial values:
\begin{equation}
  r_\mathrm{N/P} = r_\mathrm{N/P,ini} \frac{1-\mathrm{LAM}^-}{1-\mathrm{LAM}^+}, \qquad
  z^+_0 = z^+_\mathrm{0,ini} \frac{1-\mathrm{LLI}}{1-\mathrm{LAM}^+},
\end{equation}
as similarly shown by, for example,
equation (5) and (8') in \cite{dubarry_synthesize_2012}
and equation (48-49) in \cite{olson_differential_2023}.

In literature, it is not uncommon to see parametric study of an OCV model
by varying the above three loss percentages,
but this has the caveat of redundancy,
because as we have shown, the shape of an OCV curve, or the SOC-based OCV,
is only governed by two degrees of freedom,
i.e.,~the N/P and Li/P ratio.
The issue here is that a certain ratio $\mathrm{LLI}:\mathrm{LAM}^-:\mathrm{LAM}^+$
does not correspond to a unique shape of OCV,
which gives the false impression that the OCV variation has more than two degrees of freedom.
Indeed, it is the ratio $(1-\mathrm{LLI}):(1-\mathrm{LAM}^-):(1-\mathrm{LAM}^+)$,
which corresponds to a fixed ratio of
$\hat{Q}_\mathrm{max}^\mathrm{Li}:\hat{Q}_\mathrm{max}^-:\hat{Q}_\mathrm{max}^+$
according to \eqref{eq:loss-percent}
and hence fixed N/P and Li/P ratio,
that will uniquely determine the OCV shape.
Of course, if a dimensional charge-based coordinate must be used in place of
the dimensionless cell SOC $z$
when, for example, $z$ is not accessible in certain practical scenarios,
the OCV curve will depend on all three parameters of
$\hat{Q}_\mathrm{max}^\mathrm{Li}$ and $\hat{Q}_\mathrm{max}^\pm$,
which can now be equivalently re-parametrized
by $\mathrm{LLI}$ and $\mathrm{LAM}^\pm$ instead.

Another subtlety already alluded to in \cref{subsec:ocp}
lies in the dependence of the SOH parameters
$\hat{Q}_\mathrm{max}^\mathrm{Li}$ and $\hat{Q}_\mathrm{max}^\pm$
on the artificial choice of electrode cutoff potential.
For the active materials, the situation is relatively straightforward,
because when we change the upper and lower cutoff potential,
it simply changes the stoichiometry range considered usable,
and thus $\hat{Q}_\mathrm{max}^\pm$ are just scaled by the same factor
as the length of the stoichiometry range is.
For LAM percentage calculation,
since both the initial and current active materials are scaled the same way,
$\mathrm{LAM}^\pm$ will remain the same.

Things are more complicated for lithium inventory \citep{dubarry_accurate_2023}.
Since we essentially set some lower limit of stoichiometry
corresponding to some upper cutoff potential as the ``origin'' of counting lithium
(recall \cref{eq:zp0}),
changing the upper cutoff potential will alter the lithium inventory
but changing the lower one will not,
because only the former is relevant to the origin.
Moreover, since this origin only affects $\hat{Q}_\mathrm{max}^\mathrm{Li}$ itself
but not its change,
LLI will also depend on this choice of lithium origin.

Since all of $\hat{Q}_\mathrm{max}^\mathrm{Li}$ and $\hat{Q}_\mathrm{max}^\pm$
depend on the choice of electrode cutoff potentials,
the N/P ratio $r_\mathrm{N/P}$ and Li/P ratio $z^+_0$ will also vary inevitably.
However, when assembling the full-cell OCV,
as long as the full-cell cutoff voltages remain the same,
the SOC-based OCV $U_\mathrm{OCV}(z)$ also remains constant,
because we are simply renaming the electrode SOC at the cell cutoff voltages
without altering what electrode potentials they correspond to.
Consequently, the total capacity within this voltage window is also unaffected.
Furthermore, the same N/P and Li/P ratio will still uniquely determine $U_\mathrm{OCV}(z)$.

Although defining the electrode SOC based on specifying some material-specific cutoff potential
seems to introduce certain arbitrariness,
a reasonable choice is usually available,
and it does make the SOH parameters $\hat{Q}_\mathrm{max}^\mathrm{Li}$
and $\hat{Q}_\mathrm{max}^\pm$ more relevant to practical operation.
Moreover, data under one choice of electrode cutoff potentials
can easily be transformed into their counterparts under another choice as post-processing.
Therefore, as long as we keep the choice consistent and explicitly report it
when discussing these SOH parameters,
the use of electrode SOC should do more benefit than harm.

\subsection{Parametric Sensitivity: Electrode DV Fraction and Role of Fixed Cutoff Voltage}
\label{subsec:dv-fraction}

We first establish sensitivity results for a general parameter $\alpha$
of an electrode-specific cell OCV model $U_\mathrm{OCV}(z)$,
which will then be applied to various specific SOH parameters.
For the most general case,
any input functional relation to \cref{eq:ocv-z} that must be specified
might depend on $\alpha$:
\begin{equation}\label{eq:func-alpha}
  U_\mathrm{OCP}^+(z^+, \alpha), \quad U_\mathrm{OCP}^-(z^-, \alpha), \quad
  z^+(z^-, \alpha),
\end{equation}
which together constitute the final $U_\mathrm{OCV}(z, \alpha)$.

By the chain rule, we have
(derivation details in Supporting Information)
\begin{equation}\label{eq:alpha-ocv}
  \begin{aligned}
    \Par_\alpha  U_\mathrm{OCV}(z)
    =&~ \Par_\alpha U^+_\mathrm{OCP}(z^+)
    + \ud_{z^+} U^+_\mathrm{OCP} \cdot \Par_\alpha z^+(z) \\
    &- \Par_\alpha U^-_\mathrm{OCP}(z^-)
    - \ud_{z^-} U^-_\mathrm{OCP} \cdot \Par_\alpha z^-(z),
  \end{aligned}
\end{equation}
\begin{equation}\label{eq:alpha-z+}
  \Par_\alpha z^+(z) = \Par_\alpha z^+(z^-) - r_\mathrm{N/P} \Par_\alpha z^-(z),
\end{equation}
\begin{equation}\label{eq:alpha-z-}
  \Par_\alpha z^-(z) = (1-z) \Par_\alpha z^-_\mathrm{min} + z \Par_\alpha z^-_\mathrm{max}.
\end{equation}
\begin{equation}\label{eq:alpha-z--ocv}
  \Par_\alpha z^-(U_\mathrm{OCV}) 
  = \frac{\Par_\alpha U_\mathrm{OCP}^+(z^+)
  + \ud_{z^+} U_\mathrm{OCP}^+ \cdot \Par_\alpha z^+(z^-)
  - \Par_\alpha U_\mathrm{OCP}^-(z^-)}
  {r_\mathrm{N/P} \ud_{z^+} U_\mathrm{OCP}^+ + \ud_{z^-} U_\mathrm{OCP}^-}.
\end{equation}
\begin{equation}\label{eq:alpha-z-minmax}
  \Par_\alpha z^-_\mathrm{min} = \Par_\alpha z^-(U_\mathrm{min}),
  \qquad \Par_\alpha z^-_\mathrm{max} = \Par_\alpha z^-(U_\mathrm{max}).
\end{equation}
Note that we denote parametric partial derivatives by $\Par_\alpha$
while all the other state partial differentiation by $\ud_{(\cdot)}$
to emphasize the distinction.
One subtlety here is that $z^-_\mathrm{min}$ is defined implicitly
by the lower cutoff voltage as $U_\mathrm{OCV}(z^-_\mathrm{min})=U_\mathrm{min}$.
Therefore, calculating $\Par_\alpha z^-_\mathrm{min}$ reduces to
differentiating the implicit function $z^-(U_\mathrm{OCV}, \alpha)$
defined by $U_\mathrm{OCV}(z^-, \alpha)$.

Since generically, we have $\Par_\alpha z^\pm(z)\neq 0$,
the change of $\alpha$ may cause the same $z$ to correspond to a different $z^\pm$.
In particular, the electrode SOC limits $z^-_\mathrm{min}$, $z^-_\mathrm{max}$,
$z^+_\mathrm{min}$, and $z^+_\mathrm{max}$ at the upper and lower cutoff voltage
can change with $\alpha$,
causing the ``charge/discharge end-point slippage''
\citep{smith_interpreting_2011,rodrigues_capacity_2022}.
In contrast, the PE-NE relative slippage indicated by $\Par_\alpha z^+(z^-)\neq 0$
causing one $z^-$ to correspond to a different $z^+$ will also occur
unless both $r_\mathrm{N/P}$ and $z^+_0$ are fixed,
i.e.~the ratio $\hat{Q}_\mathrm{max}^- : \hat{Q}_\mathrm{max}^+ : \hat{Q}_\mathrm{max}^\mathrm{Li}$
stays the same.
A lot of researchers have studied such electrode slippage due to SEI growth during charging,
but what we have derived above reveal a bigger picture of
other potential underlying mechanisms.

The $z^-(z)$ and $z^+(z)$ slippage also causes the electrode potential $U^\pm_\mathrm{OCP}(z^\pm(z))$
and consequently the cell OCV $U_\mathrm{OCV}(z)$ to change for a particular $z$.
Therefore, a certain cell SOC $z$
does not anchor to any electrode SOC $z^\pm$ or any OCV value
(except for $z=0$ and $z=1$ in the latter case).
The cell SOC $z$ simply indicates the relative remaining discharge throughput
against total capacity $\hat{Q}_\mathrm{max}$ until fully discharged.
This is probably the only persistent physical meaning of a so-defined cell SOC $z$.

In addition to the cell OCV $U_\mathrm{OCV}(z)$ that is independent of the total capacity,
we can also characterize the parametric sensitivity of total capacity
$\hat{Q}_\mathrm{max} = (z_\mathrm{max}^- - z_\mathrm{min}^-) \hat{Q}_\mathrm{max}^-$ by
\begin{equation}\label{eq:alpha-Q}
  \Par_\alpha \hat{Q}_\mathrm{max}
  = (\Par_\alpha z_\mathrm{max}^- - \Par_\alpha z_\mathrm{min}^-) \hat{Q}_\mathrm{max}^-
  + (z_\mathrm{max}^- - z_\mathrm{min}^-) \Par_\alpha \hat{Q}_\mathrm{max}^-.
\end{equation}

Any particular parameter will usually not appear in all three functional relations
in \eqref{eq:func-alpha},
which will thus simplify
\cref{eq:alpha-ocv,eq:alpha-z+,eq:alpha-z-,eq:alpha-z--ocv,eq:alpha-z-minmax}.
There are two main types of such cases.
The first type is a parameter that only affects the two electrode OCP relations
but not $z^+(z^-)$, such as temperature \citep{plett_battery_2015}
or parameters governing the component changes in composite electrode active materials
\citep{schmitt_determination_2022} if they do not also affect the electrode capacity.
In this case, \cref{eq:alpha-z+,eq:alpha-z--ocv} reduce to
\begin{equation}
  \Par_\alpha z^+(z) = - r_\mathrm{N/P} \Par_\alpha z^-(z),
\end{equation}
\begin{equation}
  \Par_\alpha z^-(U_\mathrm{OCV}) 
  = \frac{\Par_\alpha U_\mathrm{OCP}^+(z^+)
  - \Par_\alpha U_\mathrm{OCP}^-(z^-)}
  {r_\mathrm{N/P} \ud_{z^+} U_\mathrm{OCP}^+ + \ud_{z^-} U_\mathrm{OCP}^-}.
\end{equation}
We will not look into this type of parameters in this work.

The other type includes $r_\mathrm{N/P}$ and $z^+_0$ or their re-parametrization,
which are related to LLI and LAM and affect only $z^+(z^-)$ but not the electrode OCP.
In this case,
\cref{eq:alpha-ocv,eq:alpha-z--ocv} reduce to:
\begin{equation}\label{eq:alpha-ocv2}
  \Par_\alpha U_\mathrm{OCV}(z)
  = \ud_{z^+} U^+_\mathrm{OCP} \cdot \Par_\alpha z^+(z)
  - \ud_{z^-} U^-_\mathrm{OCP} \cdot \Par_\alpha z^-(z),
\end{equation}
\begin{equation}\label{eq:alpha-z--ocv2}
  \Par_\alpha z^-(U_\mathrm{OCV}) =
  \frac{\ud_{z^+} U_\mathrm{OCP}^+}
  {r_\mathrm{N/P} \ud_{z^+} U_\mathrm{OCP}^+ + \ud_{z^-} U_\mathrm{OCP}^-}
  \Par_\alpha z^+(z^-).
\end{equation}

The physical meaning of the apparently complicated fraction coefficient
in \eqref{eq:alpha-z--ocv2} can be revealed
by defining the dimensional counterparts of the electrode and cell SOC as
\begin{equation}\label{eq:Q+--z+-}
  \hat{Q}^\pm = z^\pm \hat{Q}^\pm_\mathrm{max}, \qquad \hat{Q} = z \hat{Q}_\mathrm{max},
\end{equation}
and with charge conservation $\ud \hat{Q} = \ud \hat{Q}^- = -\ud \hat{Q}^+$,
rewriting the fraction as
\begin{equation*}
  \frac{1}{r_\mathrm{N/P}}
  \frac{r_\mathrm{N/P} \ud_{z^+} U_\mathrm{OCP}^+}
  {r_\mathrm{N/P} \ud_{z^+} U_\mathrm{OCP}^+ + \ud_{z^-} U_\mathrm{OCP}^-}
  = \frac{1}{r_\mathrm{N/P}}\frac{\ud_{\hat{Q}^+} U_\mathrm{OCP}^+}
  {\ud_{\hat{Q}^+} U_\mathrm{OCP}^+ + \ud_{\hat{Q}^-} U_\mathrm{OCP}^-}
  = \frac{1}{r_\mathrm{N/P}} \frac{\ud_{\hat{Q}} U_\mathrm{OCP}^+}
  {\ud_{\hat{Q}} U_\mathrm{OCV}}.
\end{equation*}
Since $\ud_{\hat{Q}^+} U_\mathrm{OCP}^+$ and $\ud_{\hat{Q}^-} U_\mathrm{OCP}^-$
are both negative,
the second fraction factor above can be interpreted
as the fraction of the PE contribution to the full-cell differential voltage.

\begin{figure}[!htbp]
  \centering
  \begin{subfigure}{0.48\textwidth}
    \centering
    \includegraphics[width=\textwidth]{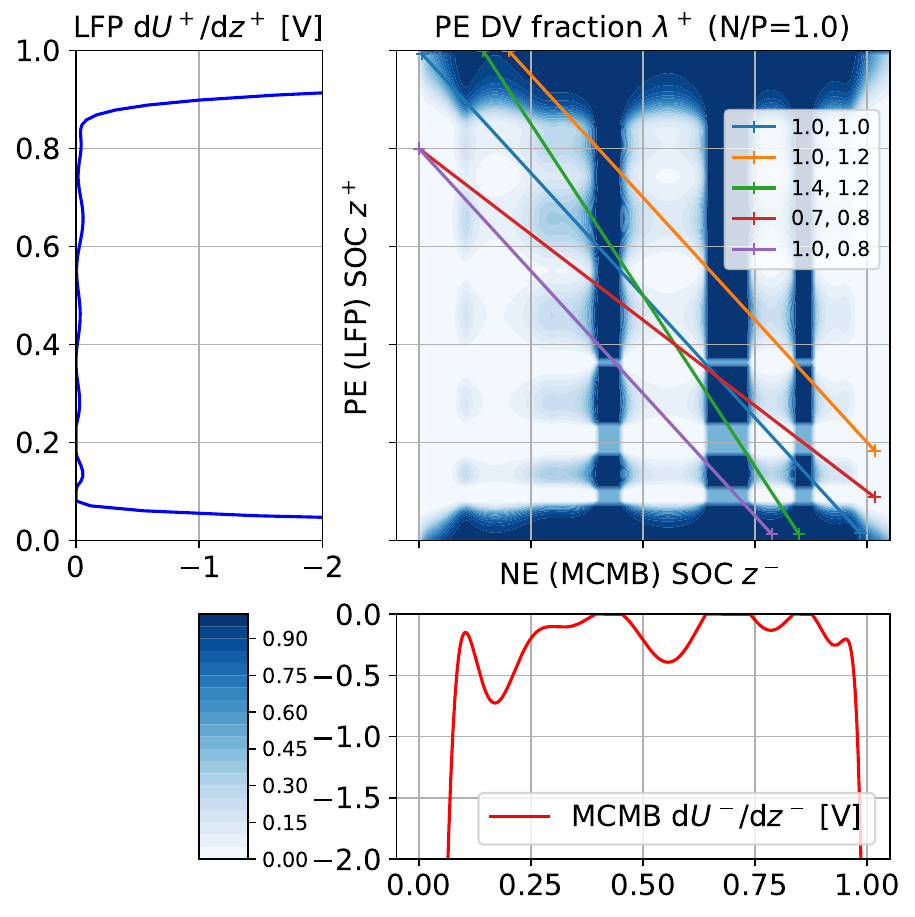}
    \caption{LFP: PE DV fraction $\lambda^+(z^-, z^+; r_\mathrm{N/P})$}
    \label{subfig:dvf-sq-lfp}
  \end{subfigure}
  \begin{subfigure}{0.48\textwidth}
    \centering
    \includegraphics[width=\textwidth]{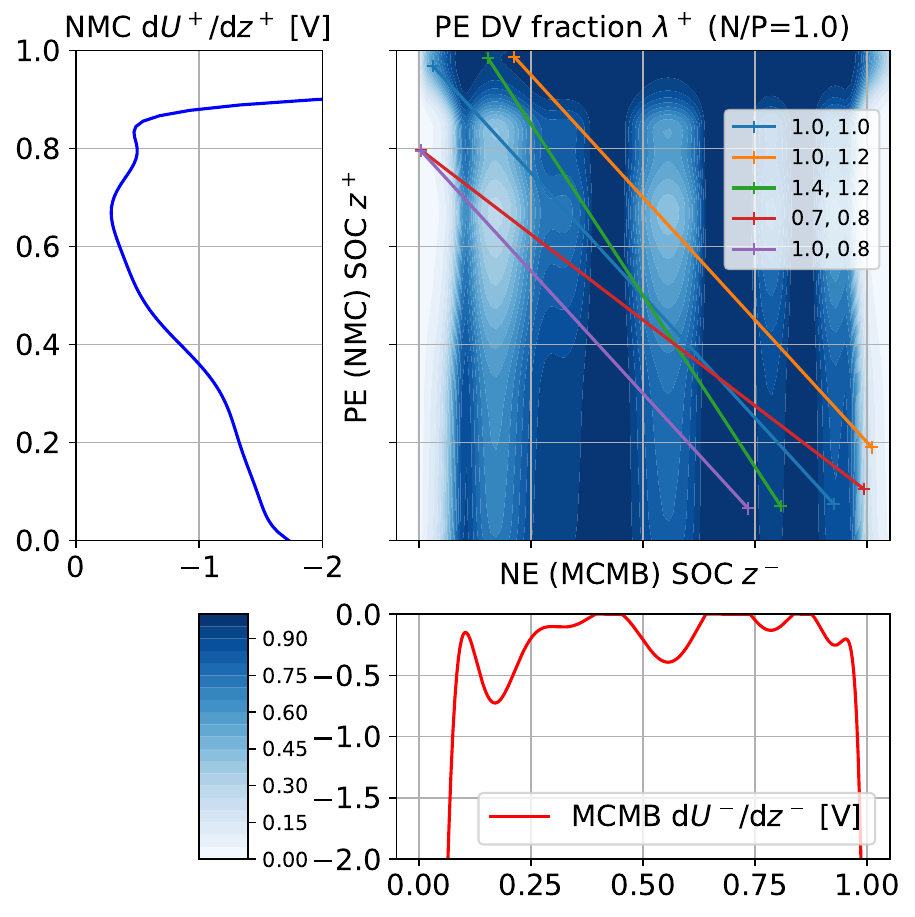}
    \caption{NMC: PE DV fraction $\lambda^+(z^-, z^+; r_\mathrm{N/P})$}
    \label{subfig:dvf-sq-nmc}
  \end{subfigure}
  \begin{subfigure}{0.48\textwidth}
    \centering
    \includegraphics[width=0.72\textwidth]{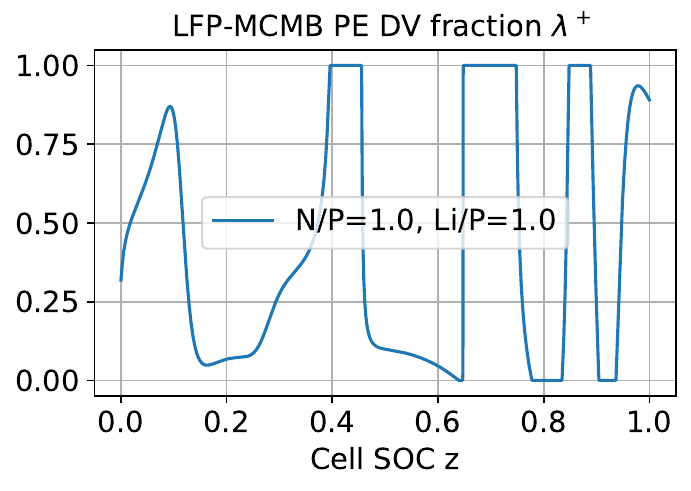}
    \caption{LFP: $\lambda^+(z; r_\mathrm{N/P}=1, z^+_0=1)$}
    \label{subfig:dvf-line-lfp}
  \end{subfigure}
  \begin{subfigure}{0.48\textwidth}
    \centering
    \includegraphics[width=0.72\textwidth]{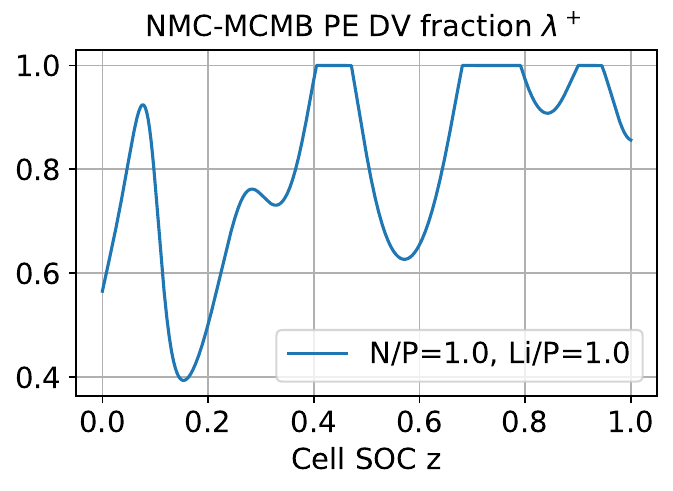}
    \caption{NMC: $\lambda^+(z; r_\mathrm{N/P}=1, z^+_0=1)$}
    \label{subfig:dvf-line-nmc}
  \end{subfigure}
  \caption{
    {\bf (a)} PE DV fraction square plots
    (inspired by but different from the DV square plots of \citet{olson_differential_2023})
    of the LFP-MCMB chemistry overlaid with five lines of $z^+(z^-)$
    corresponding to cells with different $(r_\mathrm{N/P}, z^+_0)$ pairs
    as indicated by the legends.
    Note that the color bar is associated with the contours.
    Such plots and be read analogously to
    how the OCV square plots \cref{subfig:ocv-sq-lfp,subfig:ocv-sq-nmc} are interpreted.
    {\bf (c)} The PE DV fraction with $(r_\mathrm{N/P}=1, z^+_0=1)$.
    This can again be read off from (a),
    where the peaks and troughs correspond to where the blue line in (a)
    crosses the dark and light areas, respectively.
    {\bf (b,~d)} NMC/MCMB counterparts of (a,~c).
  }
  \label{fig:dvf-sq}
\end{figure}

Due to the important role played by this fraction in subsequent sensitivity gradients
and its clear physical meaning,
we call it the PE DV (differential voltage) fraction and denote it by
\begin{equation}\label{eq:lamb+}
  \lambda^+ = \frac{r_\mathrm{N/P} \ud_{z^+} U_\mathrm{OCP}^+}
  {r_\mathrm{N/P} \ud_{z^+} U_\mathrm{OCP}^+ + \ud_{z^-} U_\mathrm{OCP}^-}
  = \frac{\ud_{\hat{Q}^+} U_\mathrm{OCP}^+}
  {\ud_{\hat{Q}^+} U_\mathrm{OCP}^+ + \ud_{\hat{Q}^-} U_\mathrm{OCP}^-} \in [0, 1]
\end{equation}
and the NE DV fraction is simply $\lambda^- = 1-\lambda^+$.
\Cref{subfig:dvf-sq-lfp,subfig:dvf-line-lfp} show $\lambda^+(z^-, z^+)$
and how $\lambda^+(z)$ is obtained analogously to
how $U_\mathrm{OCV}(z)$ is obtained from $U_\mathrm{OCV}(z^-, z^+)$
in \cref{subfig:ocv-sq-lfp,subfig:ocv-line-lfp}.
Note that compared to $U_\mathrm{OCV}(z^-, z^+)$,
which is independent of $r_\mathrm{N/P}$ or $z^+_0$,
$\lambda^+(z^-, z^+)$ as defined by \ref{eq:lamb+} does depend on $r_\mathrm{N/P}$,
although empirically this dependence is weak
and does not alter the overall pattern of $\lambda^+(z^-, z^+)$.
Therefore, we still overlay the $\lambda^+$ landscape at $r_\mathrm{N/P}=1$
in \cref{subfig:dvf-sq-lfp}
with segments of different $(r_\mathrm{N/P}, z^+_0)$ pairs for illustration purpose.

\begin{figure}[!htbp]
  \centering
  \begin{subfigure}{\textwidth}
    \centering
    \includegraphics[width=0.8\textwidth]{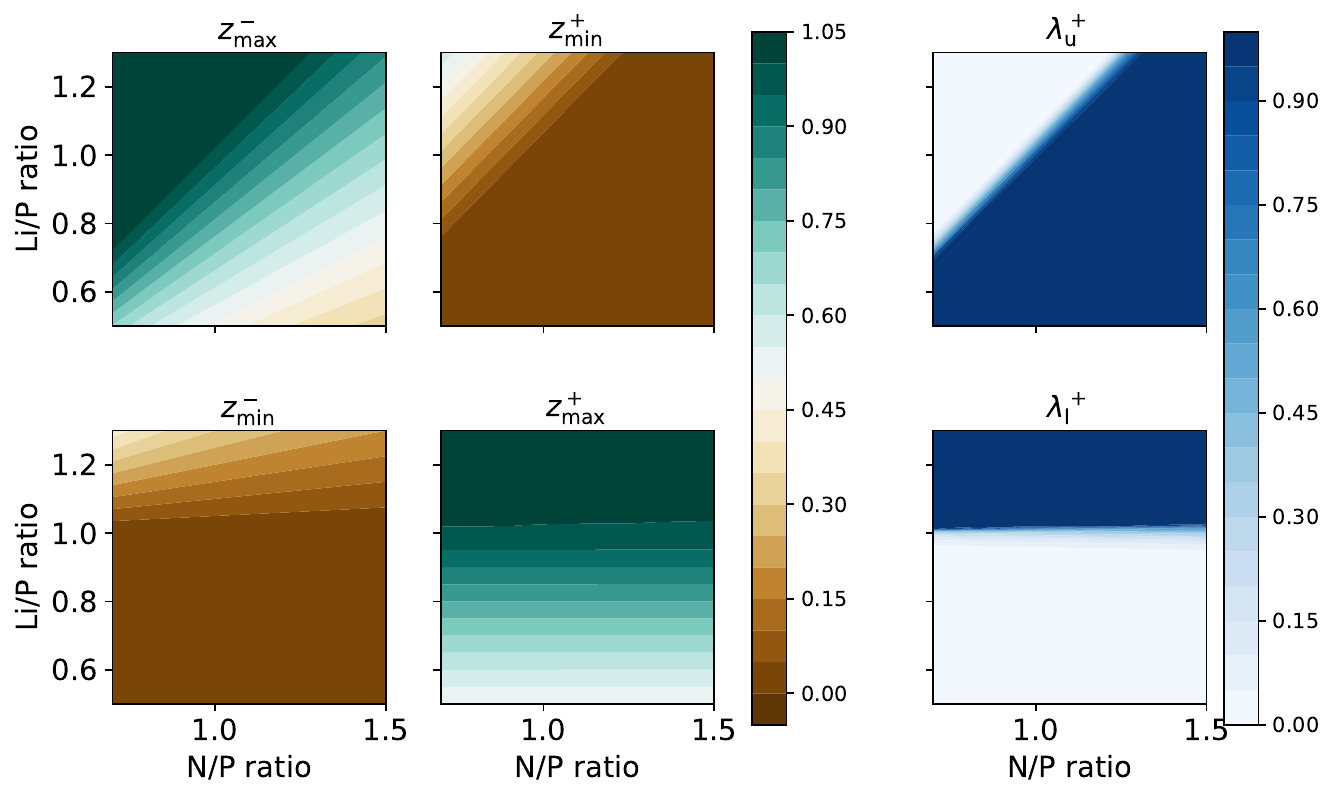}
    \caption{LFP/MCMB}
    \label{subfig:rz-zl-lfp}
  \end{subfigure}
  \begin{subfigure}{\textwidth}
    \centering
    \includegraphics[width=0.8\textwidth]{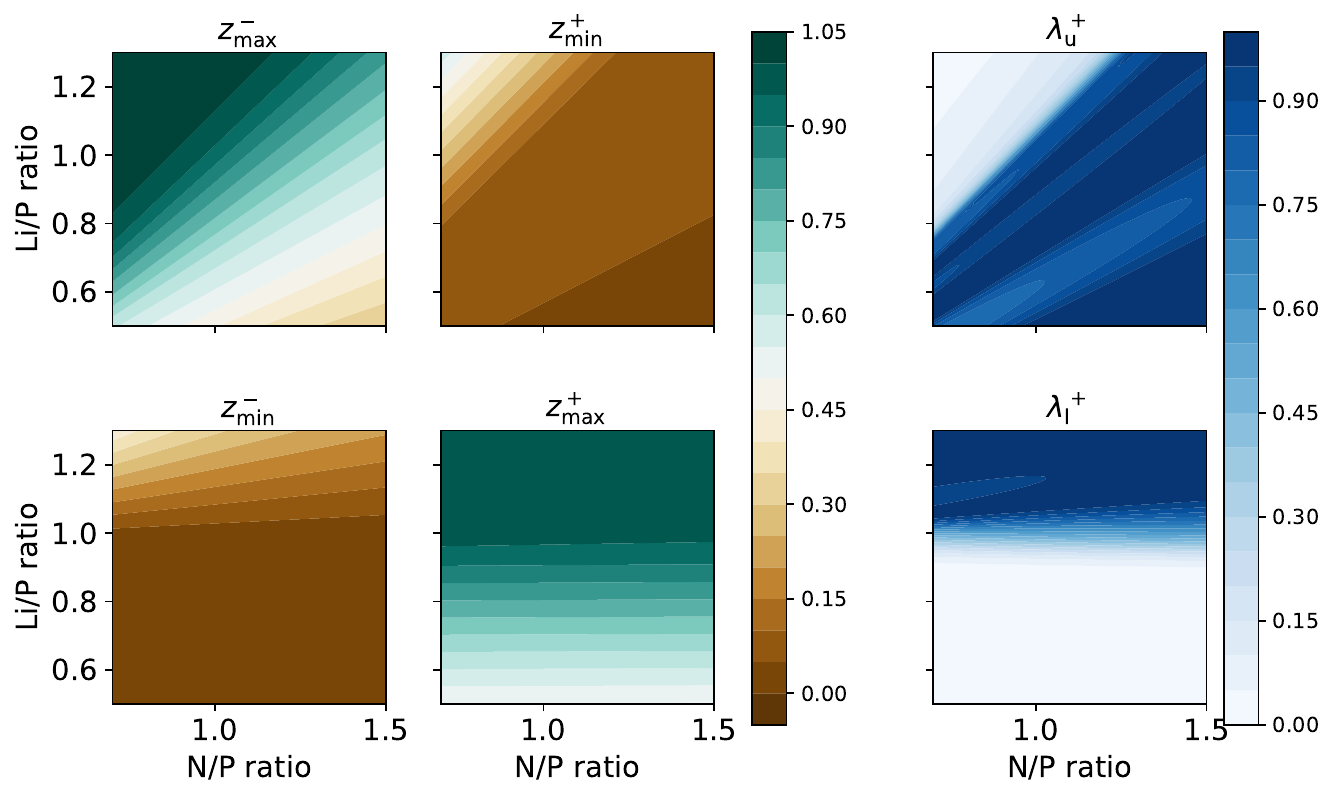}
    \caption{NMC/MCMB}
    \label{subfig:rz-zl-nmc}
  \end{subfigure}
  \caption{
    Electrode SOC limits {\bf (left)} and PE DV fractions {\bf (right)}
    at different N/P ratio and Li/P ratio.
    In either (a) or (b), the top regards
    the fully charged state $U_\mathrm{OCV}=U_\mathrm{max}$,
    where the variation mostly depends on $z^+_0/r_\mathrm{N/P}$,
    which is essentially the Li/N ratio,
    and has a sharp change at $z^+_0/r_\mathrm{N/P}=1$.
    In contrast, the bottom is at the fully discharged state $U_\mathrm{OCV}=U_\mathrm{min}$,
    whose variation mostly depends on the Li/P ratio
    and has a sharp change at $z^+_0=1$.
  }
  \label{fig:ocv-rz-zl}
\end{figure}

Since in \cref{eq:alpha-z-},
we only need the PE DV fraction \eqref{eq:alpha-z--ocv2} at the lower and upper cutoff voltage,
we specifically define
\begin{equation}
  \lambda^+_\mathrm{l}
  = \frac{r_\mathrm{N/P} \Rstr{\ud_{z^+} U_\mathrm{OCP}^+}_{z^+_\mathrm{max}}}
  {r_\mathrm{N/P} \Rstr{\ud_{z^+} U_\mathrm{OCP}^+}_{z^+_\mathrm{max}}
  + \Rstr{\ud_{z^-} U_\mathrm{OCP}^-}_{z^-_\mathrm{min}}}, \quad
  \lambda^+_\mathrm{u}
  = \frac{r_\mathrm{N/P} \Rstr{\ud_{z^+} U_\mathrm{OCP}^+}_{z^+_\mathrm{min}}}
  {r_\mathrm{N/P} \Rstr{\ud_{z^+} U_\mathrm{OCP}^+}_{z^+_\mathrm{min}}
  + \Rstr{\ud_{z^-} U_\mathrm{OCP}^-}_{z^-_\mathrm{max}}}
\end{equation}
associated with $U_\mathrm{min}$ and $U_\mathrm{max}$, respectively.
The above also implies that the sole reason why the electrode DV fractions
play a role in our sensitivity analysis is
the dependence of cell SOC $z$ and total capacity $\hat{Q}_\mathrm{max}$
on the artificially specified lower and upper cutoff voltage.

The PE DV fraction $\lambda^+_\mathrm{u}$ and $\lambda^+_\mathrm{l}$
also quantify to what extent
each electrode limits the charging or discharging towards the respective cutoff voltage,
which has been studied in the context of discerning Coulombic efficiency and capacity retention
\citep{tornheim_what_2020,rodrigues_capacity_2022}.
Note that the $\lambda$ defined in \citep{rodrigues_capacity_2022}
is just $\lambda^+_\mathrm{l}$ here,
while their $\omega$ is essentially $(\lambda^+_\mathrm{u} - 1)$.
We follow their notations to denote the PE DV fraction by letter $\lambda$,
while the definition of $\omega$ seems to
introduce an extra unnecessary quantity seemingly different from $\lambda$,
which shadows the structural symmetry between NE and PE,
and whose negative sign also makes its interpretation less straightforward.

When $\lambda^+_\mathrm{l} \approx 1$,
it implies that the cell reaches $U_\mathrm{min}$ in discharging mainly
due to the sharp drop of $U^+(z^+)$ rather than the steep increase of $U^-(z^-)$,
mostly caused by PE getting filled up with $\mathrm{Li}^+$
due to $\hat{Q}_\mathrm{max}^\mathrm{Li}>\hat{Q}_\mathrm{max}^+$,
or equivalently $z^+_0 > 1$.
This gives rise to PE-limited discharging.
In contrast, when $\lambda^+_\mathrm{l} \approx 0$,
the cell reaches $U_\mathrm{min}$ mostly due to
the $\mathrm{Li}^+$ depletion in the NE and thus the steep rise of its potential,
hence called NE-limited discharging.
We can likewise define PE- and NE-limited charging.
\Cref{fig:ocv-rz-zl} illustrates how the electrode SOC limits and PE DV fractions
vary with the N/P ratio $r_\mathrm{N/P}$ and Li/P ratio $z^+_0$,
which can be divided into four quadrants depending on
whether $z^+_0 > 1$ and $z^+_0 > r_\mathrm{N/P}$,
or more intuitively, whether $\hat{Q}_\mathrm{max}^\mathrm{Li}>\hat{Q}_\mathrm{max}^+$
and $\hat{Q}_\mathrm{max}^\mathrm{Li}>\hat{Q}_\mathrm{max}^-$.
The NMC case has a more gradual transition across these regimes than the LFP case.
We will go into more depth concerning these four regimes in \cref{subsec:dq-ideal,subsec:dq}.

With the introduction of the PE DV fraction $\lambda^+$,
\cref{eq:alpha-z-minmax} simplifies to
\begin{equation}\label{eq:alpha-z--minmax3}
  \Par_\alpha z^-_\mathrm{min} = \frac{\lambda^+_\mathrm{l}}{r_\mathrm{N/P}}
  \Par_\alpha z^+(z^-_\mathrm{min}), \quad
  \Par_\alpha z^-_\mathrm{max} = \frac{\lambda^+_\mathrm{u}}{r_\mathrm{N/P}}
  \Par_\alpha z^+(z^-_\mathrm{max}),
\end{equation}
which paves the way for deriving the sensitivity gradients with respect to
$r_\mathrm{N/P}$ and $z^+_0$.

\subsection{SOH-Sensitivity Gradients of Electrode SOC Limits and Cell OCV}
\label{subsec:dzpm-du}

Before proceeding,
we must point out a subtlety omitted in previous parts.
A partial derivative with respect to a particular parameter $\alpha$ itself is ambiguous,
because it also depends on the complete parametrization,
i.e.~choices of parameters other than $\alpha$.
All the previous results still hold,
but we must specify the full set of parameters before carrying out concrete calculations.

We know that $U_\mathrm{OCV}(z)$ can be parametrized by $r_\mathrm{N/P}$ and $z^+_0$,
so $(r_\mathrm{N/P}, z^+_0)$ serves as the complete parameter set for
the following partial differentiation.
From \cref{eq:z+z-}, we can easily obtain
\begin{equation}
  \Par_{r_\mathrm{N/P}} z^+(z^-) = -z^-, \qquad
  \Par_{z^+_0} z^+(z^-) = 1.
\end{equation}

We first derive the counterparts of
\cref{eq:alpha-ocv,eq:alpha-z+,eq:alpha-z-,eq:alpha-z--ocv,eq:alpha-z-minmax}
for $r_\mathrm{N/P}$:
\begin{equation}
  \PDer{z^-(U_\mathrm{OCV})}{r_\mathrm{N/P}} =
  \Rstr{\frac{-z^- \lambda^+}{r_\mathrm{N/P}}}_{U_\mathrm{OCV}},
\end{equation}
\begin{equation}\label{eq:rn2p-z-minmax}
  \PDer{z^-_\mathrm{min}}{r_\mathrm{N/P}} =
  \frac{-z^-_\mathrm{min} \lambda^+_\mathrm{l}}{r_\mathrm{N/P}}, \qquad
  \PDer{z^-_\mathrm{max}}{r_\mathrm{N/P}} =
  \frac{-z^-_\mathrm{max} \lambda^+_\mathrm{u}}{r_\mathrm{N/P}},
\end{equation}
\begin{equation}
  \PDer{z^-(z)}{r_\mathrm{N/P}}
  = -\frac{(1-z) z^-_\mathrm{min} \lambda^+_\mathrm{l}
  + z z^-_\mathrm{max} \lambda^+_\mathrm{u}}
  {r_\mathrm{N/P}}, \qquad
  \PDer{z^+(z)}{r_\mathrm{N/P}}
  = -(1-z) z^-_\mathrm{min} \lambda^-_\mathrm{l}
  - z z^-_\mathrm{max} \lambda^-_\mathrm{u},
\end{equation}
\begin{equation}
  \PDer{U_\mathrm{OCV}(z)}{r_\mathrm{N/P}}
  = \Der{U^+_\mathrm{OCP}}{z^+} \PDer{z^+(z)}{r_\mathrm{N/P}}
  - \Der{U^-_\mathrm{OCP}}{z^-} \PDer{z^-(z)}{r_\mathrm{N/P}}.
\end{equation}
As we can see, both $\Par_{r_\mathrm{N/P}} z^\pm(z)$ vary linearly with $z$,
which is not surprising since $z^\pm(z)$ have to remain linear by definition of $z$.

\begin{figure}[!htbp]
  \centering
  \begin{subfigure}{0.4\textwidth}
    \centering
    \includegraphics[width=\textwidth]{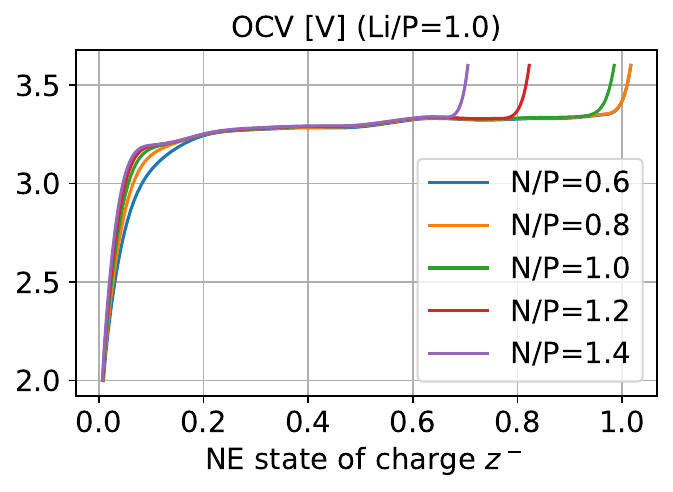}
    \caption{LFP: $U_\mathrm{OCV}(z^-; r_\mathrm{N/P}, z^+_0=1)$}
    \label{subfig:ocv-n2p-a}
  \end{subfigure}
  \begin{subfigure}{0.4\textwidth}
    \centering
    \includegraphics[width=\textwidth]{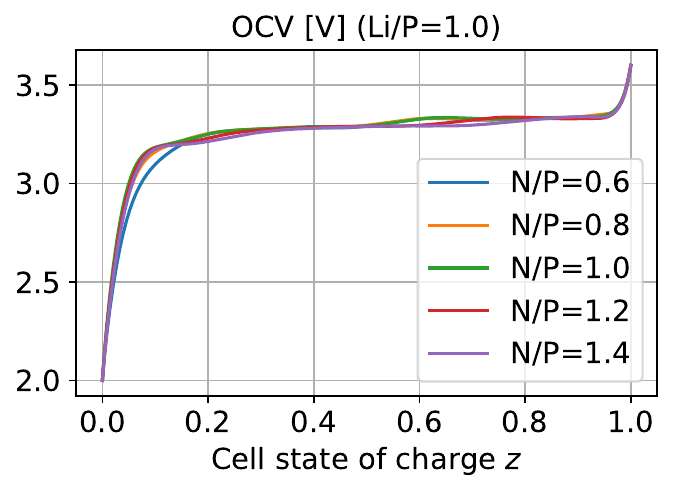}
    \caption{LFP: $U_\mathrm{OCV}(z; r_\mathrm{N/P}, z^+_0=1)$}
    \label{subfig:ocv-n2p-b}
  \end{subfigure}
  \begin{subfigure}{0.4\textwidth}
    \centering
    \includegraphics[width=\textwidth]{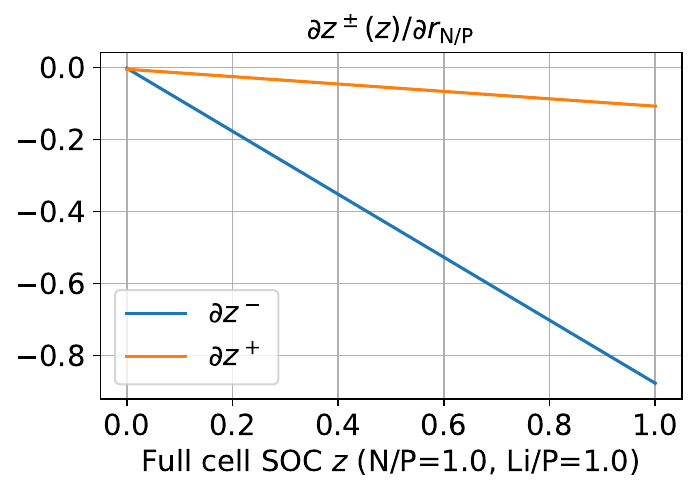}
    \caption{LFP: $\Par_{r_\mathrm{N/P}} z^\pm(z; r_\mathrm{N/P}=1, z^+_0=1)$}
    \label{subfig:ocv-n2p-c}
  \end{subfigure}
  \begin{subfigure}{0.4\textwidth}
    \centering
    \includegraphics[width=\textwidth]{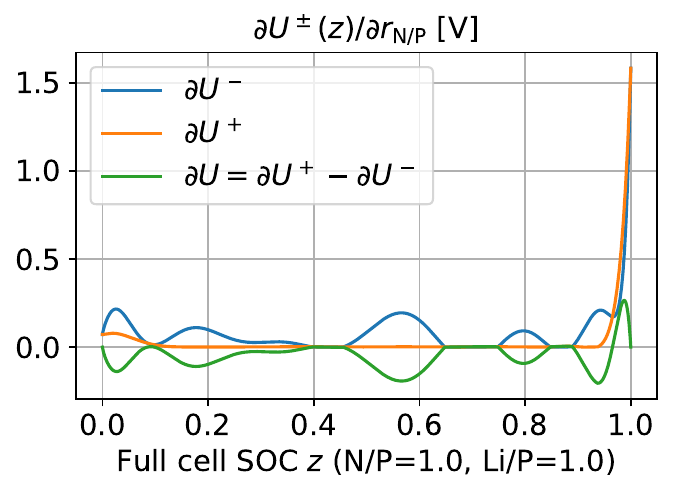}
    \caption{
      LFP: $\Par_{r_\mathrm{N/P}} U_\mathrm{OCP}^\pm(z)$
      and $\Par_{r_\mathrm{N/P}} U_\mathrm{OCV}(z)$
    }
    \label{subfig:ocv-n2p-d}
  \end{subfigure}
  \begin{subfigure}{0.4\textwidth}
    \centering
    \includegraphics[width=\textwidth]{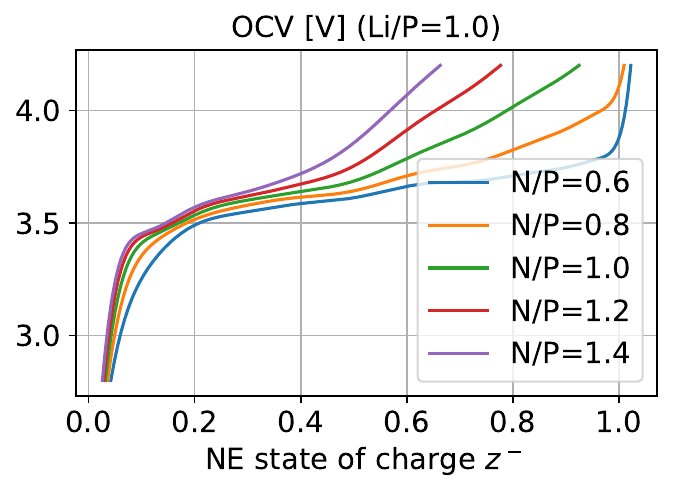}
    \caption{NMC: $U_\mathrm{OCV}(z^-; r_\mathrm{N/P}, z^+_0=1)$}
    \label{subfig:ocv-n2p-e}
  \end{subfigure}
  \begin{subfigure}{0.4\textwidth}
    \centering
    \includegraphics[width=\textwidth]{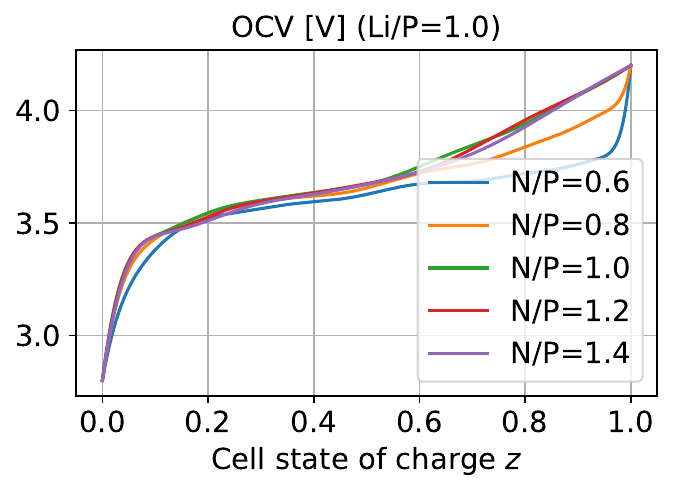}
    \caption{NMC: $U_\mathrm{OCV}(z; r_\mathrm{N/P}, z^+_0=1)$}
    \label{subfig:ocv-n2p-f}
  \end{subfigure}
  \caption{
    {\bf (a,~b)} Variation of NE SOC $z^-$-based (left)
    and cell SOC $z$-based (right) cell OCV when $r_\mathrm{N/P}$ changes and $z^+_0=1$.
    {\bf (c,~d)} Derivative-based local sensitivity of $z^\pm(z)$ (left)
    and of $z$-based electrode OCPs and cell OCV (right)
    with respect to $r_\mathrm{N/P}$.
    All the sensitivity derivatives are evaluated at $r_\mathrm{N/P}=1$ and $z^+_0=1$.
    {\bf (e,~f)} NMC/MCMB counterparts of (a,~b).
  }
  \label{fig:ocv-n2p}
\end{figure}

\Cref{fig:ocv-n2p} demonstrates empirically how the $z^-$-based OCV $U_\mathrm{OCV}(z^-)$
and $z$-based OCV $U_\mathrm{OCV}(z)$ vary with the N/P ratio $r_\mathrm{N/P}$ on the top,
and the $r_\mathrm{N/P}$-derivatives $\Par_{r_\mathrm{N/P}} z^\pm(z)$,
$\Par_{r_\mathrm{N/P}} U_\mathrm{OCP}^\pm(z)$,
and $\Par_{r_\mathrm{N/P}} U_\mathrm{OCV}(z)$ evaluated at $r_\mathrm{N/P}=1$.
\Cref{subfig:ocv-n2p-a} shows that
$z^-_\mathrm{max}$ quickly decreases with increasing $r_\mathrm{N/P}$,
while $z^-_\mathrm{min}$ is insensitive to $r_\mathrm{N/P}$ in this case,
which is consistent with $\Par_{r_\mathrm{N/P}} z^-(z=1)<-0.8$
and $\Par_{r_\mathrm{N/P}} z^-(z=0)\approx 0$ as indicated by \cref{subfig:ocv-n2p-c}.
On the other hand,
due to the OCP curves in both electrodes being rather flat for a large mid-SOC range,
$U_\mathrm{OCV}(z)$ is not very sensitive to $r_\mathrm{N/P}$ overall (\cref{subfig:ocv-n2p-b})
and the several narrow SOC windows with relatively high sensitivity
are also aligned with what the $r_\mathrm{N/P}$-derivatives imply in \cref{subfig:ocv-n2p-d}.

Next, we carry out the same sensitivity analysis with respect to $z^+_0$:
\begin{equation}
  \PDer{z^-(U_\mathrm{OCV})}{z^+_0} =
  \Rstr{\frac{\lambda^+}{r_\mathrm{N/P}}}_{U_\mathrm{OCV}},
\end{equation}
\begin{equation}\label{eq:z+0-z-minmax}
  \PDer{z^-_\mathrm{min}}{z^+_0} =
  \frac{\lambda^+_\mathrm{l}}{r_\mathrm{N/P}}, \qquad
  \PDer{z^-_\mathrm{max}}{z^+_0} =
  \frac{\lambda^+_\mathrm{u}}{r_\mathrm{N/P}},
\end{equation}
\begin{equation}
  \PDer{z^-(z)}{z^+_0}
  = \frac{(1-z) \lambda^+_\mathrm{l} + z \lambda^+_\mathrm{u}}
  {r_\mathrm{N/P}}, \qquad
  \PDer{z^+(z)}{z^+_0}
  = (1-z)\lambda^-_\mathrm{l} + z \lambda^-_\mathrm{u},
\end{equation}
\begin{equation}
  \PDer{U_\mathrm{OCV}(z)}{z^+_0}
  = \Der{U^+_\mathrm{OCP}}{z^+} \PDer{z^+(z)}{z^+_0}
  - \Der{U^-_\mathrm{OCP}}{z^-} \PDer{z^-(z)}{z^+_0}.
\end{equation}

\begin{figure}[!htbp]
  \centering
  \begin{subfigure}{0.4\textwidth}
    \centering
    \includegraphics[width=\textwidth]{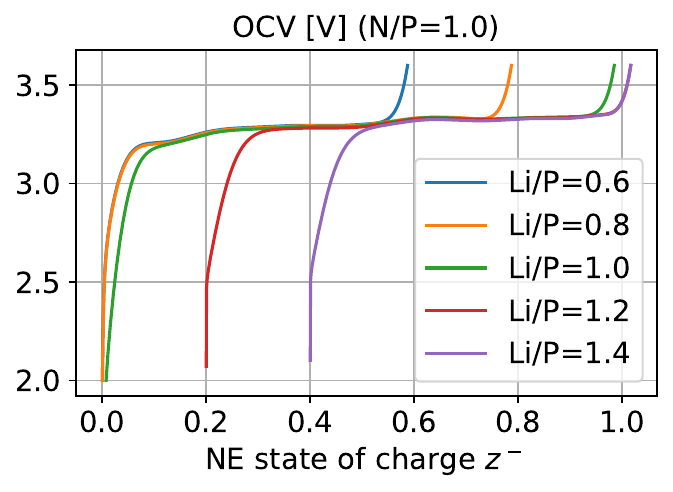}
    \caption{LFP: $U_\mathrm{OCV}(z^-; r_\mathrm{N/P}=1, z^+_0)$}
    \label{subfig:ocv-li2p-a}
  \end{subfigure}
  \begin{subfigure}{0.4\textwidth}
    \centering
    \includegraphics[width=\textwidth]{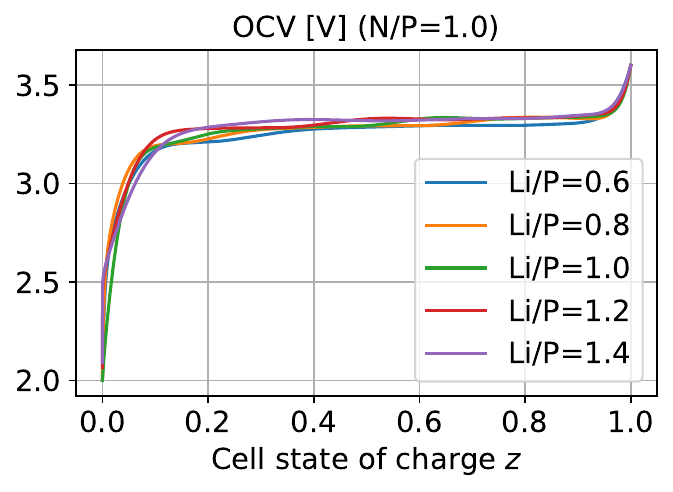}
    \caption{LFP: $U_\mathrm{OCV}(z; r_\mathrm{N/P}=1, z^+_0)$}
    \label{subfig:ocv-li2p-b}
  \end{subfigure}
  \begin{subfigure}{0.4\textwidth}
    \centering
    \includegraphics[width=\textwidth]{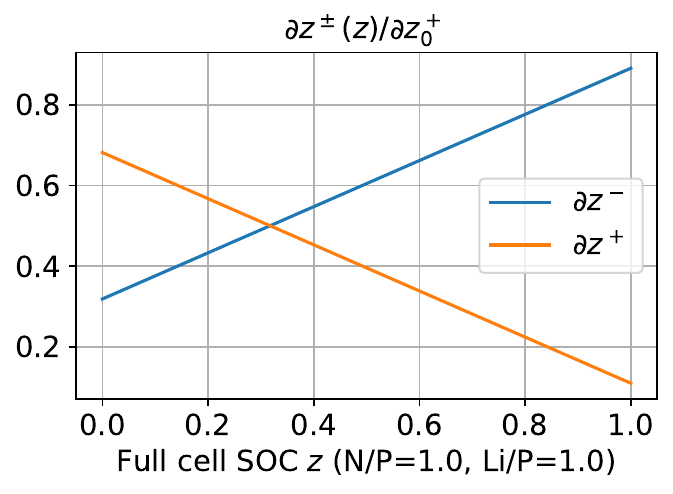}
    \caption{LFP: $\Par_{z^+_0} z^\pm(z; r_\mathrm{N/P}=1, z^+_0=1)$}
    \label{subfig:ocv-li2p-c}
  \end{subfigure}
  \begin{subfigure}{0.4\textwidth}
    \centering
    \includegraphics[width=\textwidth]{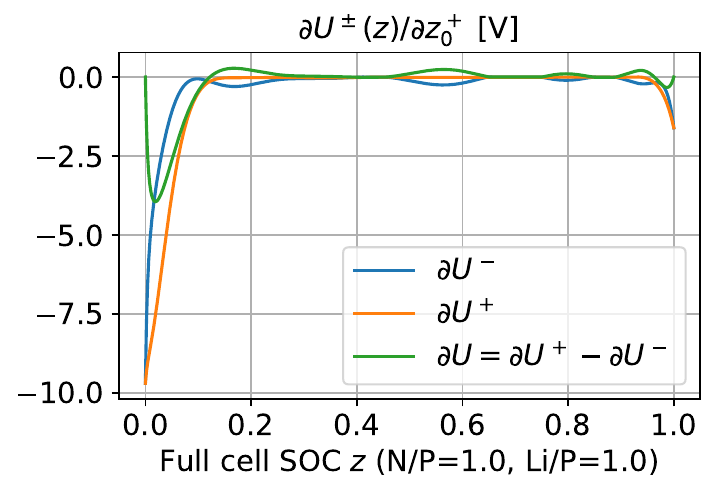}
    \caption{
      LFP: $\Par_{z^+_0} U_\mathrm{OCP}^\pm(z)$
      and $\Par_{z^+_0} U_\mathrm{OCV}(z)$
    }
    \label{subfig:ocv-li2p-d}
  \end{subfigure}
  \begin{subfigure}{0.4\textwidth}
    \centering
    \includegraphics[width=\textwidth]{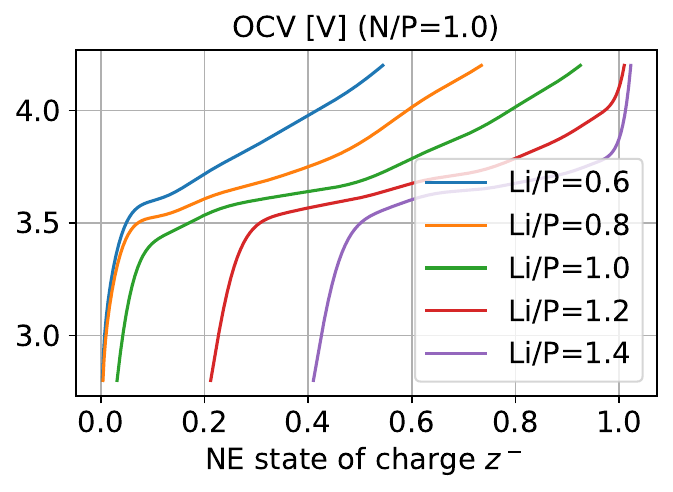}
    \caption{NMC: $U_\mathrm{OCV}(z^-; r_\mathrm{N/P}=1, z^+_0)$}
    \label{subfig:ocv-li2p-e}
  \end{subfigure}
  \begin{subfigure}{0.4\textwidth}
    \centering
    \includegraphics[width=\textwidth]{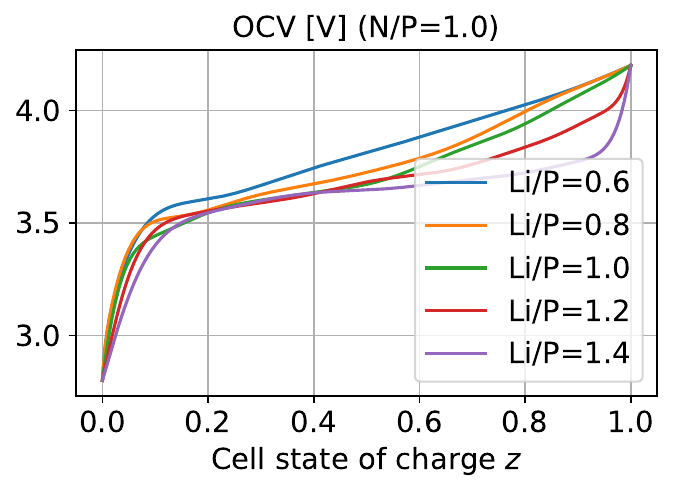}
    \caption{NMC: $U_\mathrm{OCV}(z; r_\mathrm{N/P}=1, z^+_0)$}
    \label{subfig:ocv-li2p-f}
  \end{subfigure}
  \caption{
    {\bf (a,~b)} Variation of NE SOC $z^-$-based (left)
    and cell SOC $z$-based (right) cell OCV when Li/P ratio $z^+_0$ changes and $r_\mathrm{N/P}=1$.
    {\bf (c,~d)} Derivative-based local sensitivity of $z^\pm(z)$ (left)
    and of $z$-based electrode OCPs and cell OCV (right)
    with respect to $z^+_0$.
    All the sensitivity derivatives are evaluated at $r_\mathrm{N/P}=1$ and $z^+_0=1$.
    {\bf (e,~f)} NMC/MCMB counterparts of (a,~b).
  }
  \label{fig:ocv-li2p}
\end{figure}

Again, \cref{fig:ocv-li2p}, which is the $z^+_0$-counterpart of \cref{fig:ocv-n2p},
shows the consistency between empirical and analytical sensitivity results.
Note that the electrode SOC limits always increase with the lithium inventory
as implied by \cref{subfig:ocv-li2p-a,subfig:ocv-li2p-c},
which is also consistent with \cref{fig:ocv-rz-zl}.

\subsection{Total Capacity without Cutoff Voltage: Four Characteristic Regimes}
\label{subsec:dq-ideal}

The cell total capacity $\hat{Q}_\mathrm{max}$ is typically associated with an OCV window,
so its sensitivity to SOH parameters such as $\hat{Q}_\mathrm{max}^\mathrm{Li}$
and $\hat{Q}_\mathrm{max}^\pm$ will inevitably be complicated
by the relative contributions of the two electrodes to the OCV slope,
as the previous sensitivity gradients are.
Before diving into this case,
here we will first look into an ideal total capacity $Q_\mathrm{max}$
defined solely by exhausting the usable electrode SOC range, irrespective of the voltage,
which will be compared later to the case with a voltage window.

To exhaust the usable SOC range without over-lithiation or over-delithiation,
a cell is discharged until either $z^+$ reaches 1 or $z^-$ reaches 0, whichever is earlier.
Likewise, a cell is charged until one of $z^+$ reaching 0 and $z^-$ reaching 1 occurs.
By \cref{eq:z+z-},
we can work out the criterion determining which alternative is the case
as listed in \cref{tab:zpn-limit-ideal}.

\begin{table}[!htbp]
  \centering
  \begin{tabular}{ccc|ccc}
  \toprule
  & \multicolumn{2}{c|}{Fully discharged} & & \multicolumn{2}{c}{Fully charged} \\
  & $z^-_\mathrm{min}$ & $z^+_\mathrm{max}$ & & $z^-_\mathrm{max}$ & $z^+_\mathrm{min}$ \\
  \midrule
  $\mathrm{Li/P} > 1$ & $\mathrm{(Li-P)}/\mathrm{N}$ & $1$
  & $\mathrm{Li/N} > 1$ & $1$ & $\mathrm{(Li-N)}/\mathrm{P}$ \\
  $\mathrm{Li/P} \le 1$ & $0$ & $\mathrm{Li}/\mathrm{P}$
  & $\mathrm{Li/N} \le 1$ & $\mathrm{Li}/\mathrm{N}$ & $0$ \\
  \bottomrule
  \end{tabular}
  \caption{
    Electrode SOC limits at the ideal case.
    Note that the Li/P ratio is $z^+_0$
    and the Li/N ratio is $z^+_0/r_\mathrm{N/P}$.
    We also use Li, N, and P as shorthand notations
    for $\hat{Q}_\mathrm{max}^\mathrm{Li}$, $\hat{Q}_\mathrm{max}^-$,
    and $\hat{Q}_\mathrm{max}^+$, respectively.
  }
  \label{tab:zpn-limit-ideal}
\end{table}

The combinations of the two alternatives in each state
naturally give rise to four regimes
where the ideal total capacity
$Q_\mathrm{max}=\hat{Q}_\mathrm{max}^-(z^-_\mathrm{max} - z^-_\mathrm{min})$
translates to different expressions that
depend on $\hat{Q}_\mathrm{max}^\mathrm{Li}$ and $\hat{Q}_\mathrm{max}^\pm$ differently,
as tabulated in \cref{tab:4-regimes-ideal}.

\begin{table}[!htbp]
  \centering
  \begin{tabular}{ccc}
  \toprule
  & {\bf Li $\ge$ N} & {\bf Li $<$ N} \\
  \midrule
  \multirow[c]{2}{*}{\bf Li $\ge$ P}
  & $\mathrm{Li}>\mathrm{N,~P}$
  & $\mathrm{N > Li > P}$ \\
  & $Q_\mathrm{max}=\mathrm{N + P - Li}$
  & $Q_\mathrm{max}=\mathrm{P}$ \\
  \midrule
  \multirow[c]{2}{*}{\bf Li $<$ P}
  & $\mathrm{P > Li > N}$
  & $\mathrm{Li < N,~P}$ \\
  & $Q_\mathrm{max}=\mathrm{N}$
  & $Q_\mathrm{max}=\mathrm{Li}$ \\
  \bottomrule
  \end{tabular}
  \caption{
    Four regimes of dependence of ideal cell total capacity $Q_\mathrm{max}$
    on lithium inventory $\hat{Q}_\mathrm{max}^\mathrm{Li}$
    and active materials $\hat{Q}_\mathrm{max}^\pm$.
    The four regimes are divided by the relative amount
    between $\hat{Q}_\mathrm{max}^\mathrm{Li}$ and $\hat{Q}_\mathrm{max}^-$
    and between $\hat{Q}_\mathrm{max}^\mathrm{Li}$ and $\hat{Q}_\mathrm{max}^+$.
    The Li/N ratio determines whether the end-of-charge is limited
    by NE filling up or PE being emptied,
    and $\hat{Q}_\mathrm{max}^-$ is only identifiable when NE is limiting end-of-charge.
    Likewise, the Li/P ratio determines whether the end-of-discharge is limited
    by PE filling up or NE being emptied,
    and $\hat{Q}_\mathrm{max}^+$ is only identifiable when PE is limiting end-of-discharge.
    In contrast, $\hat{Q}_\mathrm{max}^\mathrm{Li}$ is only identifiable
    when the two ratio are both larger or smaller than 1,
    where the surplus and lack of lithium become the capacity bottleneck, respectively.
  }
  \label{tab:4-regimes-ideal}
\end{table}

\citet{dubarry_best_2022} discuss in their figure 6 the relative capacity loss expressions
when the start and end state fall into each of these four regimes.
Here we further elucidate the criteria that determine which regime a cell is in
and derive a simple expression directly relating the total capacity to the SOH parameters.
Our results also reveal the counter-intuitive negative dependence of total capacity
on lithium inventory when there is a surplus of the latter,
which is missed in previous work and will be discussed in more detail later.

\subsection{SOH-Sensitivity Gradients of Cutoff-Voltage-Based Total Capacity: Four Characteristic Regimes}
\label{subsec:dq}

Besides for the cell OCV $U_\mathrm{OCV}(z)$,
we can also compute the sensitivity derivatives for the total capacity
$\hat{Q}_\mathrm{max} = (z_\mathrm{max}^- - z_\mathrm{min}^-)
r_\mathrm{N/P} \hat{Q}_\mathrm{max}^+$
by \cref{eq:alpha-Q}.
Note, however, that besides the dimensionless $r_\mathrm{N/P}$ and $z^+_0$,
$\hat{Q}_\mathrm{max}$ also depends on the dimensional $\hat{Q}_\mathrm{max}^+$.
Therefore, for $\hat{Q}_\mathrm{max}(r_\mathrm{N/P}, z^+_0, \hat{Q}_\mathrm{max}^+)$,
we can obtain the following parameter-derivatives:
\begin{equation}\label{eq:param-Q}
  \PDer{\hat{Q}_\mathrm{max}}{r_\mathrm{N/P}} =
  (z^-_\mathrm{max} \lambda^-_\mathrm{u} - z^-_\mathrm{min} \lambda^-_\mathrm{l})
   \hat{Q}_\mathrm{max}^+,
  \quad \PDer{\hat{Q}_\mathrm{max}}{z^+_0} =
  (\lambda^+_\mathrm{u} - \lambda^+_\mathrm{l})
  \hat{Q}_\mathrm{max}^+,
  \quad \PDer{\hat{Q}_\mathrm{max}}{\hat{Q}_\mathrm{max}^+} =
  (z_\mathrm{max}^- - z_\mathrm{min}^-)r_\mathrm{N/P}
\end{equation}
where we have used \cref{eq:rn2p-z-minmax,eq:z+0-z-minmax}.

In contrast to $U_\mathrm{OCV}(z)$,
the total capacity $\hat{Q}_\mathrm{max}$ can be alternatively parametrized
by $\hat{Q}_\mathrm{max}^\mathrm{Li}$, $\hat{Q}_\mathrm{max}^-$, and $\hat{Q}_\mathrm{max}^+$,
all of which are of the same dimension of charges.
To derive the parametric derivatives for this case,
one subtlety is that
we need to translate $z^+(z^-; r_\mathrm{N/P}, z^+_0)$ to
\begin{equation}
  z^+(z^-; \hat{Q}_\mathrm{max}^\mathrm{Li}, \hat{Q}_\mathrm{max}^-, \hat{Q}_\mathrm{max}^+)
  = \frac{\hat{Q}_\mathrm{max}^\mathrm{Li}}{\hat{Q}_\mathrm{max}^+}
  - \frac{\hat{Q}_\mathrm{max}^-}{\hat{Q}_\mathrm{max}^+} z^-,
\end{equation}
which yields
\begin{equation}
  \PDer{z^+(z^-)}{\hat{Q}_\mathrm{max}^\mathrm{Li}}
  = \frac{1}{\hat{Q}_\mathrm{max}^+}, \qquad
  \PDer{z^+(z^-)}{\hat{Q}_\mathrm{max}^-} = \frac{- z^-}{\hat{Q}_\mathrm{max}^+}, \qquad
  \PDer{z^+(z^-)}{\hat{Q}_\mathrm{max}^+} = \frac{- z^+}{\hat{Q}_\mathrm{max}^+}.
\end{equation}
Note that in deriving \eqref{eq:param-Q},
we have implicitly used
\begin{equation*}
  \PDer{z^+(z^-; r_\mathrm{N/P}, z^+_0, \hat{Q}_\mathrm{max}^+)}{\hat{Q}_\mathrm{max}^+}
  = 0 \neq
  \PDer{z^+(z^-; \hat{Q}_\mathrm{max}^\mathrm{Li}, \hat{Q}_\mathrm{max}^-, \hat{Q}_\mathrm{max}^+)}
  {\hat{Q}_\mathrm{max}^+},
\end{equation*}
which exemplifies the partial differentiation ambiguity
without specifying the complete parametrization as mentioned.
Now using \cref{eq:alpha-z--minmax3}, we have
\begin{equation}\label{eq:z-maxmin-QLi}
  \PDer{z^-_\mathrm{min}}{\hat{Q}_\mathrm{max}^\mathrm{Li}} =
  \frac{\lambda^+_\mathrm{l}}{\hat{Q}_\mathrm{max}^-}, \qquad
  \PDer{z^-_\mathrm{max}}{\hat{Q}_\mathrm{max}^\mathrm{Li}} =
  \frac{\lambda^+_\mathrm{u}}{\hat{Q}_\mathrm{max}^-},
\end{equation}
\begin{equation}\label{eq:z-maxmin-Q-}
  \PDer{z^-_\mathrm{min}}{\hat{Q}_\mathrm{max}^-} =
  \frac{-z^-_\mathrm{min} \lambda^+_\mathrm{l}}{\hat{Q}_\mathrm{max}^-}, \qquad
  \PDer{z^-_\mathrm{max}}{\hat{Q}_\mathrm{max}^-} =
  \frac{-z^-_\mathrm{max} \lambda^+_\mathrm{u}}{\hat{Q}_\mathrm{max}^-},
\end{equation}
\begin{equation}\label{eq:z-maxmin-Q+}
  \PDer{z^-_\mathrm{min}}{\hat{Q}_\mathrm{max}^+} =
  \frac{-z^+_\mathrm{max} \lambda^+_\mathrm{l}}{\hat{Q}_\mathrm{max}^-}, \qquad
  \PDer{z^-_\mathrm{max}}{\hat{Q}_\mathrm{max}^+} =
  \frac{-z^+_\mathrm{min} \lambda^+_\mathrm{u}}{\hat{Q}_\mathrm{max}^-},
\end{equation}
which by \cref{eq:alpha-Q}, finally give the sensitivity derivatives:
\begin{equation}\label{eq:QLiQ-Q+-Q}
  \PDer{\hat{Q}_\mathrm{max}}{\hat{Q}_\mathrm{max}^\mathrm{Li}}
  = \lambda^+_\mathrm{u} - \lambda^+_\mathrm{l}, \qquad
  \PDer{\hat{Q}_\mathrm{max}}{\hat{Q}_\mathrm{max}^-}
  = z^-_\mathrm{max} \lambda^-_\mathrm{u}
  - z^-_\mathrm{min} \lambda^-_\mathrm{l}, \qquad
  \PDer{\hat{Q}_\mathrm{max}}{\hat{Q}_\mathrm{max}^+}
  = z^+_\mathrm{max} \lambda^+_\mathrm{l} -z^+_\mathrm{min} \lambda^+_\mathrm{u}.
\end{equation}

There are several advantages of the above all-capacity parametrization.
First, it is conceptually straightforward
because the cell total capacity $\hat{Q}_\mathrm{max}$ is the direct result of
the synergy between the lithium inventory $\hat{Q}_\mathrm{max}^\mathrm{Li}$,
the anode capacity $\hat{Q}_\mathrm{max}^-$,
and the cathode capacity $\hat{Q}_\mathrm{max}^+$.
Moreover, all of they are of the same dimension and thus directly comparable,
which also makes the parametric derivatives dimensionless and intuitive
with values being readily interpretable.
For example, $\Par_{\hat{Q}_\mathrm{max}^-} \hat{Q}_\mathrm{max}$ indicates
the fraction of NE capacity change that indeed manifests in cell total capacity.
As a result, the three derivatives in \cref{eq:QLiQ-Q+-Q}
can be used to indicate to what extent each factor is the bottleneck
dominating the change of total capacity.
A derivative $\Par_Q \hat{Q}_\mathrm{max}$ close to $1$ implies the factor being limiting,
while a value close to $0$ hints the opposite.

Interestingly, if we examine the sign and range of these three derivatives carefully,
we can find that they are not necessarily positive.
We know that $0 \le \lambda^+ \le 1$,
and if each electrode potential window covers the whole active range in operation
as is typically the case,
we also have $0 \le z_\mathrm{min}^\pm < z_\mathrm{max}^\pm \le 1$.
Under these conditions, we have the following bounds:
\begin{equation}
  -1 \le \Par_{\hat{Q}_\mathrm{max}^\mathrm{Li}} \hat{Q}_\mathrm{max},~
  \Par_{\hat{Q}_\mathrm{max}^-} \hat{Q}_\mathrm{max},~
  \Par_{\hat{Q}_\mathrm{max}^+} \hat{Q}_\mathrm{max} \le 1.
\end{equation}
Moreover, $z_\mathrm{min}^\pm$ is often close to $0$
while $z_\mathrm{max}^\pm$ close to $1$,
so typically we should have
\begin{equation*}
  \Par_{\hat{Q}_\mathrm{max}^-} \hat{Q}_\mathrm{max},
  ~\Par_{\hat{Q}_\mathrm{max}^+} \hat{Q}_\mathrm{max} \ge 0,
\end{equation*}
and they will hardly be much smaller than $0$ even if they ever become negative.
A negative derivative of this kind simply means
loss of active materials or lithium inventory will, counterintuitively,
increase the total capacity.

For lithium inventory $\hat{Q}_\mathrm{max}^\mathrm{Li}$,
it is actually not hard to imagine such a case.
Cell total capacity indicates the amount of $\mathrm{Li}^+$ that can be shuttled
from one electrode to the other,
so too few $\mathrm{Li}^+$ will definitely limit the capacity,
but so will too many $\mathrm{Li}^+$
because there will be no space for them to move, so both electrodes easily get filled up.
In that case, discharging is PE-limited so $\lambda^+_\mathrm{l}\approx 1$,
while charging is NE-limited so $\lambda^+_\mathrm{u}\approx 0$,
and the two together yield
$\Par_{\hat{Q}_\mathrm{max}^\mathrm{Li}} \hat{Q}_\mathrm{max} \approx -1$.
It is therefore intuitive why loss of lithium inventory in such scenarios
will boost the total capacity.
However, it should also be noted that a normal cell rarely falls into this regime
because in typical cell assembly,
only the PE is filled with $\mathrm{Li}^+$
while the NE is totally $\mathrm{Li}^+$-free.
A cell with surplus lithium is usually manufactured deliberately
by prelithiating the NE before the assembly \citep{birkl_degradation_2017}.

\begin{figure}[!htbp]
  \centering
  \begin{subfigure}{0.4\textwidth}
    \centering
    \includegraphics[width=\textwidth]{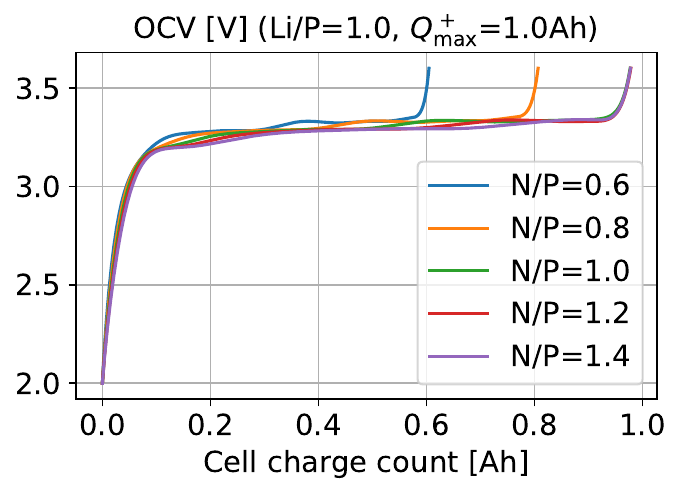}
    \caption{
      LFP: $U_\mathrm{OCV}(Q_\mathrm{c};
      r_\mathrm{N/P}, z^+_0=1, \hat{Q}_\mathrm{max}^+ =1\mathrm{Ah})$
    }
    \label{subfig:ocv-q-r-lfp}
  \end{subfigure}
  \begin{subfigure}{0.4\textwidth}
    \centering
    \includegraphics[width=\textwidth]{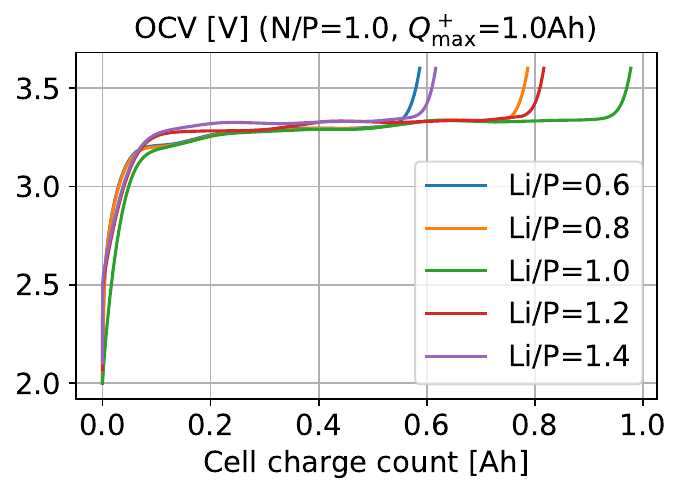}
    \caption{
      LFP: $U_\mathrm{OCV}(Q_\mathrm{c};
      r_\mathrm{N/P}=1, z^+_0, \hat{Q}_\mathrm{max}^+ =1\mathrm{Ah})$
    }
    \label{subfig:ocv-q-z-lfp}
  \end{subfigure}
  \begin{subfigure}{0.8\textwidth}
    \centering
    \includegraphics[width=\textwidth]{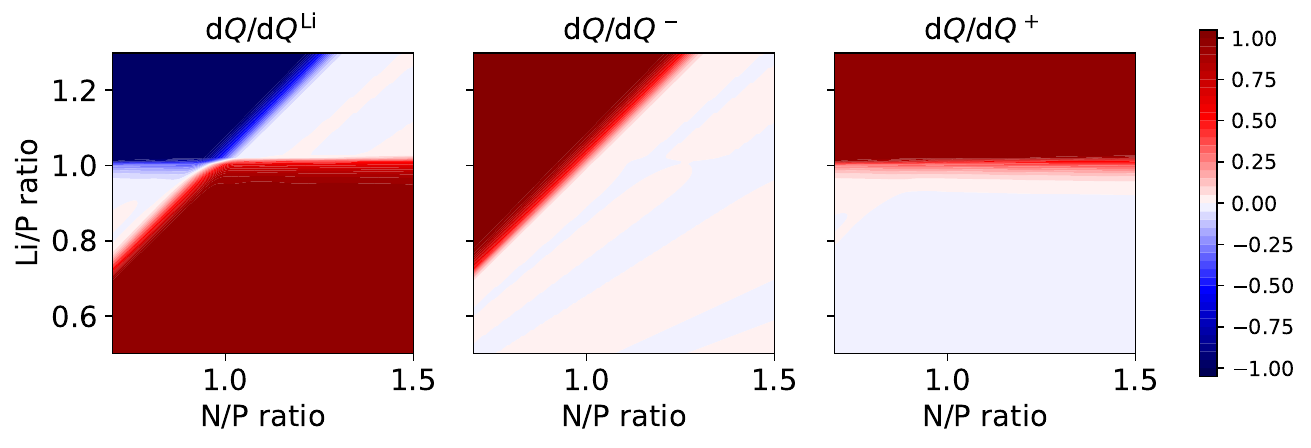}
    \caption{
      LFP: $\Par_{\hat{Q}_\mathrm{max}^\mathrm{Li}} \hat{Q}_\mathrm{max}$,
      $\Par_{\hat{Q}_\mathrm{max}^-} \hat{Q}_\mathrm{max}$,
      and $\Par_{\hat{Q}_\mathrm{max}^+} \hat{Q}_\mathrm{max}$
    }
    \label{subfig:ocv-dq-lfp}
  \end{subfigure}
  \begin{subfigure}{0.4\textwidth}
    \centering
    \includegraphics[width=\textwidth]{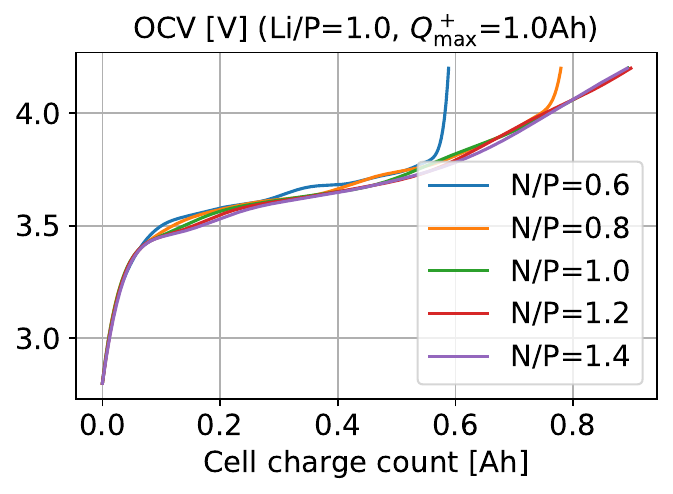}
    \caption{
      NMC: $U_\mathrm{OCV}(Q_\mathrm{c};
      r_\mathrm{N/P}, z^+_0=1, \hat{Q}_\mathrm{max}^+ =1\mathrm{Ah})$
    }
    \label{subfig:ocv-q-r-nmc}
  \end{subfigure}
  \begin{subfigure}{0.4\textwidth}
    \centering
    \includegraphics[width=\textwidth]{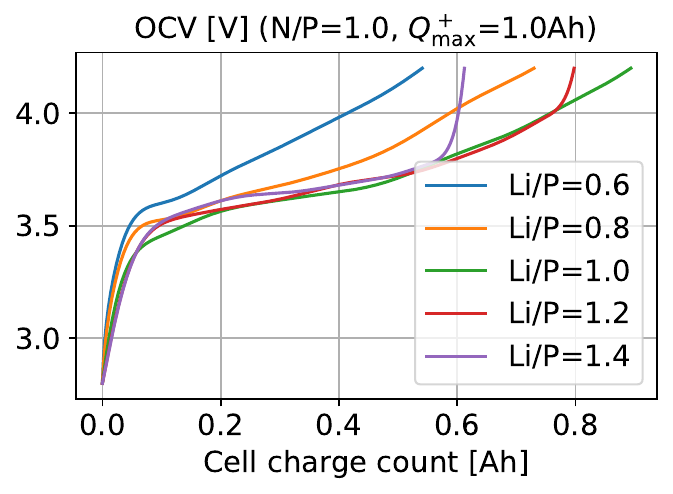}
    \caption{
      NMC: $U_\mathrm{OCV}(Q_\mathrm{c};
      r_\mathrm{N/P}=1, z^+_0, \hat{Q}_\mathrm{max}^+ =1\mathrm{Ah})$
    }
    \label{subfig:ocv-q-z-nmc}
  \end{subfigure}
  \begin{subfigure}{0.8\textwidth}
    \centering
    \includegraphics[width=\textwidth]{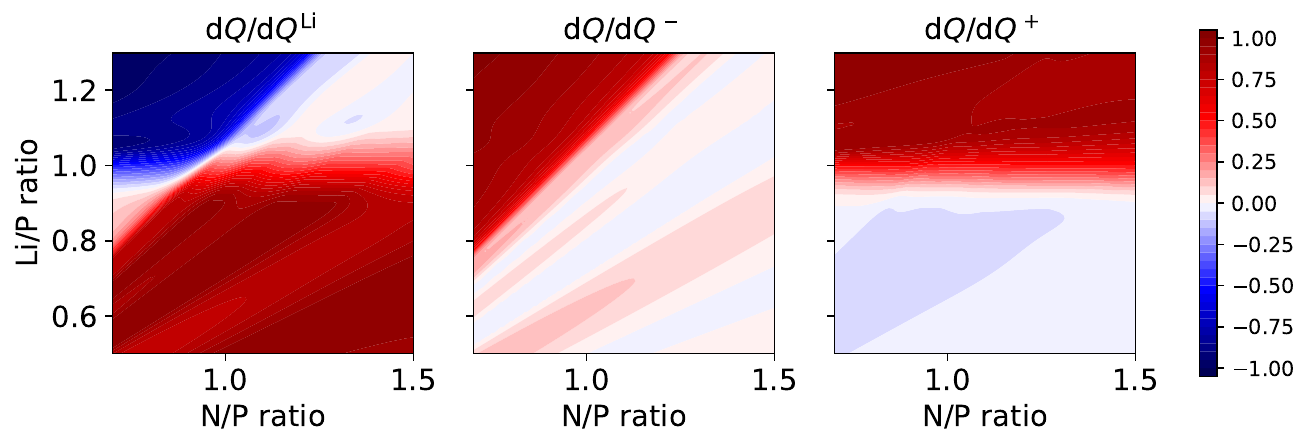}
    \caption{
      NMC: $\Par_{\hat{Q}_\mathrm{max}^\mathrm{Li}} \hat{Q}_\mathrm{max}$,
      $\Par_{\hat{Q}_\mathrm{max}^-} \hat{Q}_\mathrm{max}$,
      and $\Par_{\hat{Q}_\mathrm{max}^+} \hat{Q}_\mathrm{max}$
    }
    \label{subfig:ocv-dq-nmc}
  \end{subfigure}
  \caption{
    {\bf (a,~b)} Variation of $Q_\mathrm{c}$-based cell OCV and total capacity $\hat{Q}_\mathrm{max}$
    when $r_\mathrm{N/P}$ changes and $z^+_0=1$ (left)
    and when $z^+_0$ changes and $r_\mathrm{N/P}=1$ (right),
    both assuming $\hat{Q}^+_\mathrm{max} = 1\mathrm{Ah}$.
    For the former case, this implies that $\hat{Q}_\mathrm{max}^-$ changes
    while $\hat{Q}_\mathrm{max}^\mathrm{Li}=\hat{Q}^+_\mathrm{max} = 1\mathrm{Ah}$.
    {\bf (c)} Gradient-based sensitivity of total capacity
    with respect to the amount of lithium inventory, NE, and PE active materials
    at different $(r_\mathrm{N/P}, z^+_0)$ pairs.
    {\bf (d,~e,~f)} NMC/MCMB counterparts of (a,~b,~c).
  }
  \label{fig:ocv-q}
\end{figure}

Note that although $\hat{Q}_\mathrm{max}$ depends on all three of
$\hat{Q}_\mathrm{max}^\mathrm{Li}$, $\hat{Q}_\mathrm{max}^-$, and $\hat{Q}_\mathrm{max}^+$,
the dimensionless derivatives in \eqref{eq:QLiQ-Q+-Q} only depend on
the also dimensionless $r_\mathrm{N/P}$ and $z^+_0$.
Therefore, we can visualize the different regimes of a cell
where the total capacity is limited by the three components to different extents
by plotting the three profiles of
$\Par_{\hat{Q}_\mathrm{max}^{(\cdot)}} \hat{Q}_\mathrm{max} (r_\mathrm{N/P}, z^+_0)$
as shown in \cref{fig:ocv-q}.
If we look at \cref{fig:ocv-q,fig:ocv-rz-zl} together,
we can see how the four-quadrant pattern from
$z^\pm_\mathrm{min/max}$ and $\lambda^+_\mathrm{l/u}$
propagates into $\Par_{\hat{Q}_\mathrm{max}^{(\cdot)}} \hat{Q}_\mathrm{max}$
through \eqref{eq:QLiQ-Q+-Q}.

The characteristics of the four SOH regimes are summarized in \cref{tab:4-regimes}.
In conclusion, the four regimes are divided by the relative amount
between $\hat{Q}_\mathrm{max}^\mathrm{Li}$ and $\hat{Q}_\mathrm{max}^-$
and between $\hat{Q}_\mathrm{max}^\mathrm{Li}$ and $\hat{Q}_\mathrm{max}^+$.
The Li/N ratio determines whether the end-of-charge is limited
by NE filling up or PE being emptied,
and $\hat{Q}_\mathrm{max}^-$ is highly identifiable when NE is limiting end-of-charge.
Likewise, the Li/P ratio determines whether the end-of-discharge is limited
by PE filling up or NE being emptied,
and $\hat{Q}_\mathrm{max}^+$ is highly identifiable when PE is limiting end-of-discharge.
In contrast, $\hat{Q}_\mathrm{max}^\mathrm{Li}$ is highly identifiable
when the two ratio are both larger or smaller than 1,
where the surplus and lack of lithium become the capacity bottleneck, respectively.

\begin{table}[!htbp]
  \centering
  \begin{tabular}{ccc}
  \toprule
  & 
  {\bf Li $>$ N}
  & 
  {\bf Li $<$ N} \\
  & $\lambda^-_\mathrm{u} \rightarrow 1, \quad
  \Par_{\hat{Q}_\mathrm{max}^-} \hat{Q}_\mathrm{max} \rightarrow 1$
  & $\lambda^-_\mathrm{u} \rightarrow 0, \quad
  \Par_{\hat{Q}_\mathrm{max}^-} \hat{Q}_\mathrm{max} \rightarrow 0$ \\
  \midrule
  {\bf Li $>$ P}
  & $\hat{Q}_\mathrm{max}^\mathrm{Li} > \hat{Q}_\mathrm{max}^\pm$
  & $\hat{Q}_\mathrm{max}^- > \hat{Q}_\mathrm{max}^\mathrm{Li} > \hat{Q}_\mathrm{max}^+$ \\
  $\lambda^+_\mathrm{l} \rightarrow 1, \quad
  \Par_{\hat{Q}_\mathrm{max}^+} \hat{Q}_\mathrm{max} \rightarrow 1$
  & $\Par_{\hat{Q}_\mathrm{max}^\mathrm{Li}} \hat{Q}_\mathrm{max} \rightarrow -1$
  & $\Par_{\hat{Q}_\mathrm{max}^\mathrm{Li}} \hat{Q}_\mathrm{max} \rightarrow 0$ \\
  \midrule
  {\bf Li $<$ P}
  & $\hat{Q}_\mathrm{max}^+ > \hat{Q}_\mathrm{max}^\mathrm{Li} > \hat{Q}_\mathrm{max}^-$
  & $\hat{Q}_\mathrm{max}^\mathrm{Li} < \hat{Q}_\mathrm{max}^\pm$ \\
  $\lambda^+_\mathrm{l} \rightarrow 0, \quad
  \Par_{\hat{Q}_\mathrm{max}^+} \hat{Q}_\mathrm{max} \rightarrow 0$
  & $\Par_{\hat{Q}_\mathrm{max}^\mathrm{Li}} \hat{Q}_\mathrm{max} \rightarrow 0$
  & $\Par_{\hat{Q}_\mathrm{max}^\mathrm{Li}} \hat{Q}_\mathrm{max} \rightarrow 1$ \\
  \bottomrule
  \end{tabular}
  \caption{
    Four regimes of sensitivity of cell total capacity $\hat{Q}_\mathrm{max}$
    to lithium inventory $\hat{Q}_\mathrm{max}^\mathrm{Li}$
    and active materials $\hat{Q}_\mathrm{max}^\pm$.
    Compare to \cref{tab:4-regimes-ideal}.
  }
  \label{tab:4-regimes}
\end{table}

On the other hand, the other parametrization
$\hat{Q}_\mathrm{max}(r_\mathrm{N/P}, z^+_0, \hat{Q}_\mathrm{max}^+)$
is not without advantage.
We know the $z$-based cell OCV $U_\mathrm{OCV}(z)$
depends on $r_\mathrm{N/P}$ and $z^+_0$ only,
and $\hat{Q}^+_\mathrm{max}$ only comes into the picture
when the cell total capacity $\hat{Q}_\mathrm{max}$ is concerned.
Therefore, the original parametrization confines all the dimensional contributions
to $\hat{Q}_\mathrm{max}$ within the single $\hat{Q}^+_\mathrm{max}$,
which is structurally more aligned with the underlying parametric dependence.

\section{Applications of Analytic Sensitivity}
\label{sec:applications}

To quantify identifiability is essentially
to revert the parameter-to-measurement forward sensitivity,
and there are different approaches to this inversion.
A straightforward approach is to devise an estimator
and calibrate the estimation error by feeding measurements coming from a known ground truth
\citep{birkl_degradation_2017,dubarry_state_2017}.
The drawback of this approach is that
it entangles the identifiability intrinsic to the problem
with the error incurred by the estimator itself.
If real data are used,
the imperfectness of the model used will further complicate the apparent identifiability.

To study only the identifiability intrinsic to a problem,
one simple approach is based on the sensitivity gradients,
as we shall adopt here.
However, this approach only concerns how the measurements vary with the parameters locally.
To quantify global identifiability,
more sophisticated techniques such as those based on Bayesian inversion are needed
\cite{berliner_nonlinear_2021}.

\subsection{N/P and Li/P Identifiability from SOC-Based OCV Measurements by Fisher Information}
\label{subsec:fisher-z}

One immediate application of the sensitivity gradients is
to quantify the minimal estimation errors in the latent SOH parameters
from noisy voltage measurements.
The simple idea is
if we measure $y=f(\theta)$ with standard error $\sigma_y$ to estimate $\theta$,
and we know $\theta$ and $y$ are close to some $\theta_0$ and $y_0=f(\theta_0)$, respectively,
we can linearize $f(\theta)$ around $\theta_0$
and thus $\sigma_y/\sigma_\theta=\ud_\theta f(\theta_0)$,
which yields a standard error $\sigma_\theta = \sigma_y/\ud_\theta f(\theta_0)$.
When this is generalized to the vector case with $m$ measurements
$\mbs{y}=\mbs{f}(\mbs{\theta})\in \R^m$
and $d$ underlying parameters $\mbs{\theta}\in \R^d$,
the derivative becomes the Jacobi matrix $\nabla_{\mbs{\theta}}\mbs{f}\in\R^{m\times d}$
and the $\sigma_{\mbs{y}}^2$ and $\sigma_\theta^2$ become
the error covariance $\mbs{C}_{\mbs{y}}\in\R^{m\times m}$
and $\mbs{C}_{\mbs{\theta}}\in\R^{d\times d}$, respectively.
Aided by a square root of the symmetric positive definite $\mbs{C}_{\mbs{y}}$,
we can obtain the vector counterpart of $\sigma_\theta = \sigma_y/\ud_\theta f(\theta_0)$ as
\begin{equation}\label{eq:FImat}
  \mbs{C}_{\mbs{\theta}}^{-1}(\mbs{\theta}_0) = \nabla_{\mbs{\theta}}\T\mbs{f}(\mbs{\theta}_0)
  \mbs{C}_{\mbs{y}}^{-1} \nabla_{\mbs{\theta}}\mbs{f}(\mbs{\theta}_0).
\end{equation}

If we solve the weighted NLS (nonlinear least squares) to obtain
\begin{equation}\label{eq:NLS}
  \theta_*(\hat{\mbs{y}}) = \argmin_{\mbs{\theta}} (\mbs{f}(\mbs{\theta}) - \hat{\mbs{y}})\T
  \mbs{C}_{\mbs{y}}^{-1} (\mbs{f}(\mbs{\theta}) - \hat{\mbs{y}}),
\end{equation}
where $\hat{\mbs{y}}$ are actual measured values of $\mbs{y}$,
substituting $\mbs{\theta}_*$ to $\mbs{\theta}_0$ in \eqref{eq:FImat}
yields a standard error covariance for the optimizer $\mbs{\theta}_*$.
Note, however, that such gradient-based identifiability results are only valid locally
and are based on the premise that $\mbs{\theta}_*$ is near
the ground truth $\mbs{\theta}_\mathrm{tr}$
and $\hat{\mbs{y}}$ are noisy unbiased measurements
of $\mbs{f}(\mbs{\theta}_\mathrm{tr})$ with covariance $\mbs{C}_{\mbs{y}}$.
If the nonlinear optimization does not ever get close to $\mbs{\theta}_\mathrm{tr}$,
of course the above results are irrelevant to
how close $\mbs{\theta}_*$ is to $\mbs{\theta}_\mathrm{tr}$.

When the weighted NLS optimizer $\mbs{\theta}_*$ is equivalently formulated
as the maximum likelihood estimator
$\mbs{\theta}_*(\hat{\mbs{y}})=\argmax_{\mbs{\theta}} p(\hat{\mbs{y}}|\mbs{\theta})$
with a Gaussian likelihood
$\mbs{y}|\mbs{\theta}\sim \mathcal{N}(\mbs{f}(\mbs{\theta}), \mbs{C}_{\mbs{y}})$,
\eqref{eq:FImat} turns out to be the well known Fisher information matrix \citep{ly_tutorial_2017}
\begin{equation}\label{eq:FImat2}
  \mbs{F}_{\mbs{y}}(\theta_*) = \E_{\mbs{y}|\mbs{\theta}_*}
  [\nabla_{\mbs{\theta}}\T \ln p(\mbs{y}|\mbs{\theta_*})
  \nabla_{\mbs{\theta}} \ln p(\mbs{y}|\mbs{\theta_*})]
  = \nabla_{\mbs{\theta}}\T\mbs{f}(\mbs{\theta}_*)
  \mbs{C}_{\mbs{y}}^{-1} \nabla_{\mbs{\theta}}\mbs{f}(\mbs{\theta}_*).
\end{equation}
Under this probabilistic framework,
the inverse Fisher information matrix is the Cram\'{e}r-Rao lower bound
of the covariance of all unbiased estimator $\hat{\mbs{\theta}}(\hat{y})$,
i.e.~$(\mathrm{Cov}(\hat{\mbs{\theta}}(\hat{y})) - \mbs{F}_{\mbs{y}}^{-1}(\theta_\mathrm{tr}))$
is semi-positive definite \citep[sec.~11.10]{cover_elements_2006}.
Although we do not know $\theta_\mathrm{tr}$ in practice,
and the maximum likelihood estimator $\mbs{\theta}_*(\hat{\mbs{y}})$ is not necessarily unbiased,
we can still use the inverse of \eqref{eq:FImat2} as a semi-heuristic error covariance,
as \citep{mohtat_towards_2019,lee_estimation_2020} have done
for their OCV model parametrization \eqref{eq:ocv-Qd}.
Due to the cutoff-voltage constraint between the two electrode SOC limits in their parameters,
extra treatments are needed to evaluate the Fisher information matrix.
The independent and cutoff-voltage-agnostic parametrization introduced in this work
allows this to be done with more ease.

In practice, we can usually assume that the errors in different measurements are independent,
which translates to $\mbs{C}_{\mbs{y}} = \mathrm{Diag}(\sigma_{y_i}^2)$ being diagonal.
In this case, the contributions of different measurements to $\mbs{C}_{\mbs{\theta}}^{-1}$
become additive rank-1 updates:
\begin{equation}\label{eq:NLS2}
  \mbs{C}_{\mbs{\theta}}^{-1} = \sum_{i=1}^{m} \frac{1}{\sigma_{y_i}^2}
  \nabla_{\mbs{\theta}}\T f_i(\mbs{\theta}_0)
  \nabla_{\mbs{\theta}} f_i(\mbs{\theta}_0).
\end{equation}

If a complete OCV curve has been observed,
naturally the total capacity $\hat{Q}_\mathrm{max}$ is also directly measured by Coulomb counting.
What remains to be done is identifying the N/P and Li/P ratio $\mbs{\theta}= [r_\mathrm{N/P}, z^+_0]$
from the $(z_i, \hat{U}_i)$ pairs.
Assume $z_i$ is accurate and every $\hat{U}_i$ has a standard error $\sigma_U$.
Since $f_i(\theta) = U_\mathrm{OCV}(z_i;r_\mathrm{N/P}, z^+_0)$ here, we have
\begin{equation}
  \nabla_{\mbs{\theta}} U_\mathrm{OCV}(z_i) =
  [\Par_{r_\mathrm{N/P}} U_\mathrm{OCV}(z_i), \Par_{z^+_0} U_\mathrm{OCV}(z_i)],
\end{equation}
which together with \eqref{eq:NLS2} evaluated at the NLS optimizer $\mbs{\theta}_*$
yields error covariance for $[r_\mathrm{N/P}, z^+_0]$.

\begin{figure}[!htbp]
  \centering
  \begin{subfigure}{0.48\textwidth}
    \centering
    \includegraphics[width=\textwidth]{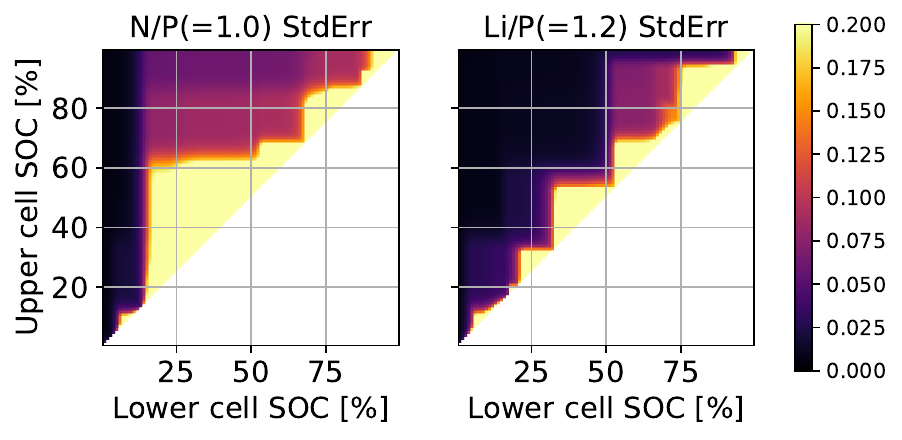}
    \caption{LFP: $r_\mathrm{N/P}=1$ and $z^+_0=1.2$}
  \end{subfigure}
  \begin{subfigure}{0.48\textwidth}
    \centering
    \includegraphics[width=\textwidth]{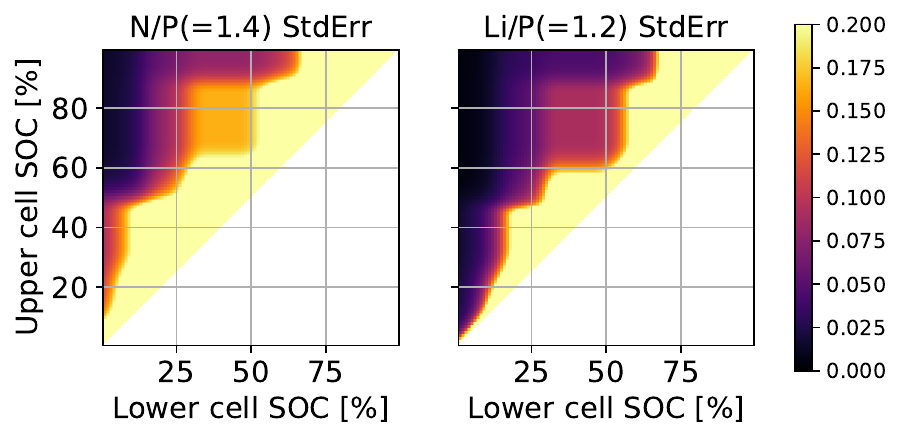}
    \caption{LFP: $r_\mathrm{N/P}=1.4$ and $z^+_0=1.2$}
  \end{subfigure}
  \begin{subfigure}{0.48\textwidth}
    \centering
    \includegraphics[width=\textwidth]{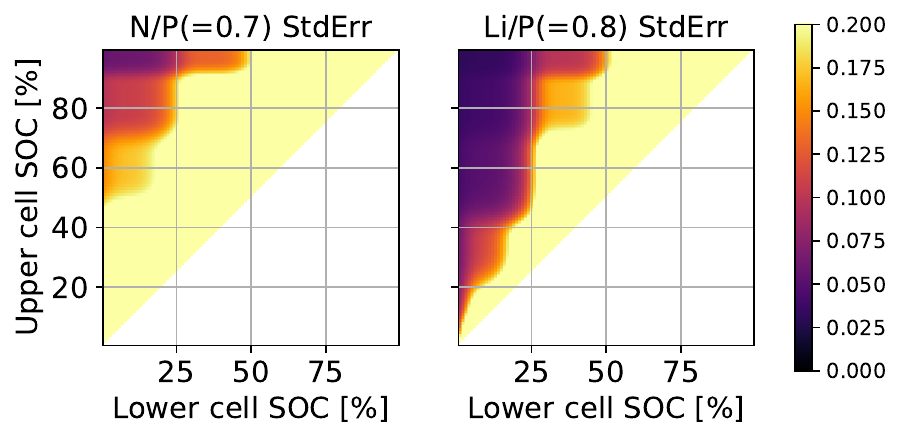}
    \caption{LFP: $r_\mathrm{N/P}=0.7$ and $z^+_0=0.8$}
  \end{subfigure}
  \begin{subfigure}{0.48\textwidth}
    \centering
    \includegraphics[width=\textwidth]{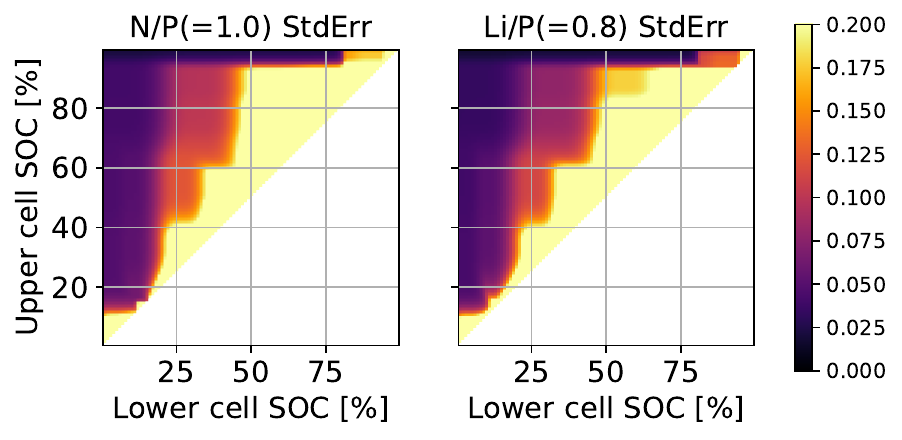}
    \caption{LFP: $r_\mathrm{N/P}=1$ and $z^+_0=0.8$}
  \end{subfigure}
  \begin{subfigure}{0.48\textwidth}
    \centering
    \includegraphics[width=\textwidth]{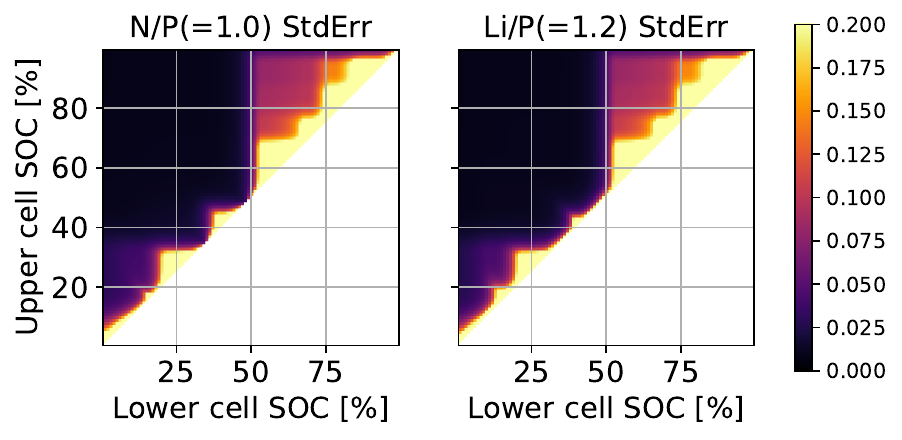}
    \caption{NMC: $r_\mathrm{N/P}=1$ and $z^+_0=1.2$}
  \end{subfigure}
  \begin{subfigure}{0.48\textwidth}
    \centering
    \includegraphics[width=\textwidth]{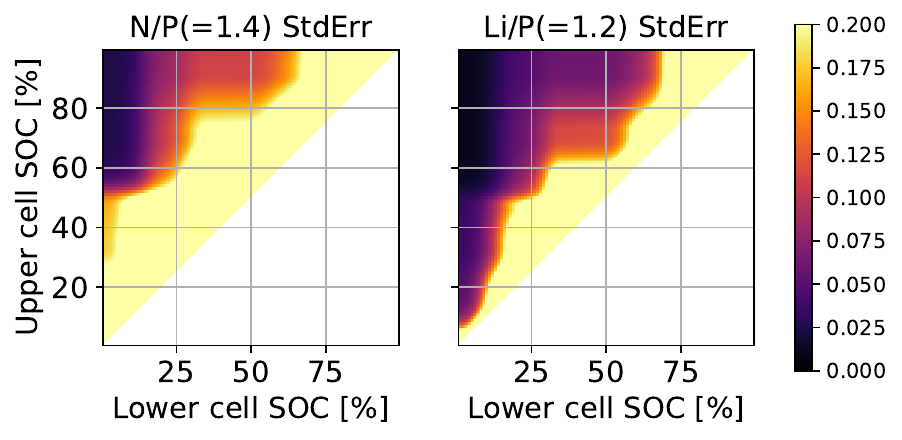}
    \caption{NMC: $r_\mathrm{N/P}=1.4$ and $z^+_0=1.2$}
  \end{subfigure}
  \begin{subfigure}{0.48\textwidth}
    \centering
    \includegraphics[width=\textwidth]{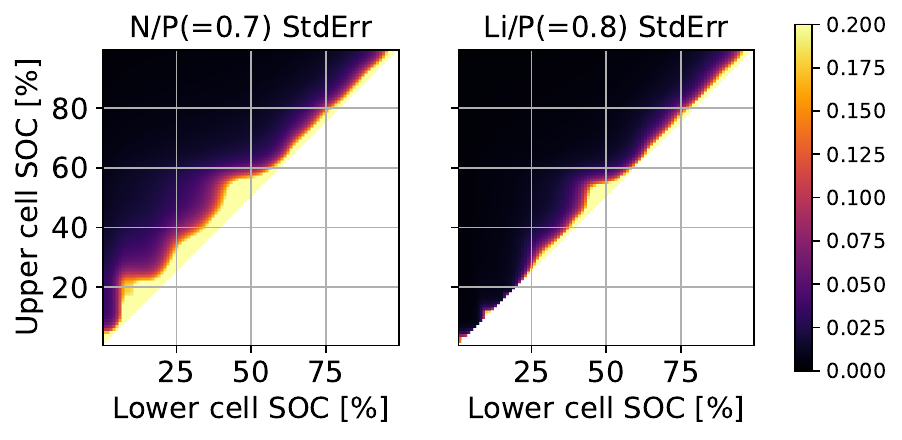}
    \caption{NMC: $r_\mathrm{N/P}=0.7$ and $z^+_0=0.8$}
  \end{subfigure}
  \begin{subfigure}{0.48\textwidth}
    \centering
    \includegraphics[width=\textwidth]{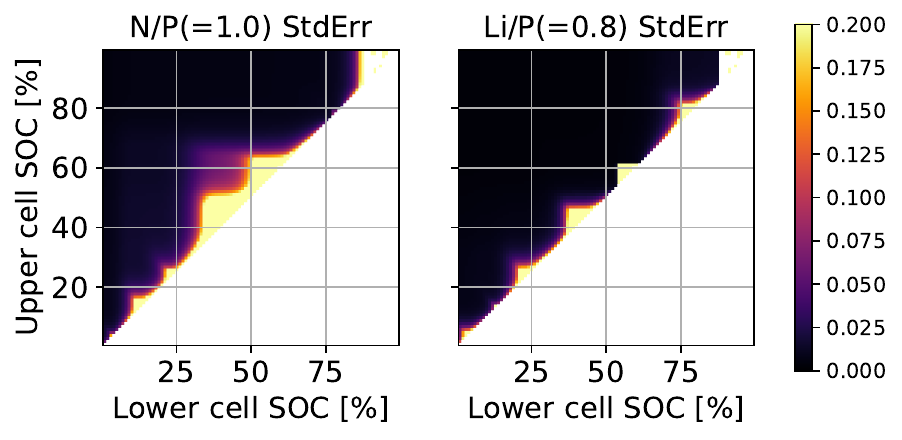}
    \caption{NMC: $r_\mathrm{N/P}=1$ and $z^+_0=0.8$}
  \end{subfigure}
  \caption{
    Standard estimation errors of the N/P ratio $r_\mathrm{N/P}$
    and Li/P ratio $z^+_0$ based on OCV measurements sampled at every $1\%$ SOC
    within any cell SOC window $[z_\mathrm{lower}, z_\mathrm{upper}]$ at least $1\%$ apart.
    The errors are obtained as the square root of
    the diagonals of the inverse Fisher information matrix
    evaluated at different pairs of $(r_\mathrm{N/P},z^+_0)$ values
    corresponding to the four regimes discussed in \cref{subsec:dq}.
    The standard OCV measurement error is assumed to be $\sigma_U=5\mathrm{mV}$.
    {\bf (a,~b,~c,~d)} LFP/MCMB results.
    {\bf (e,~f,~g,~h)} NMC/MCMB counterparts.
  }
  \label{fig:uz-rz-fisher}
\end{figure}

The above results are applied to the same LFP-graphite chemistry as used before
with four different $(r_\mathrm{N/P}, z^+_0)$ pairs
to examine how informative OCV measurements from certain SOC windows are
about the N/P ratio $r_\mathrm{N/P}$ and Li/P ratio $z^+_0$, as shown in \cref{fig:uz-rz-fisher}.
We choose the OCV at SOC $z=1\%,2\%,\ldots,99\%$ as candidate measurements
and traverse every SOC window $[z_\mathrm{lower}, z_\mathrm{upper}]$ with ends from them
that contains at least two measurements.
The standard error of OCV measurements is assumed to be $\sigma_U=5\mathrm{mV}$,
and the standard estimation errors of $r_\mathrm{N/P}$ and $z^+_0$ for each SOC window
obtained by the square root of $\mathrm{diag}(\mbs{C}_{\mbs{\theta}})$
are demonstrated in each sub-figure of \cref{fig:uz-rz-fisher}.
We limit the color bar to the range of $[0.0, 0.2]$ to visualize the practically relevant features,
because a standard error larger than $0.2$
for $r_\mathrm{N/P}$ and $z^+_0$ having a value around 1 means the estimation is almost useless.
On the other hand, the OCV model is not perfect in practice,
and model errors will limit the estimation accuracy in the first place,
so we consider an estimation error of under $0.01$ in this ideal setup as more than enough.
Hence, in \cref{fig:uz-rz-fisher}, a dark zone indicates small-enough errors
while a bright zone implies ineffective estimation.

Note that if one SOC window is nested in another,
the OCV measurements of the latter also contain all those in the former,
so the estimation error is monotonically decreasing vertically and increasing horizontally.
Moreover, we see the errors having a sharp change in certain SOC,
indicating the OCV measurements at such SOC to be particularly
sensitive to $r_\mathrm{N/P}$ and $z^+_0$,
agreeing with \cref{subfig:ocv-n2p-d,subfig:ocv-li2p-d}.
We can also observe that estimating $r_\mathrm{N/P}$ is harder than estimating $z^+_0$
in most cases here,
which is consistent with \cref{subfig:ocv-rz-ue-lfp,subfig:ocv-rz-ue-nmc},
and OCV values at lower SOC are overall more informative than those at high SOC.
Finally, it is also obvious that the SOH parameter identifiability
is significantly higher for NMC than for LFP,
aligned with the larger empirical sensitivity to SOH parameter perturbation
we have visualized previously.

\subsection{Lithium Inventory and Active Material Identifiability
from Charge-Based OCV Measurements by Fisher Information}
\label{subsec:fisher-qc}

The previous case of measuring the SOC-based $U_\mathrm{OCV}(z)$ requires the whole OCV curve,
which may not be available in some scenarios.
Given only partial OCV curve measurements,
we must work with the raw data of OCV versus charge throughput,
say $Q_\mathrm{c}$ counted from some initial state at $U_\mathrm{OCV}=U_\mathrm{ini}$
and is positive in the charging direction,
because relating $Q_\mathrm{c}$ to the full cell SOC $z$ requires knowing
the SOH parameters of
the lithium inventory $\hat{Q}^\mathrm{Li}_\mathrm{max}$
and the NE and PE active material amount $\hat{Q}^\pm_\mathrm{max}$,
or their equivalent re-parametrization,
which are being sought after in the first place.

Suppose $Q_\mathrm{c}=0$ corresponds to some known initial voltage $U_\mathrm{OCV}=U_\mathrm{ini}$.
The charge-based OCV relation becomes
\begin{equation}\label{eq:ocv-Qc-ini}
  U_\mathrm{OCV}(Q_\mathrm{c})
  = U^+_\mathrm{OCP}(z^+_\mathrm{ini}(z^-_\mathrm{ini}(U_\mathrm{ini}))
  - Q_\mathrm{c}/\hat{Q}_\mathrm{max}^+)
  - U^-_\mathrm{OCP}(z^-_\mathrm{ini}(U_\mathrm{ini}) + Q_\mathrm{c}/\hat{Q}_\mathrm{max}^-),
\end{equation}
where the $z^+_\mathrm{ini}(z^-_\mathrm{ini})$ dependence comes from \cref{eq:z+z-}
while the $z^-_\mathrm{ini}(U_\mathrm{ini})$ dependence is inverted from \cref{eq:full-ocv-z-},
both of which parametrized by the N/P ratio $r_\mathrm{N/P}$ and Li/P ratio $z^+_0$.

Note that if a quantity $f$ is parametrized by
$r_\mathrm{N/P}$ and $z^+_0$,
since
\[
  f(r_\mathrm{N/P}, z^+_0)
  = f(\hat{Q}^-_\mathrm{max}/\hat{Q}^+_\mathrm{max},
  \hat{Q}^\mathrm{Li}_\mathrm{max}/\hat{Q}^+_\mathrm{max}),
\]
$f$ can also be viewed as being, though less parsimoniously, parametrized by
$\hat{Q}^\mathrm{Li}_\mathrm{max}$ and $\hat{Q}^\pm_\mathrm{max}$.
Hence,
\begin{equation}
  \PDer{f}{\hat{Q}^\mathrm{Li}_\mathrm{max}}
  = \frac{1}{\hat{Q}^+_\mathrm{max}}\PDer{f}{z^+_0}, \quad
  \frac{\Par f}{\Par \hat{Q}^-_\mathrm{max}}
  = \frac{1}{\hat{Q}^+_\mathrm{max}}\frac{\Par f}{\Par r_\mathrm{N/P}}, \quad
  \frac{\Par f}{\Par \hat{Q}^\mathrm{Li}_\mathrm{max}}
  = -\frac{\hat{Q}^-_\mathrm{max}}{(\hat{Q}^+_\mathrm{max})^2}\frac{\Par f}{\Par r_\mathrm{N/P}}
  -\frac{\hat{Q}^\mathrm{Li}_\mathrm{max}}{(\hat{Q}^+_\mathrm{max})^2}\frac{\Par f}{\Par z^+_0}.
\end{equation}

With the aids of the above relations,
we can obtain the sensitivity gradient of $U_\mathrm{OCV}(Q_\mathrm{c})$
with respect to $\hat{Q}^\mathrm{Li}_\mathrm{max}$ and $\hat{Q}^\pm_\mathrm{max}$:
\begin{equation}
  \PDer{U_\mathrm{OCV}(Q_\mathrm{c})}{\hat{Q}^\mathrm{Li}_\mathrm{max}}
  = \frac{\lambda^-_\mathrm{ini}}{\hat{Q}^+_\mathrm{max}} \Der{U^+_\mathrm{OCP}}{z^+}
  - \frac{\lambda^+_\mathrm{ini}}{\hat{Q}^-_\mathrm{max}} \Der{U^-_\mathrm{OCP}}{z^-},
\end{equation}
\begin{equation}
  \PDer{U_\mathrm{OCV}(Q_\mathrm{c})}{\hat{Q}^-_\mathrm{max}}
  = -\frac{z^-_\mathrm{ini} \lambda^-_\mathrm{ini}}{\hat{Q}^+_\mathrm{max}} \Der{U^+_\mathrm{OCP}}{z^+}
  + \frac{z^-_\mathrm{ini} \lambda^+_\mathrm{ini} + Q_\mathrm{c}/\hat{Q}^-_\mathrm{max}}
  {\hat{Q}^-_\mathrm{max}} \Der{U^-_\mathrm{OCP}}{z^-},
\end{equation}
\begin{equation}
  \PDer{U_\mathrm{OCV}(Q_\mathrm{c})}{\hat{Q}^+_\mathrm{max}}
  = \frac{-z^+_\mathrm{ini} \lambda^-_\mathrm{ini} + Q_\mathrm{c}/\hat{Q}^+_\mathrm{max}}
  {\hat{Q}^+_\mathrm{max}} \Der{U^+_\mathrm{OCP}}{z^+}
  + \frac{z^+_\mathrm{ini} \lambda^+_\mathrm{ini}}
  {\hat{Q}^-_\mathrm{max}} \Der{U^-_\mathrm{OCP}}{z^-}.
\end{equation}

With the sensitivity gradient calculated,
we can now use the same approach of Fisher information matrix to study
the identifiability of $\hat{Q}^\mathrm{Li}_\mathrm{max}$ and $\hat{Q}^\pm_\mathrm{max}$
from partial OCV measurements from a certain SOC window.
For testing purpose,
we choose the charge throughput increment $\Delta Q_\mathrm{c}$ to be 1\% of the ground truth
of the cell total capacity
so $\Delta Q_\mathrm{c}$ translates to $1\%$ cell SOC change.
Moreover, we choose the starting voltage $U_\mathrm{ini}$ to be the ground truth of cell OCV
at each integer cell SOC percentage.
This way, each partial segment of OCV measurements can be normalized into an SOC window
as in the previous case for more intuitive visualization.

\begin{figure}[!htbp]
  \centering
  \begin{subfigure}{0.48\textwidth}
    \centering
    \includegraphics[width=\textwidth]{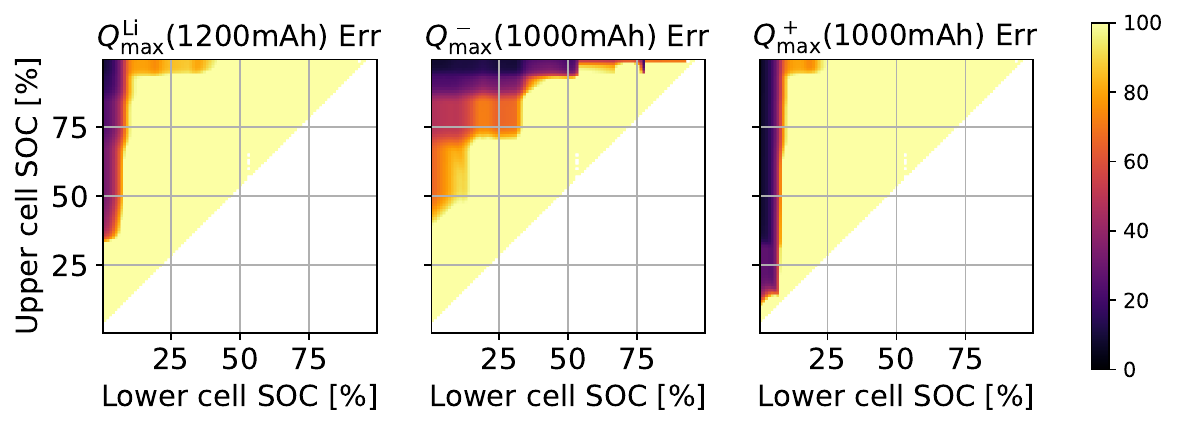}
    \caption{LFP: $r_\mathrm{N/P}=1$ and $z^+_0=1.2$}
  \end{subfigure}
  \begin{subfigure}{0.48\textwidth}
    \centering
    \includegraphics[width=\textwidth]{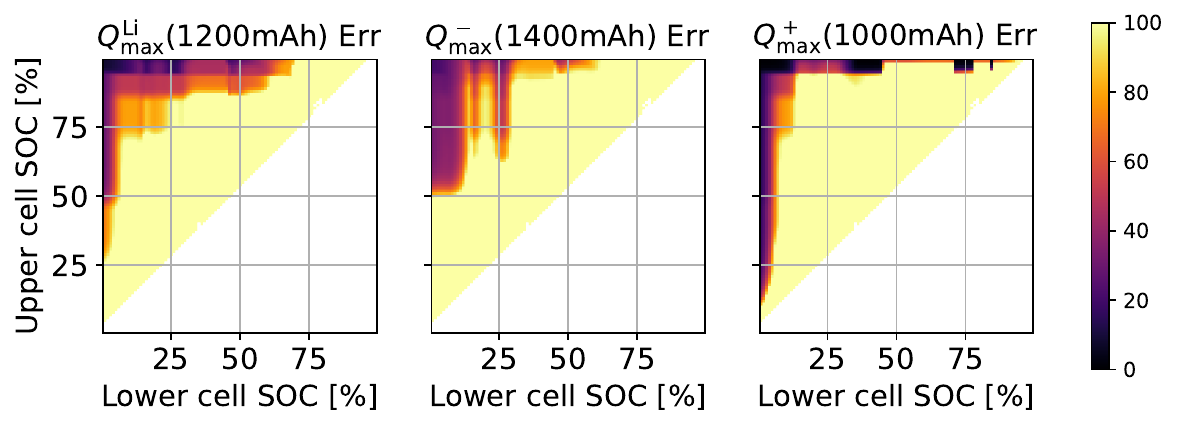}
    \caption{LFP: $r_\mathrm{N/P}=1.4$ and $z^+_0=1.2$}
  \end{subfigure}
  \begin{subfigure}{0.48\textwidth}
    \centering
    \includegraphics[width=\textwidth]{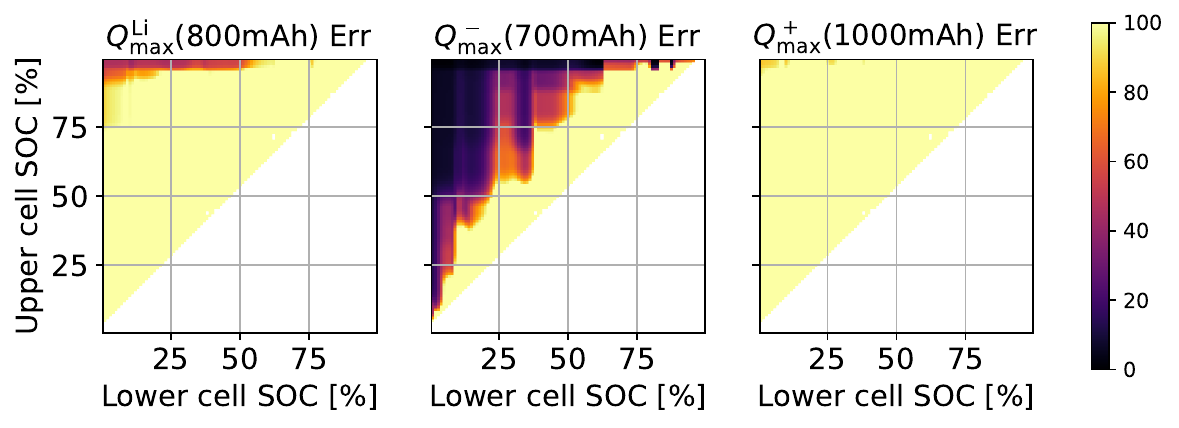}
    \caption{LFP: $r_\mathrm{N/P}=0.7$ and $z^+_0=0.8$}
  \end{subfigure}
  \begin{subfigure}{0.48\textwidth}
    \centering
    \includegraphics[width=\textwidth]{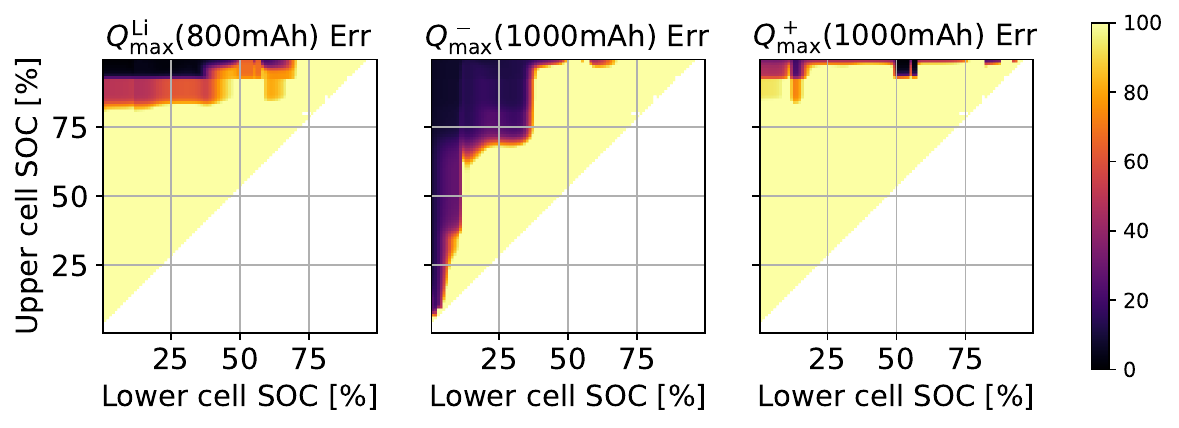}
    \caption{LFP: $r_\mathrm{N/P}=1$ and $z^+_0=0.8$}
  \end{subfigure}
  \begin{subfigure}{0.48\textwidth}
    \centering
    \includegraphics[width=\textwidth]{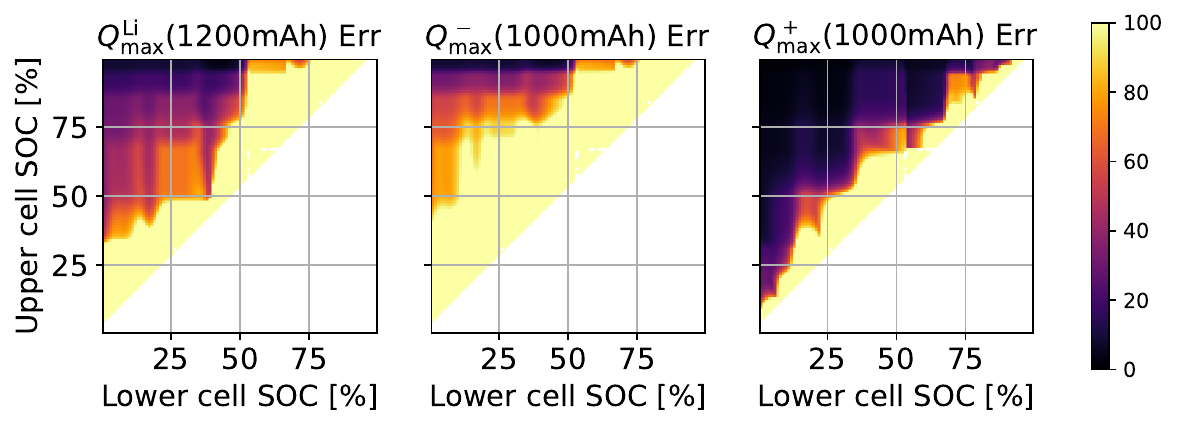}
    \caption{NMC: $r_\mathrm{N/P}=1$ and $z^+_0=1.2$}
  \end{subfigure}
  \begin{subfigure}{0.48\textwidth}
    \centering
    \includegraphics[width=\textwidth]{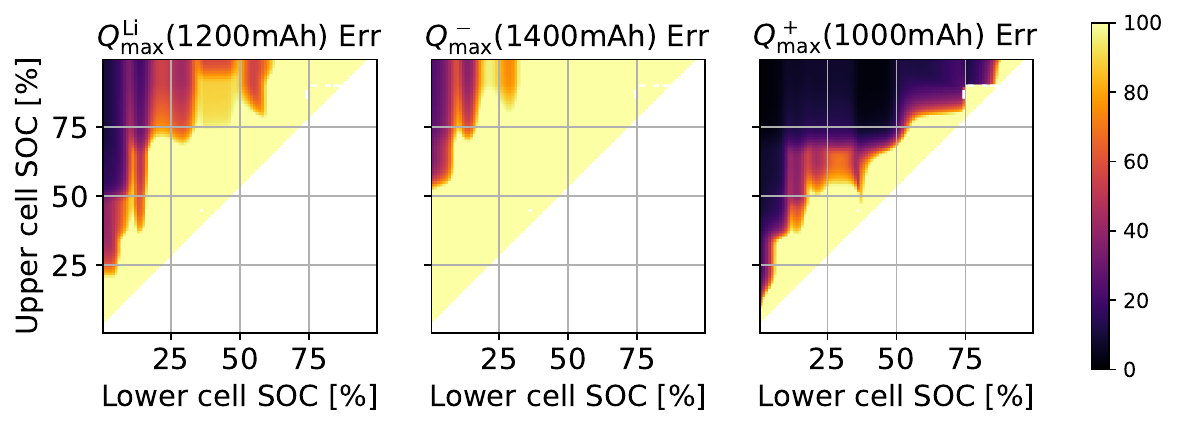}
    \caption{NMC: $r_\mathrm{N/P}=1.4$ and $z^+_0=1.2$}
  \end{subfigure}
  \begin{subfigure}{0.48\textwidth}
    \centering
    \includegraphics[width=\textwidth]{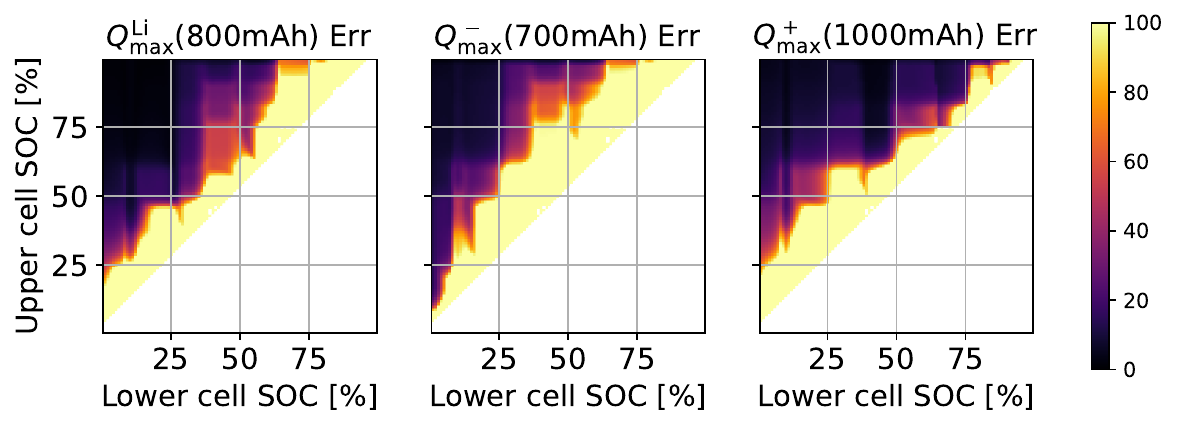}
    \caption{NMC: $r_\mathrm{N/P}=0.7$ and $z^+_0=0.8$}
  \end{subfigure}
  \begin{subfigure}{0.48\textwidth}
    \centering
    \includegraphics[width=\textwidth]{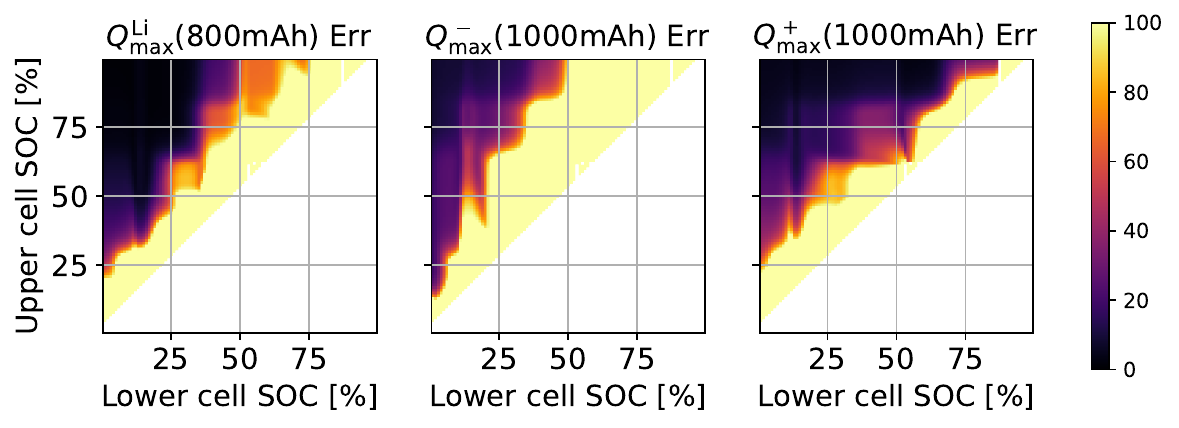}
    \caption{NMC: $r_\mathrm{N/P}=1$ and $z^+_0=0.8$}
  \end{subfigure}
  \caption{
    Standard estimation errors (in mAh shown by the color bar) of the lithium inventory $\hat{Q}^\mathrm{Li}_\mathrm{max}$,
    NE active materials $\hat{Q}^-_\mathrm{max}$,
    and PE active materials $\hat{Q}^+_\mathrm{max}$
    based on OCV measurements sampled at every charge throughput increment $\Delta Q_\mathrm{c}$
    amounting to $1\%$ cell SOC change.
    Each charge throughput window of voltage measurements
    is normalized into a cell SOC window $[z_\mathrm{lower}, z_\mathrm{upper}]$ at least $1\%$ apart.
    The errors are obtained as the square root of
    the diagonals of the inverse Fisher information matrix
    evaluated at different pairs of $(r_\mathrm{N/P},z^+_0)$ values
    corresponding to the four regimes discussed in \cref{subsec:dq},
    with $\hat{Q}^+_\mathrm{max}=1\mathrm{Ah}$ being fixed in all cases.
    The standard OCV measurement error is assumed to be $\sigma_U=5\mathrm{mV}$.
    {\bf (a,~b,~c,~d)} LFP/MCMB results.
    {\bf (e,~f,~g,~h)} NMC/MCMB counterparts.
  }
  \label{fig:uqc-allq-fisher}
\end{figure}

The standard estimation errors are shown in \cref{fig:uqc-allq-fisher}
using the same layout as in \cref{fig:uz-rz-fisher}.
The overall patterns are similar to those in \cref{fig:uz-rz-fisher},
except that the estimation errors are,
though still monotonely increasing along the upper SOC axis,
no longer strictly monotonely decreasing along the lower SOC axis,
as indicated by the pattern of some vertical downward pointing ``fingers''.
This is because the voltage measurement at a particular cell SOC $z$ is interpreted differently
when the initial voltage changes,
so it is not really the same measurement under such different scenarios.
For example, if the starting OCV is $U_\mathrm{ini}=U_\mathrm{OCV}(z=20\%)$,
an OCV measurement at $z=60\%$ is interpreted as the OCV reached
when the cell is charged from $U_\mathrm{OCV}=U_\mathrm{ini}$
by a charge throughput of $(60\% - 20\%)\hat{Q}_\mathrm{max}=40\%\hat{Q}_\mathrm{max}$.
Obviously, such meaning will change along the starting OCV,
so the information born by such measurements regarding the underlying SOH parameters are not identical.
Such quantities of voltage change across certain charge throughput
have also been extracted as features to estimate cell total capacity in literature.

Besides, we can also observe that the identifiability of the active material amount in an electrode
strongly depends on the intrinsic OCP relation so this particular material.
For example, the active material of LFP tends to be hard to identify,
likely due to the flatness and lack of features in its OCP.
In contrast, the active material of NMC is much more identifiable.
Moreover, the identifiability of graphite shares a rather similar pattern
whether it is matched with LFP or NMC to form a cell.

\section{Conclusions, Discussions, and Future Work}

In this work, we study the electrode-specific OCV model assembled from the PE and NE OCP
and re-parametrize a popular formulation in literature
by the dimensionless N/P ratio $r_\mathrm{N/P}=\hat{Q}_\mathrm{max}^-/\hat{Q}_\mathrm{max}^+$
and the Li/P ratio $z^+_0=\hat{Q}_\mathrm{max}^\mathrm{Li}/\hat{Q}_\mathrm{max}^+$,
which are independent, cutoff-voltage-agnostic,
pristine-condition-agnostic,
and intuitively linked to the lithium inventory $\hat{Q}_\mathrm{max}^\mathrm{Li}$
and active material amounts $\hat{Q}_\mathrm{max}^\pm$.
Estimating them thus gives us important information about the degradation mode of
LLI, LAMn, and LAMp.
Along the way, we also clarify several confusions regarding
electrode stoichiometry $x^\pm$ versus electrode SOC $z^\pm$,
precise meaning of cell SOC $z$ and SOH,
and their relations to the artificially specified upper and lower cutoff voltage.

We then analytically derive the sensitivity gradients of the electrode SOC limits
and $z$-based cell OCV $U_\mathrm{OCV}(z)$ with respect to the N/P and Li/P ratio.
Such analytic results not only offer a closed-form expression
to easily calculate these derivatives,
but the form of the expressions also provides insights into
how each factor affects the parametric sensitivity.
In particular,
the concept of PE and NE DV fraction naturally emerges in the derivation,
which characterize which electrode is dominating the OCV changes at each cell SOC
and are instrumental in propagating the effects of a fixed upper and lower cutoff voltage
into the sensitivity gradients.
The local sensitivity delineated by these parametric derivatives
are visualized and verified by simulations and comparison against empirical sensitivity results.

Furthermore, we derive the sensitivity gradients of cell total capacity $\hat{Q}_\mathrm{max}$
with respect to the lithium inventory $\hat{Q}_\mathrm{max}^\mathrm{Li}$
and active material amounts $\hat{Q}_\mathrm{max}^\pm$.
These dimensionless derivatives intuitively indicate
to what extent each degradation mode of LLI, LAMn, and LAMp
is limiting the apparent total capacity.
Visualizing the pattern of their dependence on the N/P and Li/P ratio
reveals four characteristic regimes regarding their identifiability.

We also demonstrate a direct application of these analytic results in practice,
which regards estimating the N/P and Li/P ratio from complete OCV curve measurements using NLS
and estimating lithium inventory and active material amounts from partial OCV measurements.
We show how to obtain an NLS estimator error covariance
from the Fisher information matrix,
which directly depends on the sensitivity gradients,
and how such information
helps us choose an SOC window that yields informative OCV measurements.

We want to emphasize that any statements based on sensitivity gradients
are only valid locally,
i.e.~valid for a small range of parameters around the point at which the gradients are evaluated.
More empirical tools from statistical inference will be needed to discern global sensitivity,
and we will report our findings in future work.

Another limitation is that we have only discussed inferring degradation modes
from OCV measurements,
which has a relatively restricted scope of application in practice.
To go beyond OCV measurements,
we need to somehow estimate the cell SOC or OCV from terminal voltage
recorded in finite-current operation.
This can either be done by simple $IR_0$ corrections
based on equivalent internal resistance $R_0$ \citep{schmitt_capacity_2023},
or by more sophisticated model-based filtering
such as in a battery management system \citep{plett_battery_2016}.



\section*{Acknowledgements}

J.L. and E.K. acknowledge funding by Agency for Science, Technology and Research (A*STAR)
under the Career Development Fund (C210112037).




\bibliographystyle{elsarticle-num-names}
\bibliography{reference}





\end{document}